\documentclass{tcibook}
\usepackage{fancyhea}
\usepackage{work}
\usepackage{bm}       %    enables bold math symbols  e.g.  \bm{\gamma}
\usepackage{graphicx}
\usepackage{hyperref}      % hypertext links %%ARXIV

\newcommand{\nc}{\newcommand}  

\def\Acknowledgements{\bigskip  \bigskip \begin{center} \begin{large}
             \bf ACKNOWLEDGEMENTS \end{large}\end{center}}

\def\beq{\begin{equation}}
\def\eeq#1{\label{#1}\end{equation}}
\def\eeqn{\end{equation}}

\newenvironment{Eqnarray}%
   {\arraycolsep 0.14em\begin{eqnarray}}{\end{eqnarray}}
\def\beqa{\begin{Eqnarray}}
\def\eeqa#1{\label{#1}\end{Eqnarray}}
\def\eeqan{\end{Eqnarray}}

\nc{\ra}{\rightarrow}  
\nc{\slsh}{\slash\hspace*{-0.22cm}}
\def\Re{{\cal R \mskip-4mu \lower.1ex \hbox{\it e}\,}}
\def\Im{{\cal I \mskip-5mu \lower.1ex \hbox{\it m}\,}}

\nc{\vev}[1]{ \left\langle {#1} \right\rangle }
\nc{\bra}[1]{ \langle {#1} | }
\nc{\ket}[1]{ | {#1} \rangle }
\nc{\fb}{\,{\rm fb}^{-1}}
\nc{\ev}{{\rm eV}}
\nc{\kev}{{\rm keV}}
\nc{\Mev}{{\rm MeV}}
\nc{\gev}{{\rm GeV}}
\nc{\tev}{{\rm TeV}}
\nc{\mev}{{\rm MeV}}

\def\del{\partial}
\def\Dslash{\not{\hbox{\kern-4pt $D$}}}
\def\dslash{\not{\hbox{\kern-2pt $\del$}}}
\def\pslash{\not{\hbox{\kern-2pt $p$}}}
\def\ETmiss{ \not{\hbox{\kern-4pt $E$}}_T }

\def\msb{{\bar{\ssstyle M \kern -1pt S}}}

\begin{document}

\def\bibname{References}
\bibliographystyle{plain}

\raggedbottom

\pagenumbering{roman}

\parindent=0pt
\parskip=8pt
\setlength{\evensidemargin}{0pt}
\setlength{\oddsidemargin}{0pt}
\setlength{\marginparsep}{0.0in}
\setlength{\marginparwidth}{0.0in}
\marginparpush=0pt

% The content begins here

\pagenumbering{arabic}

\newcommand {\met}{\ensuremath{E^{\rm miss}_{\rm T}}}
\newcommand {\pt}{\ensuremath{p_{\rm T}}}
\newcommand {\hadt}{\ensuremath{H_{\rm T}}}
\newcommand {\hgt}{\ensuremath{H^{*}_{\rm T}}}

\renewcommand{\chapname}{chap:intro_}
\renewcommand{\chapterdir}{.}
\renewcommand{\arraystretch}{1.25}
\addtolength{\arraycolsep}{-3pt}

%%%%%%%%%%%%%%%%%%%%%%%%%%%%%%%%%%%%%%%%%%%%%%%%%%%
%%%%%%%%%%%%%%%%%%%%%%%%%%%%%%%%%%%%%%%%%%%%%%%%%%%
%%%     All of your files should be in a subdirectory.  Here the
%%%     subdirectory is called Magnetism  .   The title of your
%%%     report should be   wgreport.tex in that subdirectory.  Input
%%%     that file here
%%%%%%%%%%%%%%%%%%%%%%%%%%%%%%%%%%%%%%%%%%%%%%%%%%%%
%%%%%%%%%%%%%%%%%%%%%%%%%%%%%%%%%%%%%%%%%%%%%%%%%%%

%%%%%% Simulations Chapter  %%%%%%%%%%%%%%%%
 
\chapter{Snowmass Energy Frontier Simulations}
\label{chap:Simulations}

%%%%%%%%%%%%%%%%%%%%%%%%%%%%%%%%%%%%%%%%%%%%%%%%%%%%%%%%%%%
%%%%%%%%%%%%%%%%%%%%%%%%%%%%%%%%%%%%%%%%%%%%%%%%%%%%%%%%%%%
%%%%%%%%%%%%%%%%%%%%%%%%%%%%%%%%%%%%%%%%%%%%%%%%%%%%%%%%%%%
%%%%%%%%%%%%%%%%%%%%%%%%%%%%%%%%%%%%%%%%%%%%%%%%%%%%%%%%%%%
\begin{center}\begin{boldmath}

% list of HSPAW authors

%\hyphenpenalty 10000

\begin{center}

\begin{large} {\bf Conveners: Sergei Chekanov, Sanjay Padhi} \end{large}

{Jacob Anderson$^{\bf 2}$},
{Aram Avetisyan$^{\bf 1}$},
{Raymond Brock$^{\bf 3}$},
{Sergei Chekanov$^{\bf 4}$},
{Timothy Cohen$^{\bf 5}$},
{Nitish Dhingra$^{\bf 6}$},
{James Dolen$^{\bf 14}$},
{James Hirschauer$^{\bf 2}$},
{Kiel Howe$^{\bf 7}$},
{Ashutosh Kotwal$^{\bf 8}$},
{Tom LeCompte$^{\bf 4}$},
{Sudhir Malik$^{\bf 9}$},
{Patricia Mcbride$^{\bf 2}$},
{Kalanand Mishra$^{\bf 2}$},
{Meenakshi Narain$^{\bf 10}$},
{Jim Olsen$^{\bf 11}$},
{Sanjay Padhi$^{\bf 12}$},
{Michael E. Peskin$^{\bf 5}$},
{John Stupak III$^{\bf 13}$},
and
{Jay G. Wacker$^{\bf 5}$}

%\affiliation{
$^{\bf 1}$  Boston University, Boston, MA, USA\\
$^{\bf 2}$  Fermi National Accelerator Laboratory, Batavia, USA\\
$^{\bf 3}$  Michigan State University East Lansing, MI, USA\\
$^{\bf 4}$  Argonne National Laboratory, Argonne, USA\\
$^{\bf 5}$  SLAC National Accelerator Laboratory, Menlo Park, USA\\
$^{\bf 6}$  Panjab University, Chandigarh, India\\
$^{\bf 7}$  Stanford Institute for Theoretical Physics, Stanford University, Stanford, USA\\
$^{\bf 8}$  Duke University, Durham, USA\\
$^{\bf 9}$  University of Nebraska-Lincoln, Lincoln, USA\\
$^{\bf 10}$  Brown University, Providence, USA\\
$^{\bf 11}$  Princeton University, Princeton, USA\\
$^{\bf 12}$  University of California, San Diego, USA\\
$^{\bf 13}$ Purdue University Calumet, Hammond, USA\\
$^{\bf 14}$ SUNY Buffalo, USA\\

%}

\end{center}

%\hyphenpenalty 1000

%Conveners are also listed separately in authorlist.tex

\end{boldmath}\end{center}

%%%%%%%%%%%%%%%%%%%%%%%%%%%%%%%%%%%%%%%%%%%%%%%%%%%%%%%%%%%
%%%%%%%%%%%%%%%%%%%%%%%%%%%%%%%%%%%%%%%%%%%%%%%%%%%%%%%%%%%
%%%%%%%%%%%%%%%%%%%%%%%%%%%%%%%%%%%%%%%%%%%%%%%%%%%%%%%%%%%
%%%%%%%%%%%%%%%%%%%%%%%%%%%%%%%%%%%%%%%%%%%%%%%%%%%%%%%%%%%
%\begin{abstract}
This document describes the simulation framework used in the Snowmass Energy 
Frontier studies for future Hadron Colliders. An overview of event generation 
with {\sc Madgraph}5 along with parton shower and hadronization with 
{\sc Pythia}6 is followed by a detailed description of pile-up and detector 
simulation with {\sc Delphes}3.  Details of event generation are included 
in a companion paper cited within this paper.  The input parametrization 
is chosen to reflect the best object performance  
expected from the future ATLAS and CMS experiments; this 
is referred to as the ``Combined Snowmass Detector''. We perform simulations 
of $pp$ interactions at center-of-mass energies $\sqrt{s}=$ 14, 33, and 100 
TeV with 0, 50, and 140 additional $pp$ pile-up interactions. 
The object performance with multi-TeV $pp$ collisions are studied for the 
first time using large pile-up interactions.
%\end{abstract}

\section{Introduction}
\label{sec:sim-intro}
The Large Hadron Collider (LHC) at CERN, which collided protons at $\sqrt{s}=8$ TeV center-of-mass energy in 2012, is the most powerful particle accelerator ever built and the forefront of the energy frontier.  The ATLAS~\cite{atlasdet} and CMS~\cite{cmsdet} multipurpose detectors record the $pp$ collisions 
of the LHC, reconstructing decay products with excellent efficiency and 
resolution.  For precise analysis of collision data, the ATLAS and CMS 
collaborations 
fully simulate interactions of particles with their detectors using 
GEANT4~\cite{geant}. In addition, the collaborations use sophisticated 
algorithms for reconstruction of events, particles, and physical quantities 
in simulated and collision data. 

Long-term planning studies in high energy physics, which seek to 
compare various scenarios of center-of-mass energy, numbers of 
$pp$ interactions per bunch crossing (pile-up), and integrated luminosity, 
require a fast and realistic estimation of the expected detector performance 
as a function of these variables. In this context, we use simulation
parameters that are expected to reflect the best expected 
performance from the ATLAS and CMS detectors, this is referred to as 
the ``Combined Snowmass LHC detector''.
%In this context, we use the 
%{\sc Delphes}~\cite{Delphes} framework as a pile-up and detector 
%simulation with input parameters chosen to reflect the best expected performance
%from the ATLAS and CMS sub-detectors, this is referred to as the ``Combined Snowmass LHC detector''.
%the combined best expected performance of the future ATLAS and CMS experiments
%; this is thus referred to as the ``Combined Snowmass LHC detector''.
The Combined Snowmass LHC detector and the simulations based on it
are meant to provide a single transparent fast-simulation framework for
studies of the capabilities of hadron collider experiments at
14, 33, and 100 TeV, including effects due to in-time pile-up at high luminosity.
This framework uses the {\sc Delphes}~\cite{Delphes} fast-simulation
model with inputs from publicly available detector and performance
parameters~\cite{cms-tdr,atlas-loi, cms-elec-dp, cms-muon-dp} that were expected, in the spring of 2013, to reflect the
best performance of the future ATLAS and CMS sub-detector components.
The framework does not take into account the subsequent evolution of the
ATLAS and CMS detector designs, and it is not at the level of a full detailed
Geant-based simulation.  Results derived from this framework are not
official ATLAS or CMS simulation results and should not be considered
on the same footing.

The main goal of the present long-term planning exercise, the Snowmass 
Community Summer Study, is to fully develop the long-term physics aspirations 
of the community. The Snowmass narrative will communicate the opportunities 
for discovery in high energy physics to the broader scientific community and 
to the United States government. The studies associated with the energy 
frontier address the physics potential of $pp$ interactions at $\sqrt{s} = $14, 33 and 100 TeV with 0, 50 and 140 pile-up interactions, and integrated luminosities of 300 and 3000 fb$^{-1}$. A bunch spacing of 25 ns is 
assumed for these studies.

Section~\ref{sec:sim-coll} discusses future hadron collider scenarios and the expected luminosity evolution.  
In Sec.~\ref{sec:sim-tools}, we describe simulation tools including the background processes in Sec.~\ref{sec:sim-bkg} and {\sc Delphes} in Sec.~\ref{sec:delphes}.  In Sec.~\ref{sec:sim-perf}, we describe the performance of the simulation, followed by summary and conclusion in Sec.~\ref{sec:sim-stat}.

\section{Scenarios:  Future hadron colliders and detectors}
\label{sec:sim-coll}

The energies and pile-up scenarios chosen for simulation are intended to represent real proposed hadron colliders.
The upcoming LHC run, which includes the Phase I upgrades of the CMS and ATLAS detectors, is expected to 
provide 300~fb$^{-1}$ of 14~TeV $pp$ collisions with a mean $<\mu> = $50 pile-up interactions.   The HL-LHC 
is planned to commence following a long shut down in 2022 and will provide 3000-fb$^{-1}$ at 14 TeV over the 
following decade.  Other proposed hadron colliders include the HE-LHC, which would use the same tunnel as 
the LHC with stronger dipole magnets to achieve 33~TeV $pp$ collisions, and the VHE-LHC, which would require a 
new larger ring to produce 100 TeV $pp$ collisions. Table~\ref{tab:colliders} summarizes various hadron 
collider scenarios that are used in this study.
\begin{table}[htbp]
\begin{center}
\begin{tabular}{l|ccccc}
\hline\hline
Parameter           &  LHC  &  HL-LHC  &  HE-LHC  & VLHC \\
\hline
Energy [TeV]          &  14   &   14   &   33   & 100 \\ 
Mean additional interactions per crossing ($<\mu>$) &  50   &  140   &  140   & 140 \\
Integrated Luminosity [fb$^{-1}$] & 300   & 3000   &  3000  & 3000 \\
\hline\hline
\end{tabular}
\caption{Benchmark hadron collider scenarios under study with the simulated samples described in this document.}
\label{tab:colliders}
\end{center}
\end{table}

%%%%%%%%%%%%%%%%%%%%%%%%%%%%%%%%%%%%%%%%%%%%%%%%%%%%%
\begin{figure}[htb]
\begin{center}
\includegraphics[width=0.6\hsize]{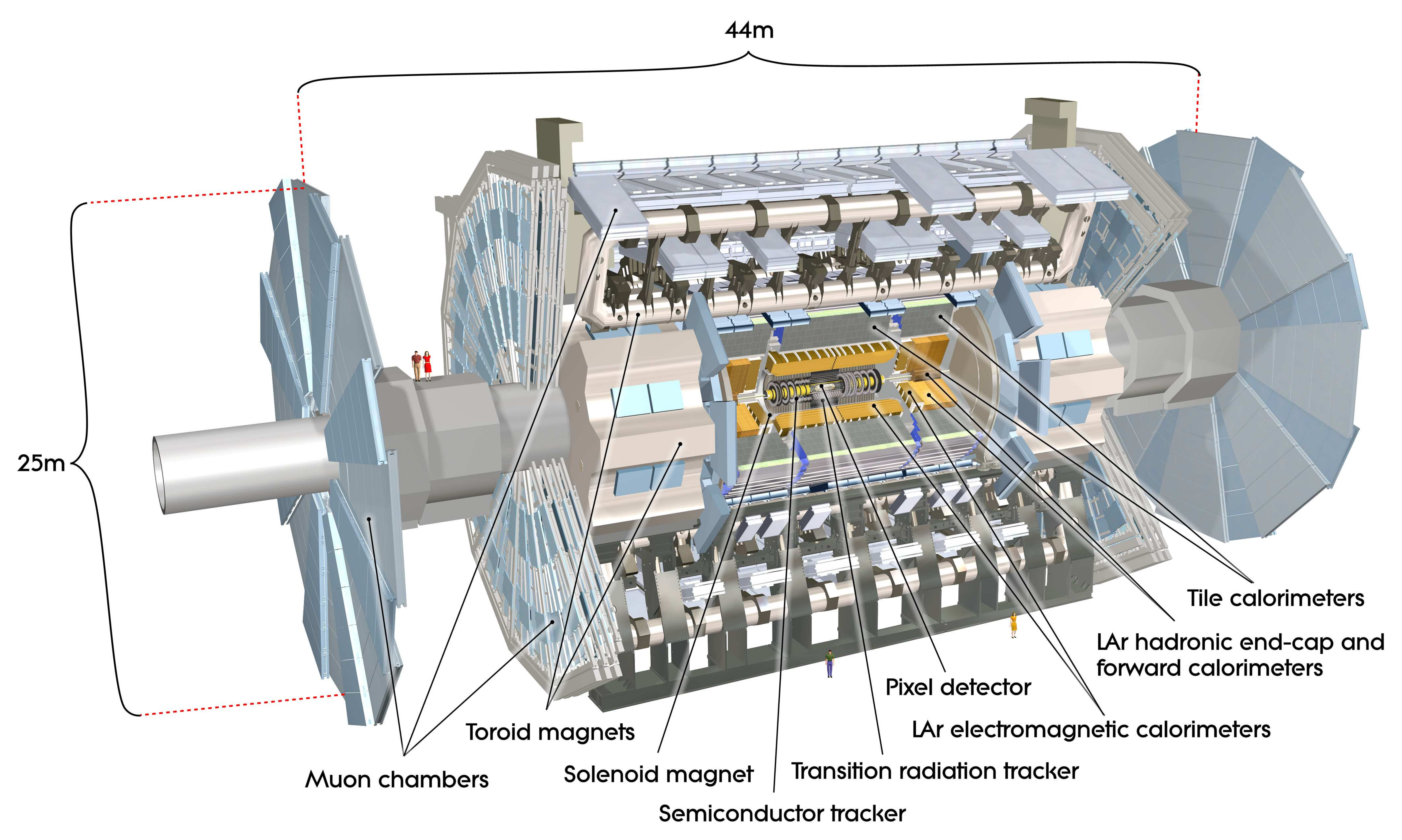}
\includegraphics[width=0.6\hsize]{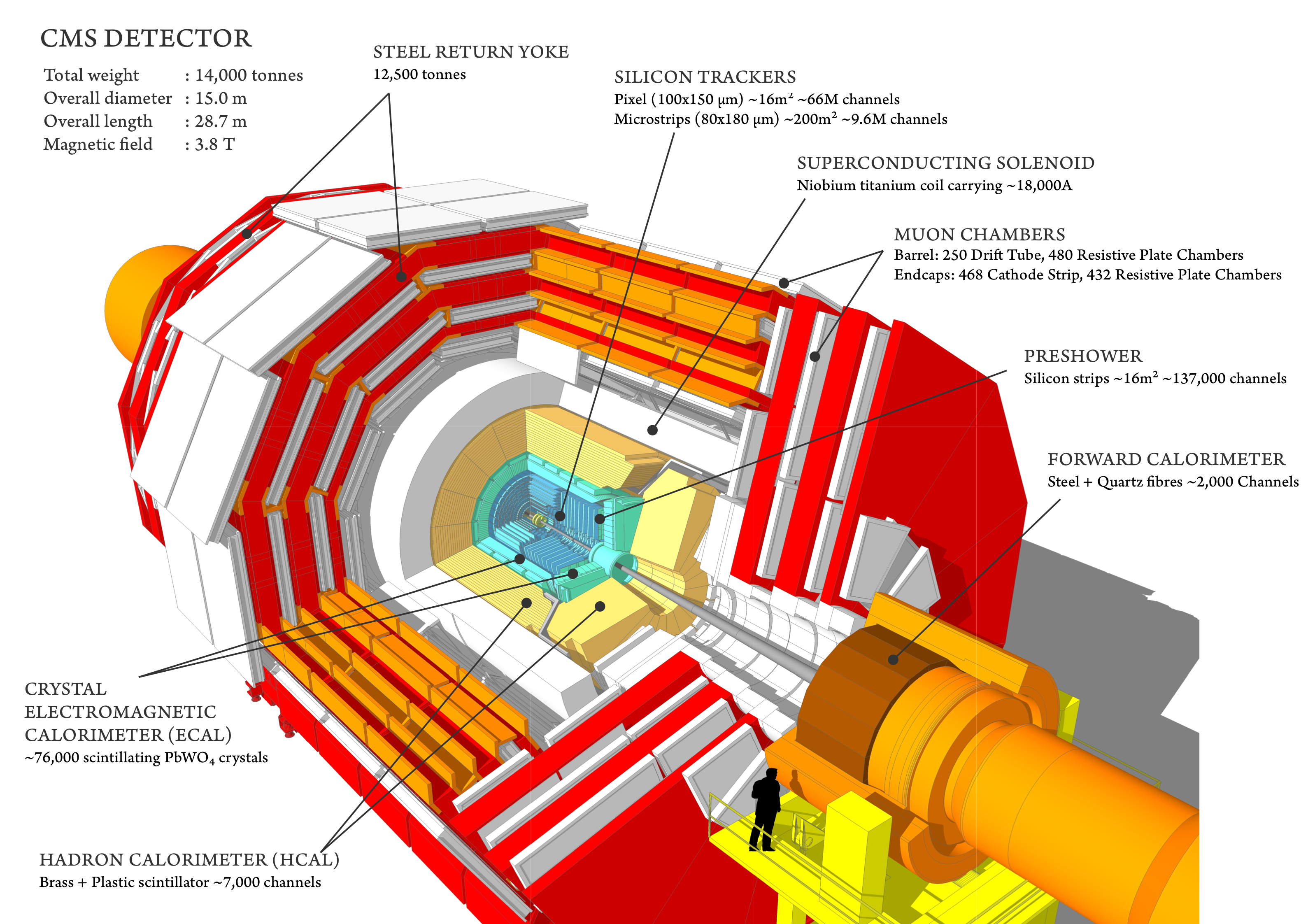}
\caption{Schematic overview of ATLAS and CMS detectors}
\label{fig:detector}
\end{center}
\end{figure}
%%%%%%%%%%%%%%%%%%%%%%%%%%%%%%%%%%%%%%%%%%%%%%%%%%%%%

We use the same basic detector parametrization for all of the collider scenarios. 
The Combined Snowmass Detector parametrization incorporates the performance of the best features of the 
existing LHC detectors and, the expected performance of the upgraded LHC detectors~\cite{cms-tdr, atlas-loi, cms-elec-dp, cms-muon-dp}.  A schematic overview of detector sub-components of the ATLAS and CMS detectors is shown in Fig.~\ref{fig:detector}. As described in Sec.~\ref{sec:delphes}, the tracking and identification efficiencies are parametrized 
and input to {\sc Delphes}, while the isolation efficiency is not an input and is determined by the simulation.

\section{Simulation tools}
\label{sec:sim-tools}

We simulate a set of standard model (SM) processes that we expect will encompass the backgrounds for 
most new phenomena searches at the future colliders described in Table~\ref{tab:colliders}.  The simulation 
procedure involves parton-level generation of events at leading order in bins of the 
scalar sum of the recoil jet $p_{\rm T}$ (\hgt), particle decay, parton showering and 
hadronization, jet-parton matching, corrections for next-to-leading order (NLO) 
contributions, and detector simulation.  In this section, we give 
an overview of the procedure and describe the detector simulation in detail; the 
details of all other stages can be found in Ref.~\cite{htbin, Snowmassosg}. 

% Other details: MG5, Bridge, Pythia, jet-parton matching, weights, k-factors
We use {\sc Madgraph5}~\cite{madgraph} for event generation.  To enhance numbers 
of events in boson decay modes, we use {\sc Bridge}~\cite{bridge} to decay the generated bosons with 
equal probability to a set of allowed final states.  For example, $Z$ bosons decay with equal probability 
to pairs of quarks, neutrinos, 
electrons, muons, or tau leptons.  Stable particles are passed to 
{\sc Pythia6}~\cite{pythia} for parton showering and hadronization. 
The {\sc Madgraph5} generated partons are appropriately matched to jets 
from {\sc Pythia6} at a given energy scale that depends on the process to avoid overlap 
between phase-space descriptions.  Each event has an associated weight that corrects 
for NLO contribution to the total cross section and the effects of branching ratio 
leveling. Because the NLO correction depends on the sub-process (e.g. $W$+jets or $Z$+jets) 
and the branching ratio on the decay mode (e.g. $Z\rightarrow \mu^{+}\mu^{-}$ or 
$Z\rightarrow q\bar{q}$), the weights are inserted at the event level and can be 
retrieved with the \verb4Weight4 method of the \verb4Event4 object in the output sample.

After generation, decay, parton showering, and hadronization, the stable particles are passed to {\sc Delphes} for pile-up mixing, detector simulation, and object reconstruction.  The {\sc Delphes} steps are described in
Sec.~\ref{sec:delphes} after a brief description of the background processes and binning.

\subsection{Background processes}
\label{sec:sim-bkg}

The background processes all involve combinations of bosons (B = $\gamma$, W$^\pm, Z$), 
Higgs bosons (H), charged and neutral leptons (L), top quarks (t), and 
light quark jets (j).  We summarize the processes in Table~\ref{tab:bkgs}.  
For all samples (except B-4p) we generate four particles at 
parton level: the $N$ particles specified in the dataset name with the remaining $4-N$ 
particles generated as hadronic jets.  The B-4p, Bj-4p, and Bjj-vbf-4p samples all include
on-shell boson+jets processes and, being exclusive samples, are all required for full 
boson+jets simulation.  The B-4p sample includes events with a boson plus zero generated 
jets; the Bj-4p sample includes events with a boson plus one to three generated jets; and 
the Bjj-vbf-4p sample includes events with a boson produced in association with at least 
two jets resulting from an off-shell electroweak process: either the usual t-channel 
vector boson fusion (VBF) diagrams or an off-shell s-channel vector boson decaying to jets.
In practice, the Bjj-vbf-4p cross section is very small and can be neglected 
for analyses not concerned with the VBF topology.  The LL samples include processes with 
off-shell $W$/$Z$ and jets.  Particles are considered on-shell if they are within 15 
natural widths of the resonance mass.

% HT binning
\subsection{Binning in \hgt}
\label{sec:sim-binning}

These samples are intended for studies with integrated luminosity
up to 3 ab$^{-1}$.  To efficiently populate the full particle spectra 
for quantities such as $p_{\rm T}$, invariant mass, 
and missing transverse energy (\met), we generate events in 
bins of \hgt.  Each process in Table~\ref{tab:bkgs} is generated in 5-7 bins of \hgt~depending on the process, and analysts must use all bins to correctly simulate the complete process.  The binning and numbers of events generated are chosen so that the statistical uncertainty at all points in the \hgt\ spectrum appropriate for studies at 3 ab$^{-1}$ is 
much less than $30\%$.  By binning in \hgt, the numbers of events required are decreased from $\sim10^{10}$ per 
process to $10^{6-7}$ per process.   This binning process, including choice of bin boundaries, is described in detail in Ref.~\cite{htbin}. 

% The Standard Model background processes in association with additional 
%jets were produced both inclusively as well as with  weighted LHE events. For studies 
%where the future datasets are expected to cover integrated luminosity in the ab$^{-1}$ 
%range, we use weighted events both during parton level generation as well as after 
%matching and hadronization.  For details on event generation see reference [xx]. 

\begin{table}[htbp]
\label{tab:bkgs}
\begin{center}
\begin{tabular}{l|l|l}
\hline\hline
Dataset name &  Physics process & Number of recoil jets \\ 
\hline
B-4p        & $\gamma$ or on-shell $W$, $Z$ & 0 \\
Bj-4p       & $\gamma$ or on-shell $W$, $Z$ & 1-3 \\
Bjj-vbf-4p  & $\gamma$ or off-shell $W$, $Z$, $H$ in VBF topology & 2-3 \\
BB-4p & Diboson ($\gamma$, $W$, $Z$) processes & 0-2 \\ 
BBB-4p & Tri-boson ($\gamma$, $W$, $Z$) processes including BH & 0-1 \\ 
LL-4p & Non-resonant dileptons (including neutrinos) with $m_{ll} > 20$ GeV & 0-2  \\
LLB-4p & Non-resonant dileptons with an on-shell boson, $m_{ll} > 20$ GeV & 0-1 \\
H-4p & Higgs & 0-3 \\
tj-4p & Single top (s- and t-channel) & 0-2 \\
tB-4p & Single top associated with a boson & 0-2\\ 
tt-4p & $t\bar{t}$ pair production & 0-2 \\
ttB-4p & $t\bar{t}$ associated with $\gamma$, $W$, $Z$, $H$ & 0-1 \\ 
\hline\hline
\end{tabular}
\caption{Table of background processes.  All processes include the particles in the dataset name plus additional recoil jets up to four generated particles.  On-shell vector bosons, off-shell dileptons, Higgs bosons, top quarks, and jets are denoted B, LL, H, t, and j, respectively.  In the Bjj-vbf-4p case, B includes Higgs.  In the BBB-4p case, BBB includes BH.  Samples are generated in bins of \hgt\ for $\sqrt{s}=$ 14, 33, and 100 TeV.}
\label{tab:bkgproc}
\end{center}
\end{table}

\subsection{{\sc Delphes} detector simulation}
\label{sec:delphes}

{\sc Delphes} is a C++ framework for parametrized simulation of a general collider experiment.  
Starting from the output of an event generator, the framework propagates particles through the 
detector with a solenoidal magnetic field; applies tracking efficiency, reconstruction efficiency, and momentum 
resolution; and clusters hadronic jets with identification of those resulting from $b$ quarks, $\tau$ leptons and decay 
products from $t$ quarks, $H, W$ and $Z$ bosons.

The detector simulation is composed of five primary components listed below.
The ultimate efficiency and resolution for each object comes from the combination of these components:

\begin{itemize}
\item input tracking efficiency (for charge hadrons, electrons, and muons),
\item input momentum resolution (for charge hadrons, electrons, and muons),
\item input calorimeter resolution (for electromagnetic and hadron calorimeters),
\item input reconstruction efficiency (for photons, electrons, and muons),
\item and isolation (for photons, electrons, and muons), which is not input by the user 
but determined from the simulation.
\end{itemize}

The specific modules of the framework, their user inputs, and the values used
for the Combined Snowmass Detector (denoted with square brackets []) are described below.

\subsubsection{{\sc Delphes} modules}

\begin{itemize}
\item \verb4PileUpMerger4\\ 
{\bf Input}: Name of file of minimum bias events to be mixed as pile-up, mean number of 
pile-up interactions per crossing [0, 50, or 140], Gaussian standard deviation of the 
bunch length (in the z direction) in meters [0.05m].\\
{\bf Function}: Stable particles from minimum bias pile-up events are mixed with the final state 
particles from the generated event with the given z distribution.

\item \verb4ParticlePropagator4:\\ 
{\bf Input}: Radius [1.2m], length [6m], and magnitude [3.8T] of solenoidal magnetic field.\\
{\bf Function}: Stable particles (including unsubtracted pile-up) are classified as charged hadrons, 
electrons, or muons and propagated through the solenoidal magnetic field.

\item \verb4ChargedHadronTrackingEfficiency4:\\ 
{\bf Input}: Efficiency for tracking of charged hadrons as a function of pseudorapidity 
$\eta$ and $p_T$ [Table \ref{tab:trkeff}].\\
{\bf Function}: Tracking efficiency is applied to propagated charged hadrons.

\item \verb4ElectronTrackingEfficiency4:\\ 
{\bf Input}: Efficiency for tracking of electrons as a function of $\eta$ and $p_{\rm T}$ [Table \ref{tab:trkeff}].\\
{\bf Function}: Tracking efficiency is applied to propagated electrons.

\item \verb4MuonTrackingEfficiency4:\\ 
{\bf Input}: Efficiency for tracking of muons as a function of $\eta$ and $p_{\rm T}$ [Table \ref{tab:trkeff}].\\
{\bf Function}: Tracking efficiency is applied to propagated muons.

\item \verb4ChargedHadronMomentumSmearing4:\\ 
{\bf Input}: Momentum resolution for charged hadron as a function of $\eta$ and $p_{\rm T}$ [Table \ref{tab:ptres}].\\
{\bf Function}: Momentum of propagated charged hadrons is smeared according to resolution.

\item \verb4ElectronMomentumSmearing4:\\ 
{\bf Input}: Momentum resolution for electrons as a function of $\eta$ and $E$ [ $\sigma_{p_{\rm T}}=1.5\% \cdot E$ for electron energy less than 25 GeV and $\sigma_{p_{\rm T}}=\sqrt{ (0.5\%)^2 \cdot E^2 + (2.7\%)^2 \cdot E + (15\%)^2 }$ for electron energy greater than 25 GeV].\\
{\bf Function}: Momentum of propagated electrons is smeared according to resolution.

\item \verb4MuonMomentumSmearing4:\\ 
{\bf Input}: Momentum resolution for muons as a function of $\eta$ and $p_T$ [Table \ref{tab:ptres}].\\
{\bf Function}: Momentum of propagated muons is smeared according to resolution.

\item \verb4Calorimeter4:\\
{\bf Input}: Energy resolution as function of $\eta$ and energy for electromagnetic and hadronic energy measurements.  Segmentation in $\eta$ and $\phi$ of calorimeter towers.\\
{\bf Function}: Charged hadrons, muons, electrons, and other stable particles are combined into \verb4EFlow4 tracks and towers.  Tower energy is smeared by the input energy resolutions.

\item \verb4TrackPileUpSubtractor4:\\
{\bf Input}: Resolution on $z$-position of primary vertex [0.1cm].\\
{\bf Function}: Charged particles with $z$-position of point of closest approach in the $x$-$y$ plane greater than the input resolution on $z$-position are rejected as pile-up.

\item \verb4PhotonEfficiency4:\\
{\bf Input}: Photon reconstruction efficiency as a function of $\eta$ and $p_T$ [Table \ref{tab:recoeff}].
\\
{\bf Function}: Reconstruction efficiency is applied to photons.

\item \verb4ElectronEfficiency4:\\
{\bf Input}: Electron reconstruction efficiency as a function of $\eta$ and $p_T$ [Table \ref{tab:recoeff}].\\
{\bf Function}: Reconstruction efficiency is applied to electrons.

\item \verb4MuonEfficiency4:\\
{\bf Input}: Muon reconstruction efficiency as a function of $\eta$ and $p_T$ [Table \ref{tab:recoeff}].\\
{\bf Function}: Reconstruction efficiency is applied to muons.

\item \verb4PhotonIsolation4:\\
{\bf Input}: Cone size [0.3] and threshold on energy ratio for photon relative isolation [0.1].\\
{\bf Function}: Isolation requirement is applied to photons.

\item \verb4ElectronIsolation4:\\
{\bf Input}: Cone size [0.3] and threshold on energy ratio for electron relative isolation [0.1].\\
{\bf Function}: Isolation requirement is applied to electrons.

\item \verb4MuonIsolation4:\\
{\bf Input}: Cone size [0.3] and threshold on energy ratio for muon relative isolation [0.1].\\
{\bf Function}: Isolation requirement is applied to muons.

\item \verb4FastJetFinder4:\\ 
{\bf Input}: Jet algorithm [anti-$k_T$~\cite{ak5}] and relevant parameters [distance parameter = 0.5] (AK5).\\
{\bf Function}: \verb4EFlow4 tower and track collections from the \verb4Calorimeter4 stage and muons are clustered into jets. Jet area is computed and stored.

\item \verb4JetPileUpSubtractor4:\\ 
{\bf Function}: The jet four-vector is corrected for contributions from neutral pile-up using FastJet $\rho$ subtraction~\cite{fastjet}.

\item \verb4BTagging4:\\
{\bf Input}: Efficiency to identify a true $b$-jet as having come from a $b$ quark ($b$-tag) as function of $\eta$ and $p_T$.  Probability to misidentify a light flavor jet as having come from a $b$ quark (mistag rate).  [Figs.~\ref{fig:btag_loose}-\ref{fig:bfake_med}].\\
{\bf Function}: Clustered jets are identified as having come from a $b$ quark.  Multiple working points can be evaluated for each jet.

\item \verb4TauTagging4:\\
{\bf Input}: Efficiency to identify a true $\tau$-jet as having come from a $\tau$ lepton ($\tau$-tag) as 
function of $\eta$ and $p_T$ [$65\%$ flat in $\eta$ and $p_T$].  Probability to misidentify a jet as having come from a $\tau$ lepton
(mistag rate) [$0.4\%$ flat in $\eta$ and $p_T$].\\
{\bf Function}: Clustered jets are identified as having come from a $\tau$.

\item \verb4UniqueObjectFinder4 : 
{\bf Function}: Overlapping particles are removed from collections as appropriate.  For instance, photons are removed from electron collections.

\item \verb4TreeWriter4 : 
{\bf Function}: Desired output is written to an output file.
\end{itemize}

%%%%%%%%%%%%%%%%%%%%%%%%%%%%%%%%%%%%%%%%%%%%%%%%%%%%%%%%%%%%%%%%%%%%%%%%%
\begin{table}[hbtp]
\begin{center}
\begin{tabular}{l|l|cc}
\hline\hline
Object         & $p_{\rm T}$ range     & $\vert\eta\vert\leq1.5$  & $1.5<\vert\eta\vert\leq2.5$ \\
\hline
Charged Hadron & $>1$ GeV     & 97.0\% &   90.0\% \\
Electron       & $1-100$ GeV  & 97.0\% &   90.0\% \\
Electron       & $>100$ GeV   & 99.0\% &   95.0\% \\
Muon           & $>1$ GeV     & 99.9\% &   98.0\% \\
\hline\hline
\end{tabular}
\caption{Tracking efficiency for charged hadrons, electrons, and muons.}
\label{tab:trkeff}
\end{center}
\end{table}
%%%%%%%%%%%%%%%%%%%%%%%%%%%%%%%%%%%%%%%%%%%%%%%%%%%%%%%%%%%%%%%%%%%%%%%%%%%

%%%%%%%%%%%%%%%%%%%%%%%%%%%%%%%%%%%%%%%%%%%%%%%%%%%%%%%%%%%%%%%%%%%%%%%%%
\begin{table}[hbtp]
\begin{center}
\begin{tabular}{l|cc}
\hline\hline
 $p_{\rm T}$ range     & $\vert\eta\vert\leq1.5$  & $1.5<\vert\eta\vert\leq2.5$ \\
\hline
$1-10 $ GeV   & 1.3\% & 1.5\% \\
$10-200$ GeV  & 2.0\% & 4.0\% \\
$>200$ GeV    & 5.0\% & 5.0\% \\
\hline\hline
\end{tabular}
\caption{$p_{\rm T}$ resolution for charged hadrons and muons.}
\label{tab:ptres}
\end{center}
\end{table}
%%%%%%%%%%%%%%%%%%%%%%%%%%%%%%%%%%%%%%%%%%%%%%%%%%%%%%%%%%%%%%%%%%%%%%%%%%%

%%%%%%%%%%%%%%%%%%%%%%%%%%%%%%%%%%%%%%%%%%%%%%%%%%%%%%%%%%%%%%%%%%%%%%%%%
\begin{table}[hbtp]
\begin{center}
\begin{tabular}{l|cc}
\hline\hline
Object     & $\vert\eta\vert\leq1.5$  & $1.5<\vert\eta\vert\leq2.5$ \\
\hline
Photon    & 96.4\% &     96.2\% \\
Electron  & 98.0\% &     90.0\% \\
Muon      & 99.0\% &     97.0\% \\
\hline\hline
\end{tabular}
\caption{Reconstruction efficiency for photons, electrons, and muons.  The upper
limit of the $\eta$ range is 2.4 for muons.}
\label{tab:recoeff}
\end{center}
\end{table}
%%%%%%%%%%%%%%%%%%%%%%%%%%%%%%%%%%%%%%%%%%%%%%%%%%%%%%%%%%%%%%%%%%%%%%%%%%%

\subsubsection{Pile-up mixing and subtraction}
\label{sec:pileup}
{\sc Delphes} simulates the pile-up expected with higher luminosity by mixing additional 
minimum bias interactions with the original generated event. We address scenarios of 
$<\mu> = $0, 50 and 140 pile-up interactions for each $\sqrt{s}=$ 14, 33, and 100 TeV.

%Raw event properties such as particle momenta, vertex position, %and ID are accessed 

Minimum bias events that are expected with a ``loose'' trigger that accepts a large
fraction of the overall inelastic $pp$ interactions are produced using CMS  
$Z2^*$ {\sc Pythia} tune. Pile-up events are randomly selected from the minimum bias sample and are mixed 
with the event from the primary interaction according to a Poisson distribution with a mean 
of 0, 50, or 140 additional interactions. These events are randomly distributed along 
the beam axis according to a Gaussian distribution with a width of 0.05 m.  

If the $z$-position of a pile-up vertex is less than the 0.1 cm vertex resolution.
the pile-up interaction cannot be separated from the additional vertices. For such vertices, all 
particles from both the pile-up and primary interactions are included in the object
reconstruction. For pile-up interactions with $z$-vertex position greater than the 
resolution, subtraction of charged pile-up particles within the tracker volume is 
applied with an efficiency of unity.  We use the FastJet area method~\cite{fastjet}
to correct measurements of jet four-vectors and isolation energy for contribution
from neutral pile-up particles and charged pile-up particles outside the tracker acceptance.

\begin{figure}[hbtp]
\begin{center}
\includegraphics[width=0.48\hsize]{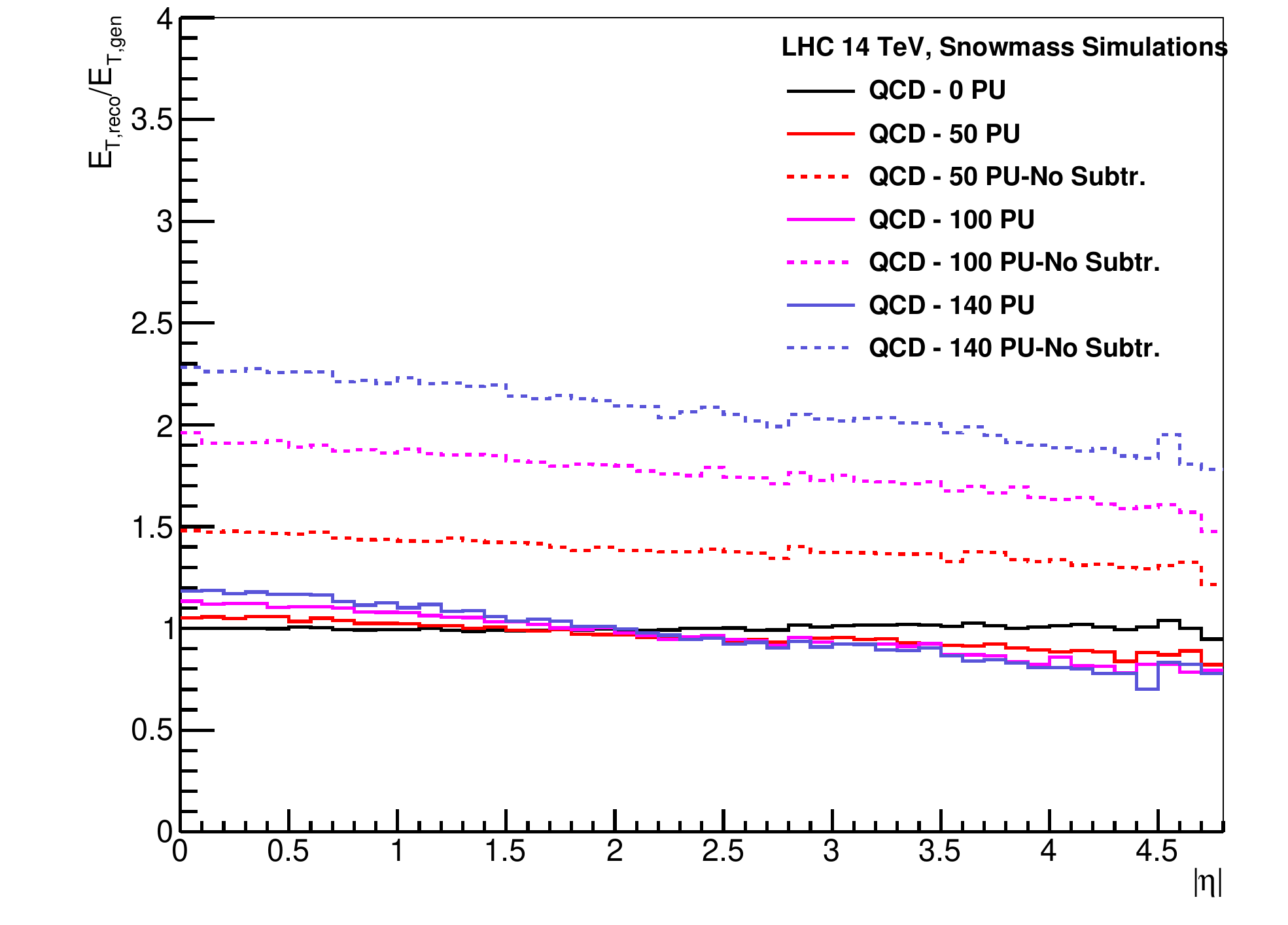}
\includegraphics[width=0.48\hsize]{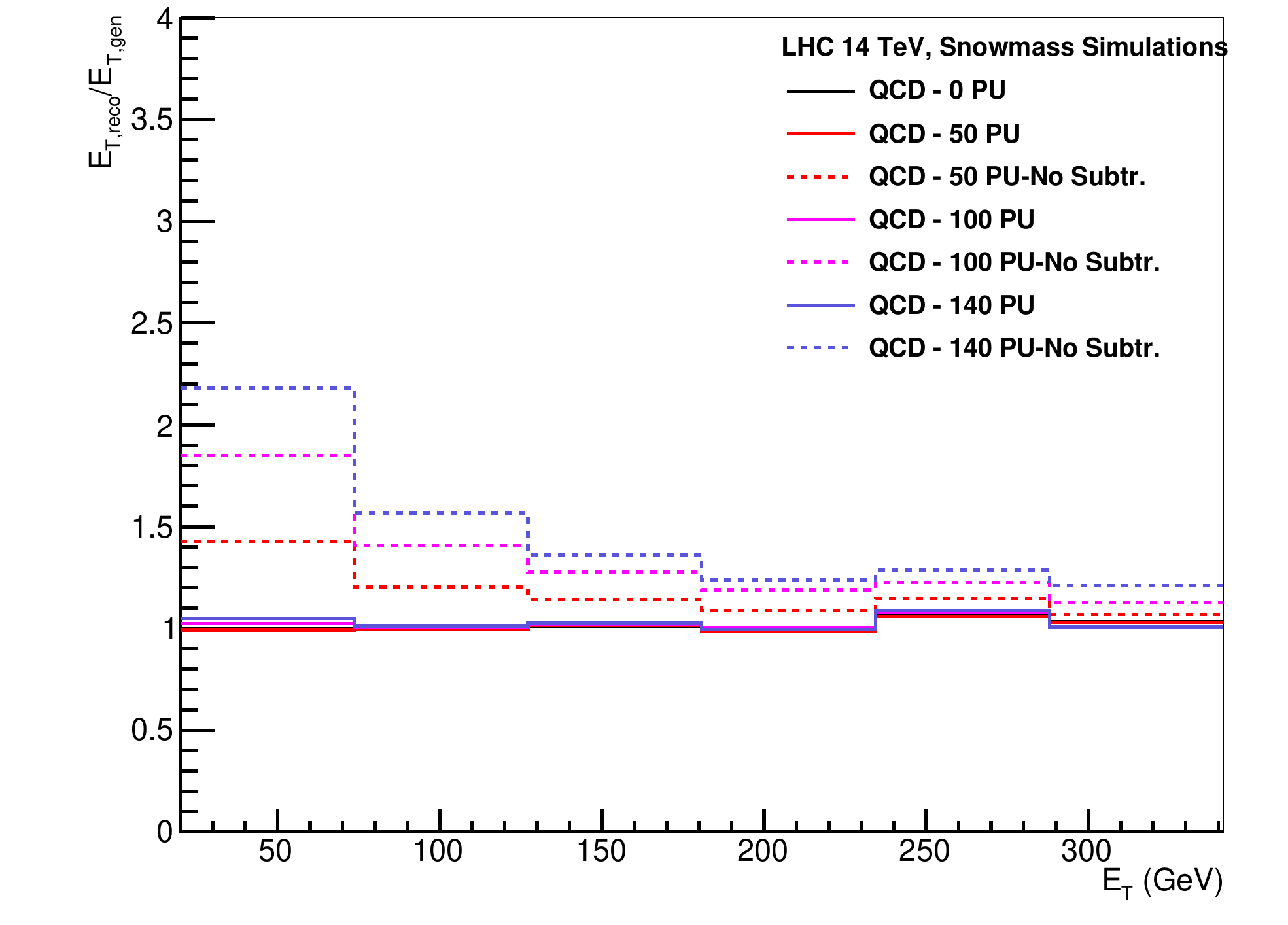}
\caption{Jet response in QCD events as a function of $|\eta|$ (left) and $p_T$ (right) for 0, 50, 100 and 140 pile-ups, before and after pile-up subtraction.}
\label{fig:jetresponse}
\end{center}
\end{figure}

In Fig.~\ref{fig:jetresponse}, we show the detector response to jets in simulated events using QCD processes.
The response is shown as a function of $\eta$ and $p_{\rm T}$ for 0, 50, 100, and 140 pile-up interactions before 
and after pile-up subtraction.  The subtraction removes the pile-up dependence on average with some 
residual $\eta$ dependence. Residual off-set corrections are applied to events with large pile-ups 
in bins of $\eta$ and $p_T$ to recover the jet energy response as in the case of 0 pile-up scenario. 
%%%%%%%%%%%%%%%%%%%%%%%%%%%%%%%%%%%%%%%%
% Jet response
%%%%%%%%%%%%%%%%%%%%%%%%%%%%%%%%%%%%%%%%
%%%%%%%%%%%%%%%%%%%%%%%%%%%%%%%%%%%%%%%%

\subsubsection{Jet substructure}
\label{sec:jetsubstructure}

In the past several years, there has been significant development of techniques for discriminating between new phenomena and the large SM background with the substructure of jets.   When particles decaying to jets (such as top quarks, $W$, $Z$, and Higgs bosons) have sufficient Lorentz boost in the lab frame, their jet daughters overlap in the detector and are more likely to be reconstructed as a single jet than as two or more jets.  Recognizing that this overlap is possible 
it is important to use discriminants for classifying boosted new phenomena and SM backgrounds.
%and affords useful discrimination between boosted new phenomena and SM background.

We have added functionality to the FastJetFinder module that \textit{(i)} computes and saves variables useful for jet substructure studies and \textit{(ii)} flags jets as having come from a top quark or $H, Z, W$ bosons.  
We compute these variables and flags for jets clustered 
with the the Cambridge-Aachen algorithm~\cite{ca} with a distance parameter of 0.8 (CA8); this jet collection is in addition to the basic anti-$k_{\rm T}$ jet collection with distance parameter of 0.5.   The saved variables are trimmed jet mass, $\tau_1$, $\tau_2$, $\tau_3$, $N_{\rm subjets}$, and mass drop.  The flags are based on trimmed jet mass and $N_{\rm subjets}$ for top identification, and trimmed jet mass and mass drop for $W$ and Higgs identification.  The trimmed jet parameter values are $p_T$ fraction of 0.5 and CA distance parameter of 0.2.

The variables used in top- and $W$-tagging were chosen because of their pile-up stability in order to minimize pile-up dependence of the requirements on these variables and the tagging efficiency itself.  The N-subjettiness variables ($\tau_3, \tau_2, \tau_1$) were not used in the taggers because they exhibit pile-up dependence, but they are included in the output \verb4Jet4 object for analysis-level jet tagging.

\begin{figure}[hbtp]
\begin{center}
\includegraphics[width=0.48\hsize]{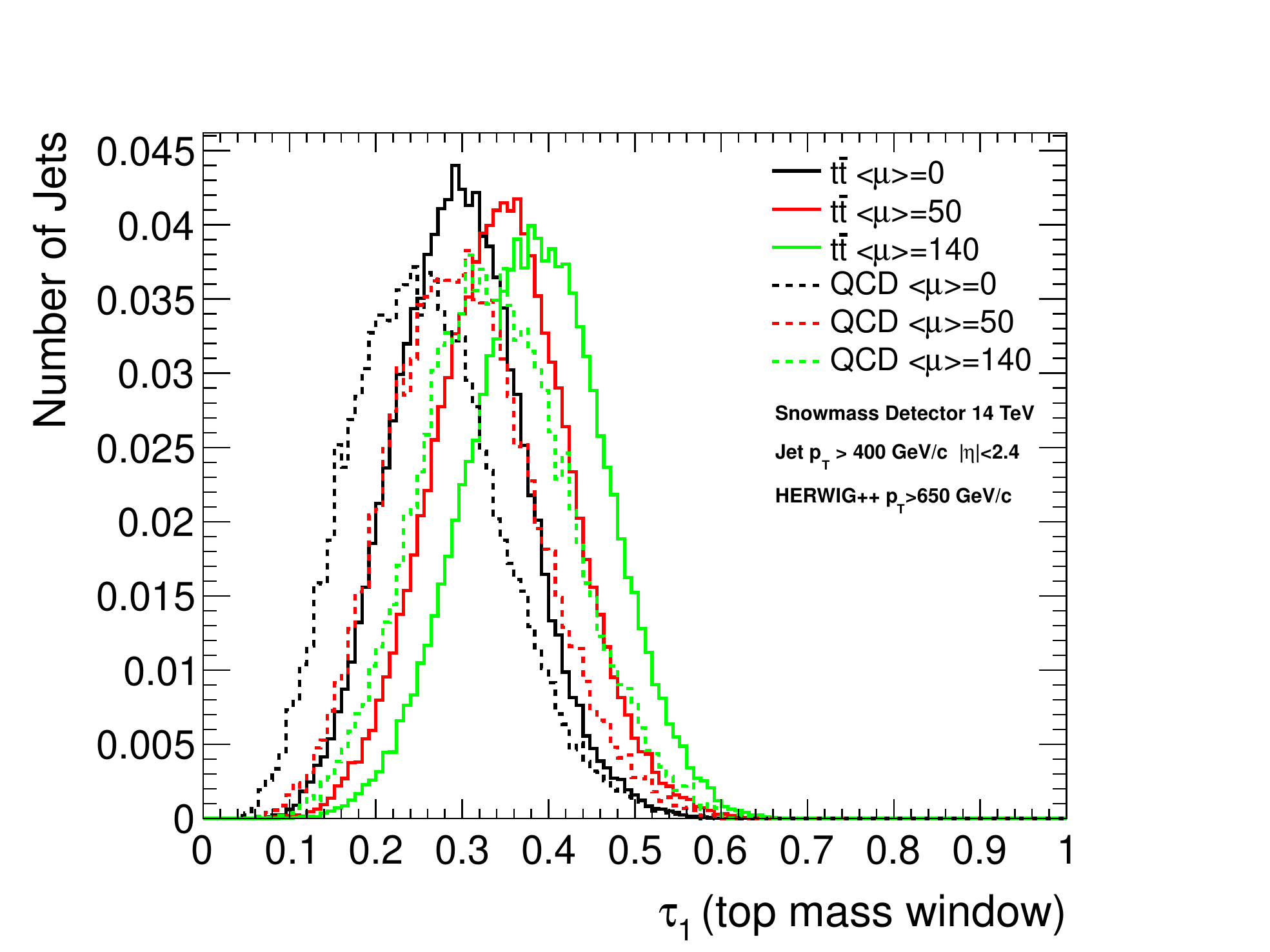}
\includegraphics[width=0.48\hsize]{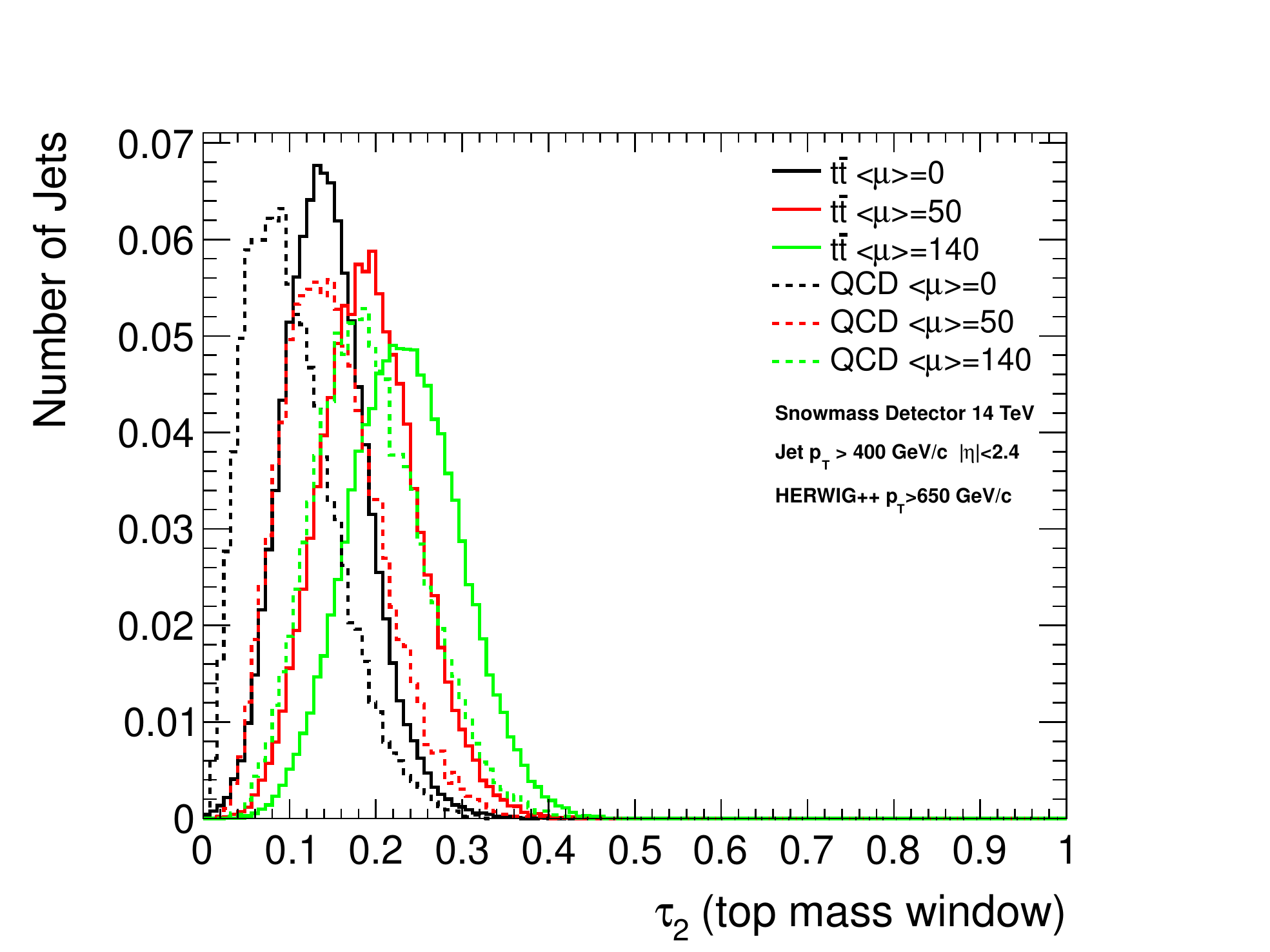}
\includegraphics[width=0.48\hsize]{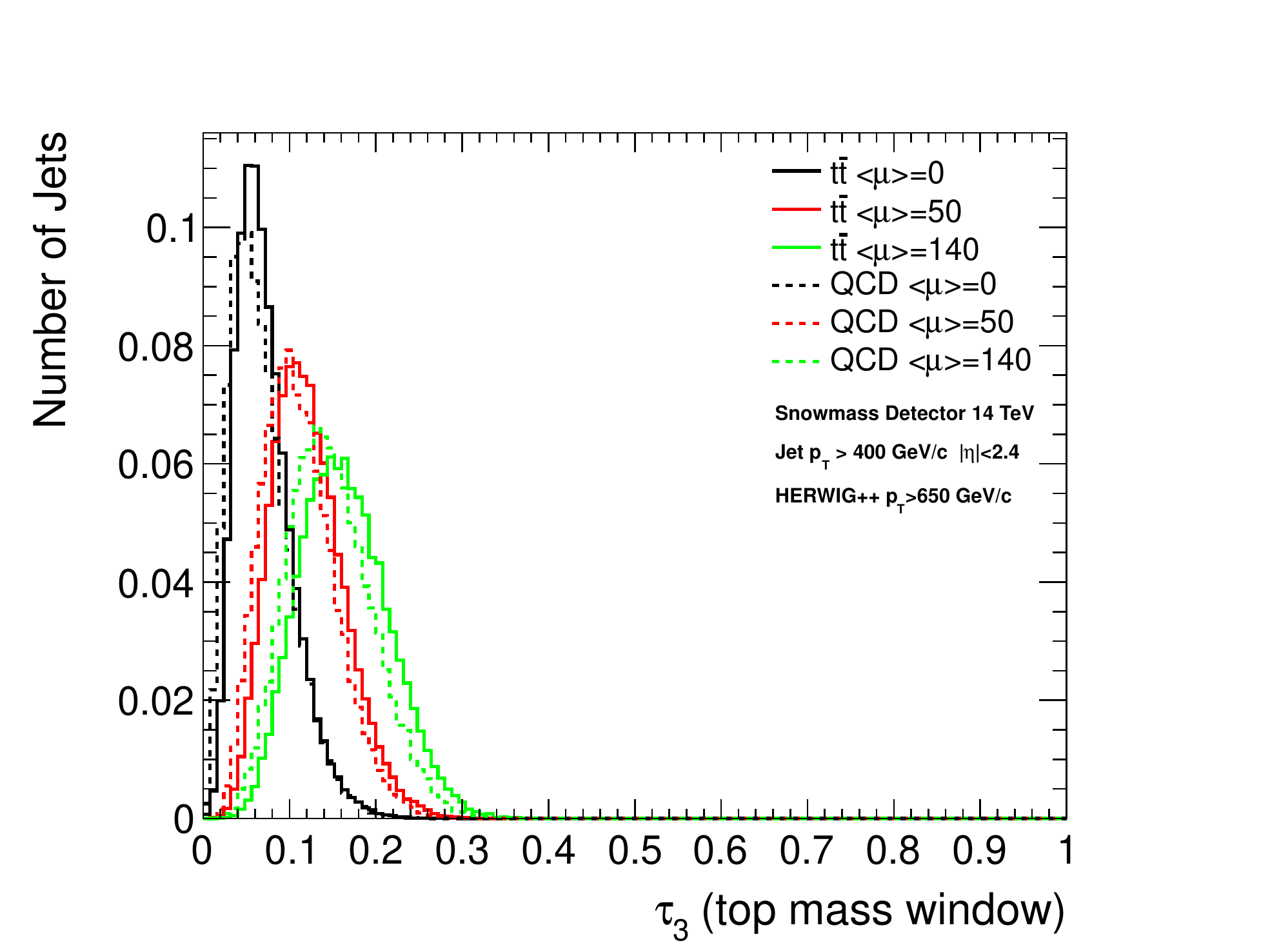}
\includegraphics[width=0.48\hsize]{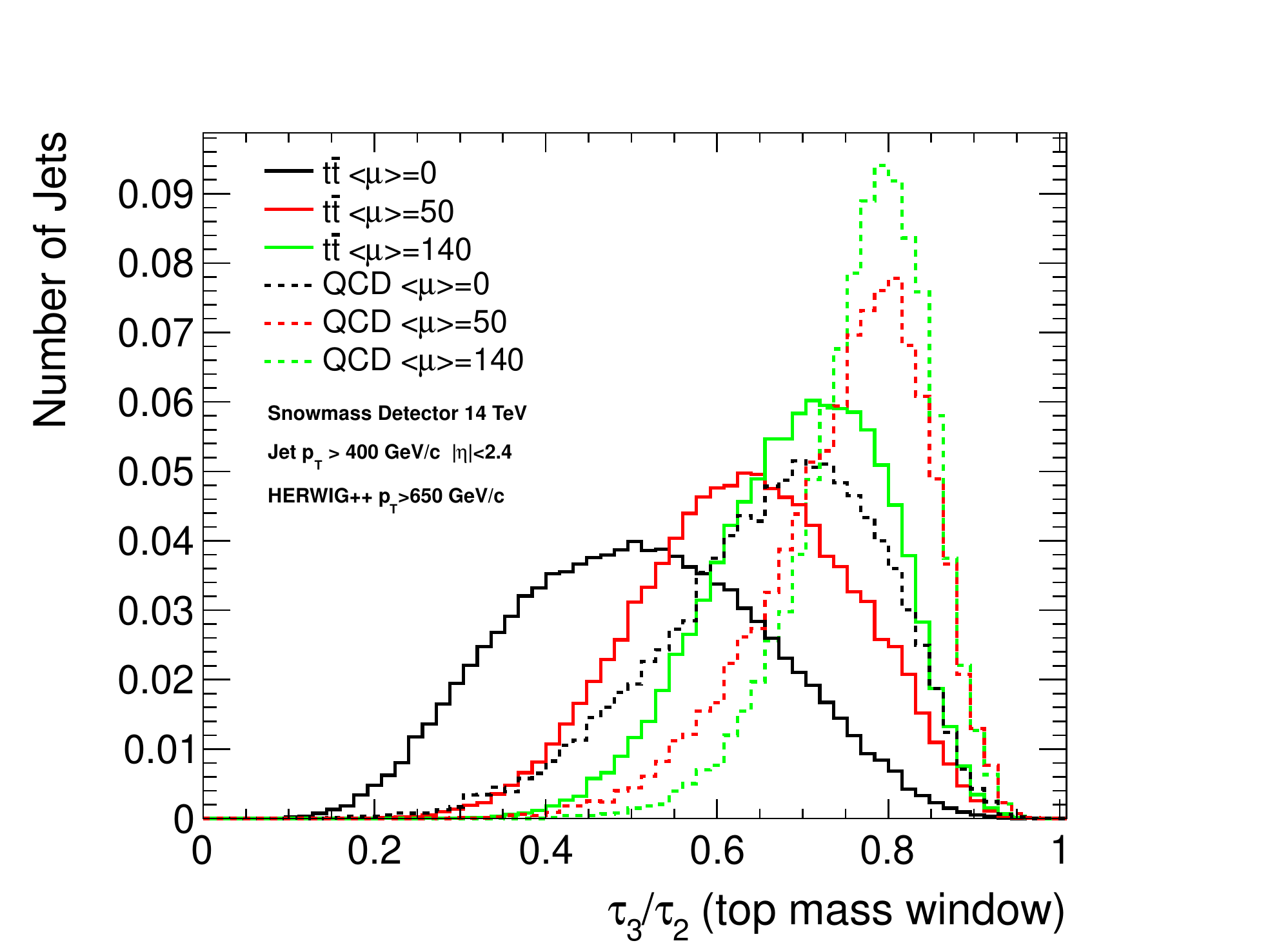}
\caption{N-subjettiness variables $\tau_1$ (top left), $\tau_2$ (top right), $\tau_3$ (bottom left), and $\tau_3/\tau_2$ (bottom right) in $t\bar{t}+$jets and QCD events with 0, 50, and 140 pile-up.}
\label{fig:nsubjettiness}
\end{center}
\end{figure}

\begin{figure}[hbtp]
\begin{center}
\includegraphics[width=0.48\hsize]{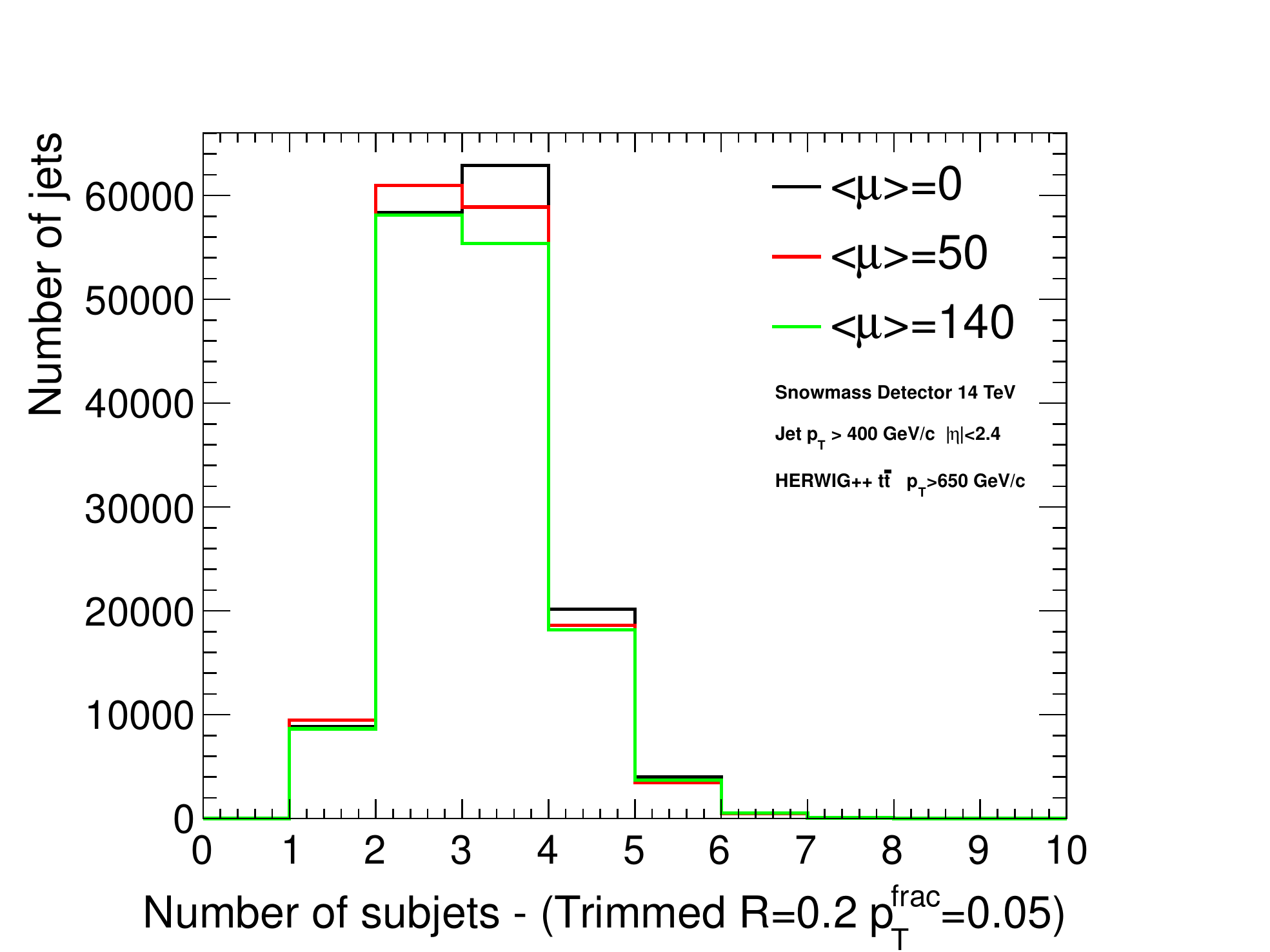}
\includegraphics[width=0.48\hsize]{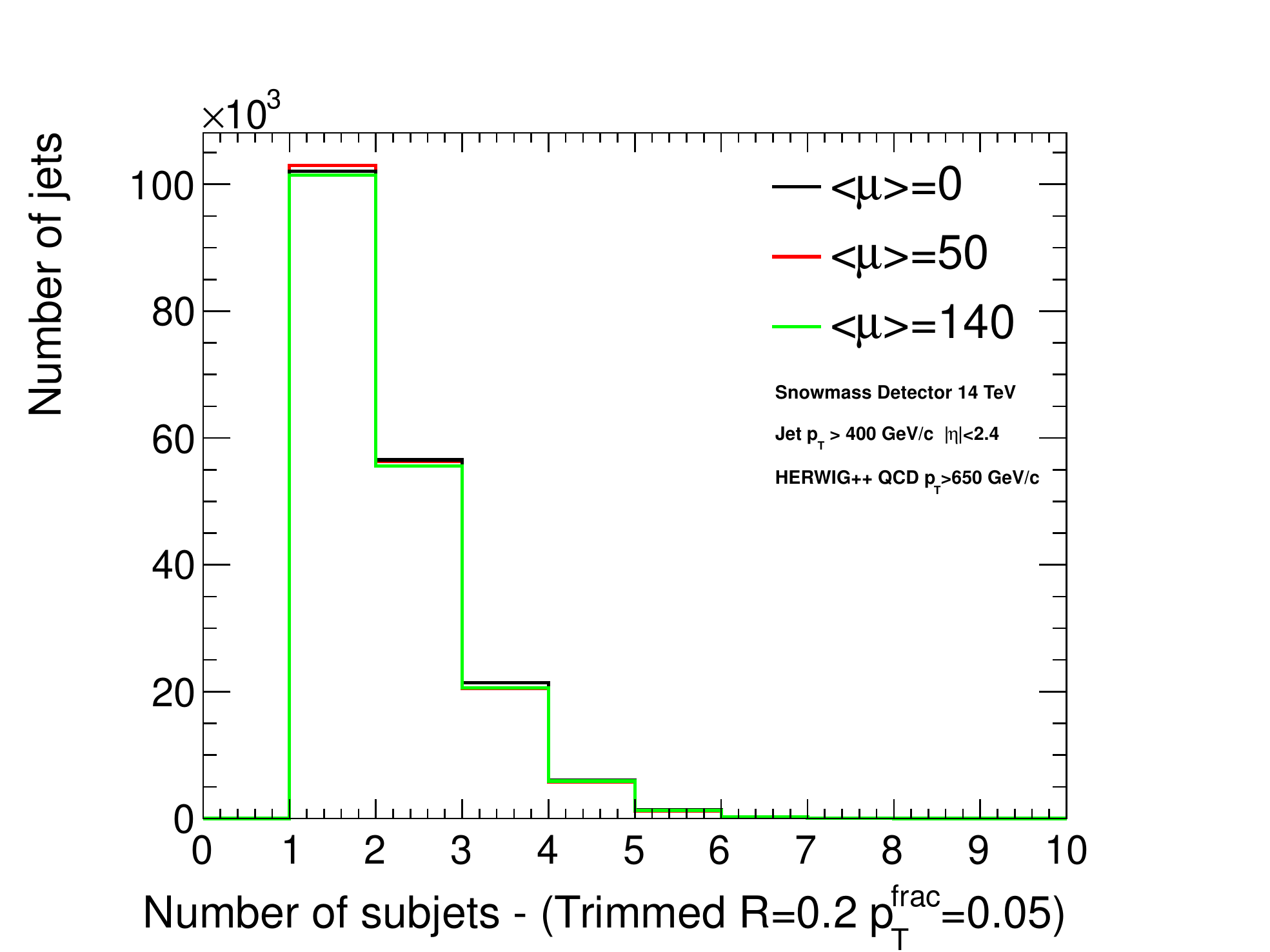}
\caption{N-subjets in $t\bar{t}+$jets (left) and QCD events (right) with 0, 50, and 140 pile-up.}
\label{fig:nsubjets}
\end{center}
\end{figure}

%DrawHistogramFrom6Files_JetMass_Trimmed_CA2_PFRAC5_compare_all.pdf
%DrawHistogramFrom6Files_JetMass_Delphes3_compare_all.pdf

%In this approach the $k_t-$ [xx] (or $CA$ [xx] for jet sub%-structure studies) jet algorithms with the R-parameter of 0.6 was used. 
%Active areas are computed which are expected to have an uniform background throughout the phase space. The soft massless 
%‘ghost’ particles within $|eta| < 5$ were used to determine the average transverse momentum per unit area ($\rho$). Average
%$\rho$ of background events are then subtracted from both the hard %jets as well as pileup events in the lepton or photon isolation %cones.

\section{Simulation performance}
\label{sec:sim-perf}

In this section we show the output of the full simulation procedure with focus on the effects of pile-up.  We show basic spectra for missing energy, \hadt\  (scalar sum $p_T$ of the jets), electrons, and muons; reconstruction efficiency for electrons, muons, and photons; jet substructure variables; and efficiencies and fake rates for tagging of $b$ quarks, $\tau$ leptons, $t$ quarks, and $W$ bosons.

The performance is evaluated in $t\bar{t}+$jets and boson+jets events with $\sqrt{s}=13$ TeV and $\sqrt{s}=14$ TeV and three pile-up scenarios.  We require at least four jets with $\pt>30$ GeV and $\vert\eta\vert<$2.5 and one electron (muon) with $\pt>30 (20)$ GeV and $\vert\eta\vert<$2.5; when evaluating photon efficiency we replace the lepton requirement with a requirement for a photon with $\pt>20$ GeV and $\vert\eta\vert<$2.5.

\subsection{Jet and \met\ performance}

%%%%%%%%%%%%%%%%%%%%%%%%%%%%%%%%%%%%%%%%
% MET and HT
%%%%%%%%%%%%%%%%%%%%%%%%%%%%%%%%%%%%%%%%

\begin{figure}[hbtp]
\begin{center}
\includegraphics[width=0.48\hsize]{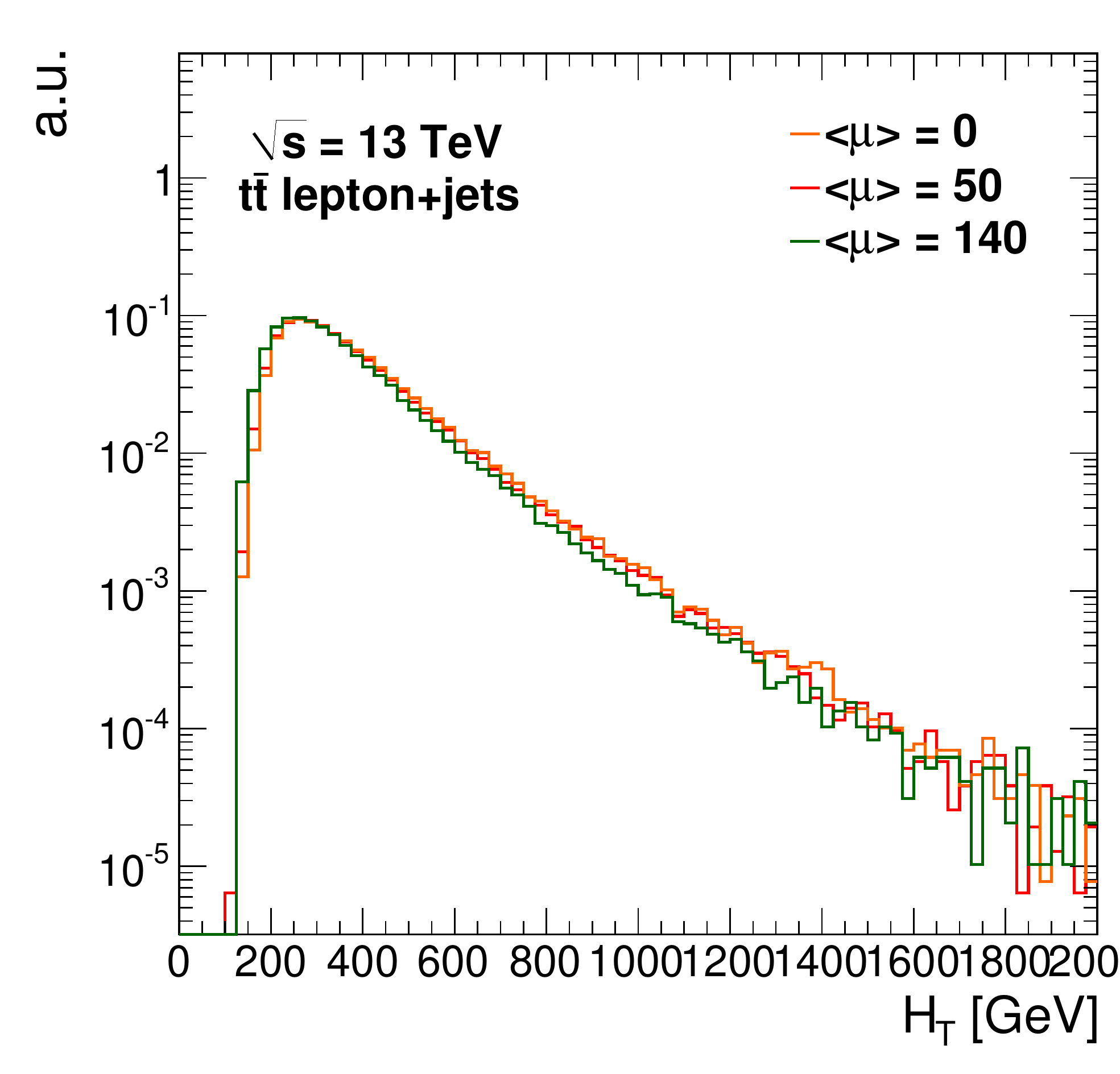}
\includegraphics[width=0.48\hsize]{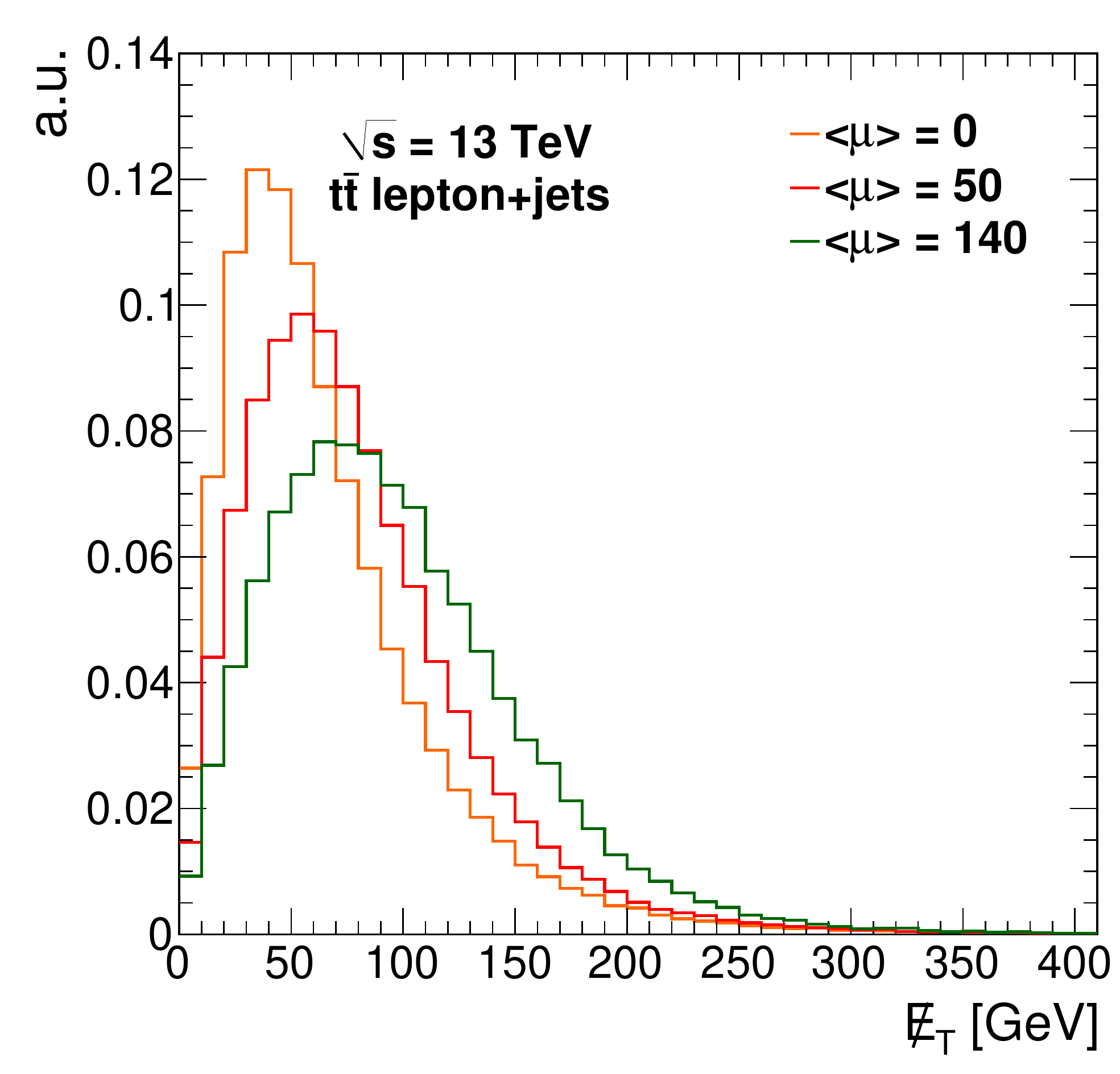}
\caption{Event \hadt\ (left) and \met\ (right).}
\label{fig:htmet}
\end{center}
\end{figure}
%\includegraphics[width=0.48\hsize]{Simulations/Plots%/MET_Delphes_Cat1.pdf}
%%%%%%%%%%%%%%%%%%%%%%%%%%%%%%%%%%%%%%%%

In Fig.~\ref{fig:htmet}, we show the scalar sum of the \pt\ of jets with \pt$>$30 GeV and $\vert\eta\vert<$2.5 (\hadt) 
and the event \met.  The \hadt\ distribution exhibits less pile-up dependence than the \met\ distribution mainly 
because \hadt\ is computed from jets that are pile-up subtracted and pass the \pt\ threshold requirement.
However, the \met\ is computed using all event objects (with minimal thresholds) in order to maintain an 
unbiased response to missing energy. Computing missing energy from jets passing a \pt\ threshold (often called 
missing \hadt) minimizes pile-up dependence, but can have bias in \met\ response at lower thresholds.

\begin{figure}[hbtp]
\begin{center}
\includegraphics[width=0.48\hsize]{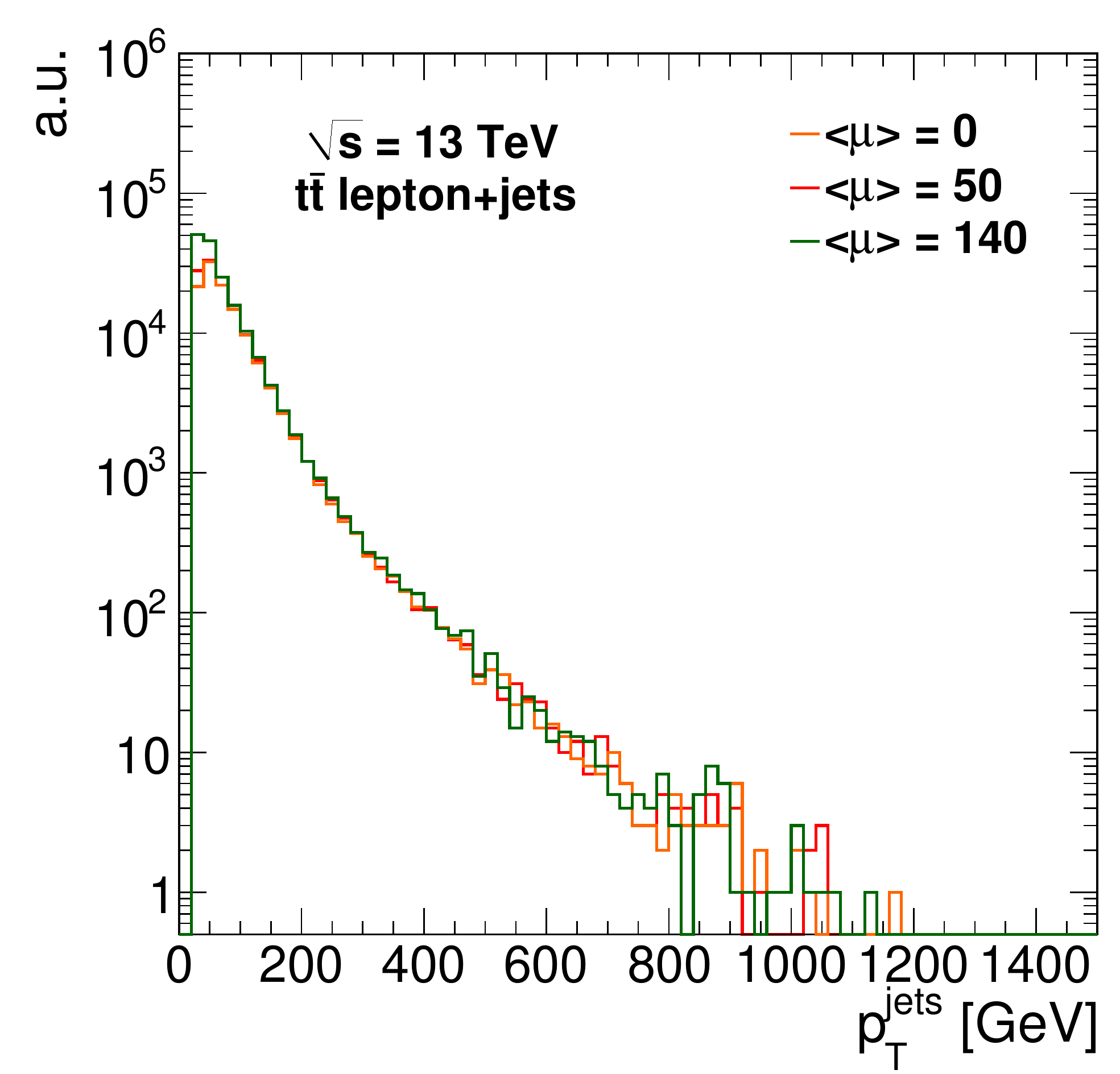}
\includegraphics[width=0.48\hsize]{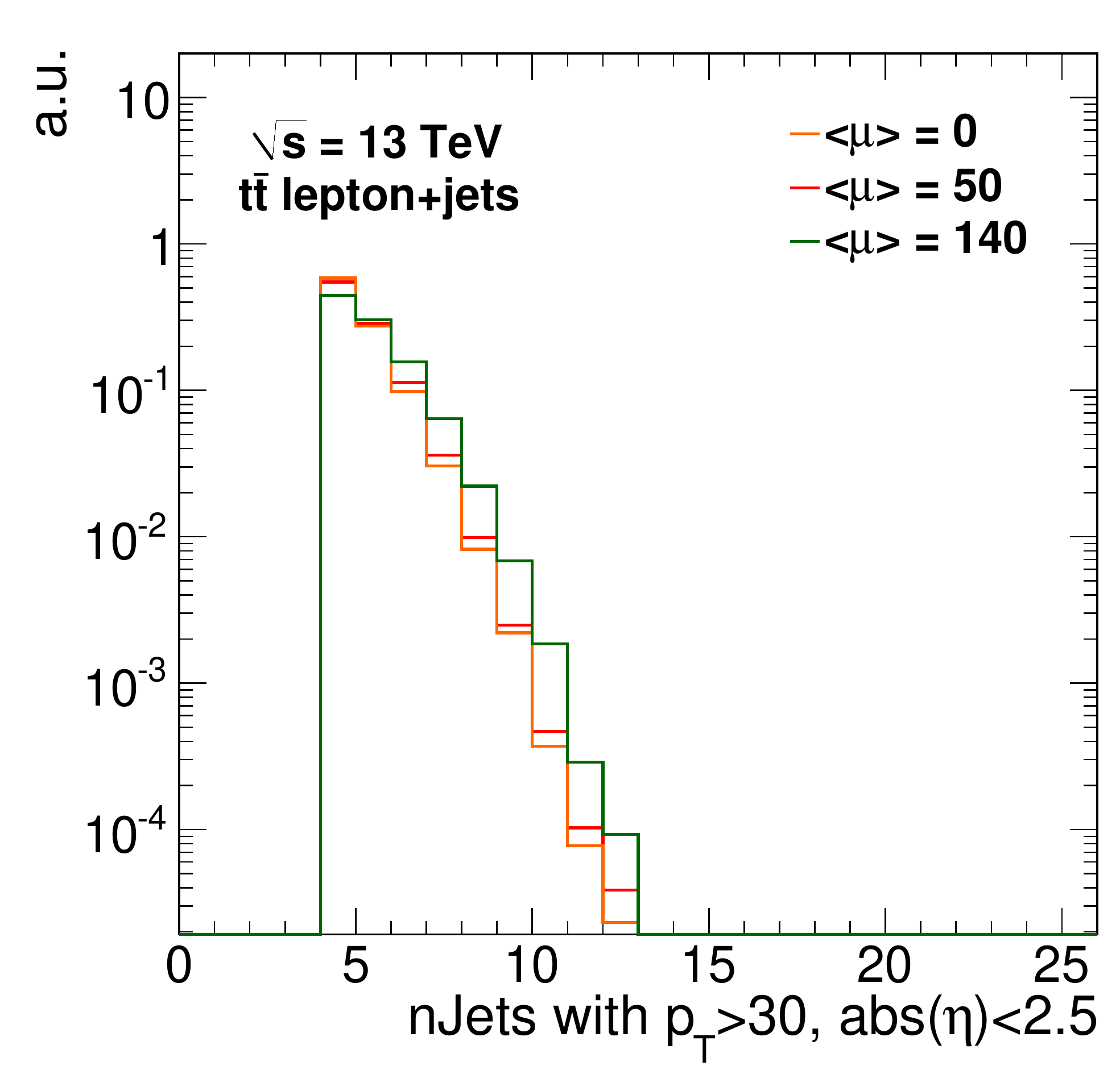}
\caption{Jet \pt\ (top left) and multiplicity (right) distributions.}
\label{fig:jetpteta}
\end{center}
\end{figure}

In Fig.~\ref{fig:jetpteta}, we show distributions for jet \pt\, and multiplicity.  
The effect of pile-up even after subtraction can be observed in the additional number of jets in the distribution.
In this forward detector region, where there is no acceptance due to tracking, both neutral and charged
pile-up interactions are subtracted using average $\rho$ area method.

\begin{figure}[hbtp]
\begin{center}
\includegraphics[width=0.8\hsize]{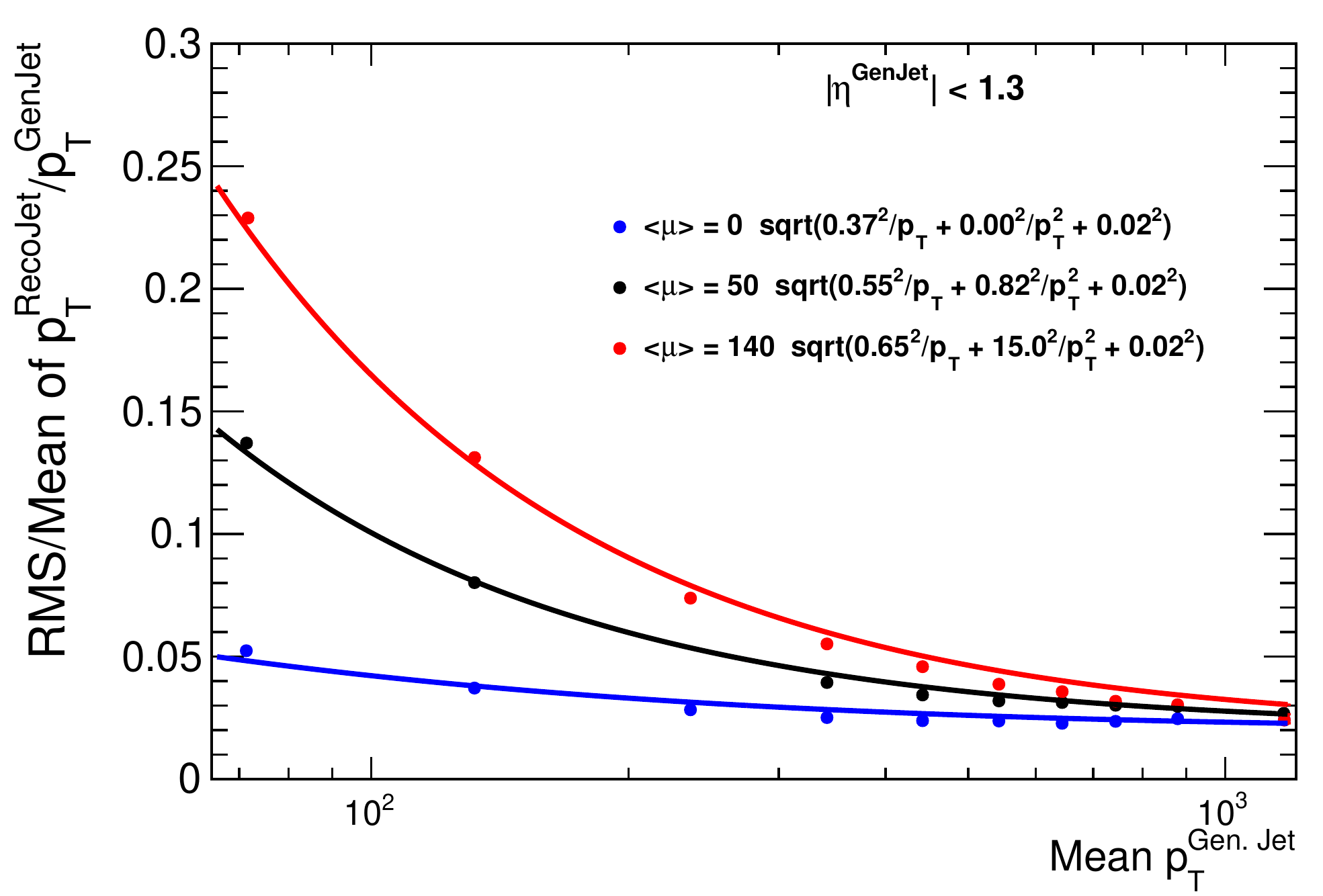}
\caption{Jet energy resolution as a function of jet $p_T$.}
\label{fig:jer}
\end{center}
\end{figure}

The jet energy resolution as a function of \pt\ and various pile-up scenarios is shown in  Fig.~\ref{fig:jer}.
The impact of pile-up is evident from the increase in the ``noise'' term in the resolution function.
%The functional form comes from [FIXME???].

\clearpage\newpage
\subsection{Lepton and photon performance}

\begin{figure}[hbtp]
\begin{center}
\includegraphics[width=0.48\hsize]{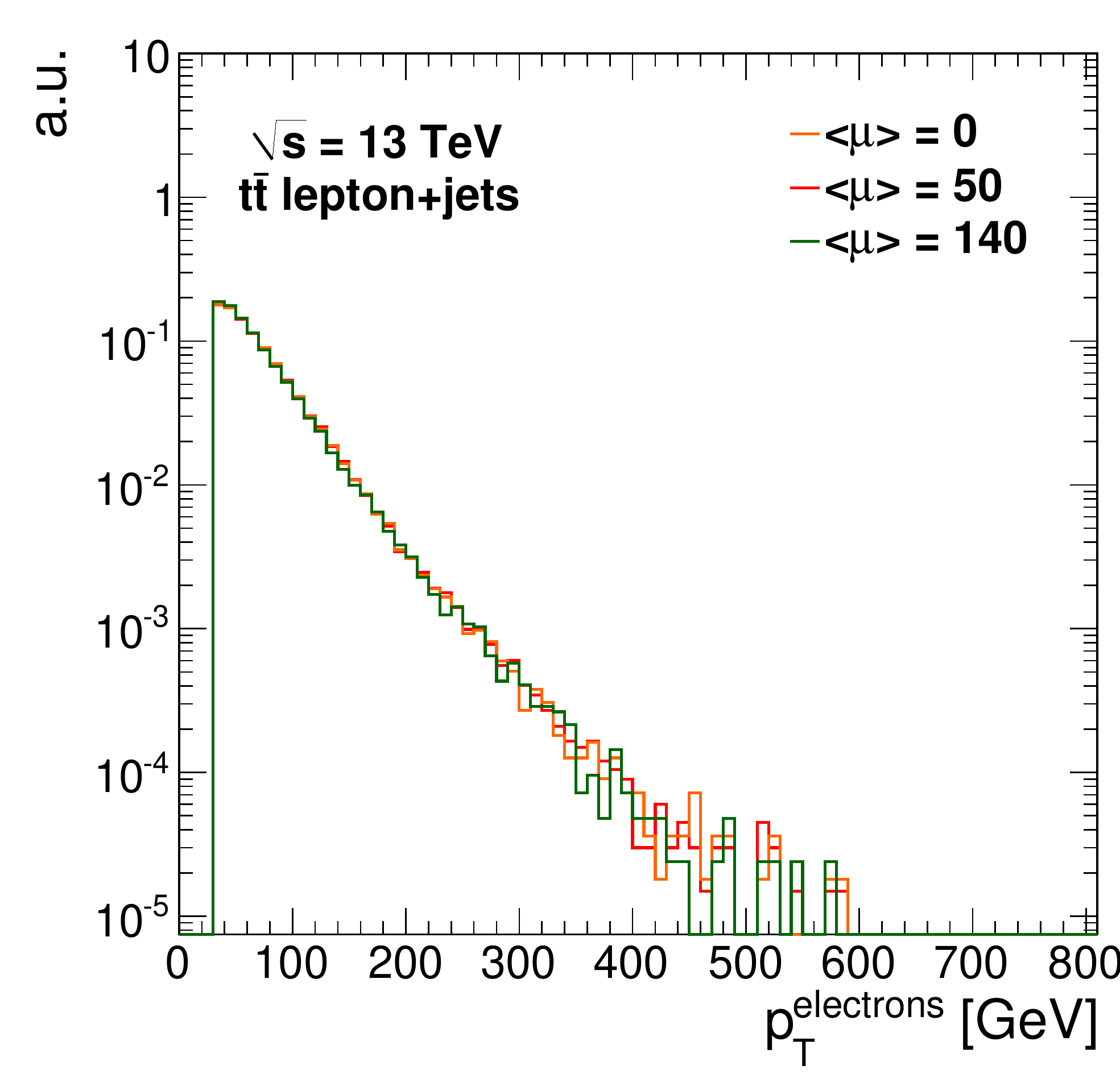}
\includegraphics[width=0.48\hsize]{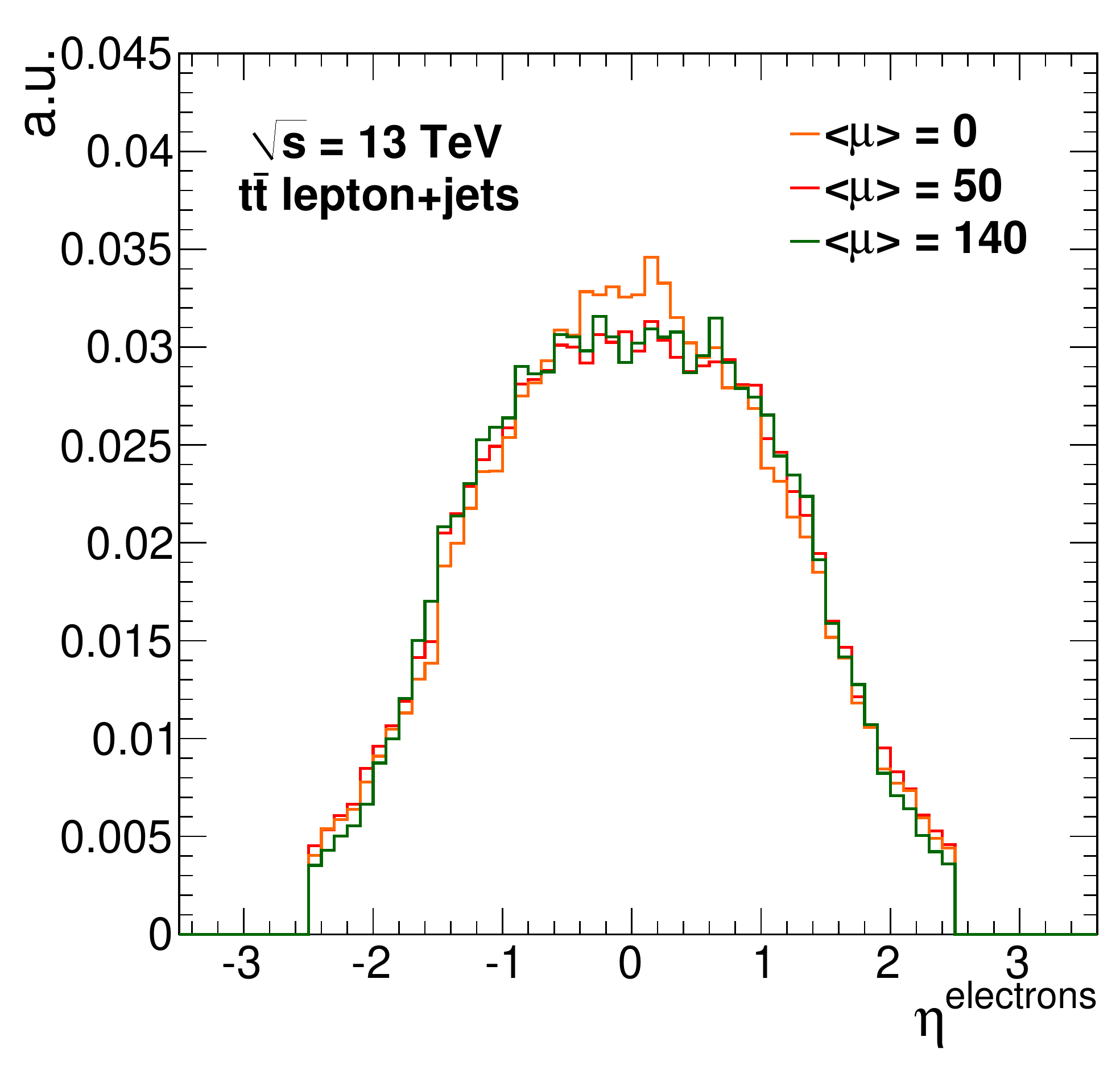}
\includegraphics[width=0.48\hsize]{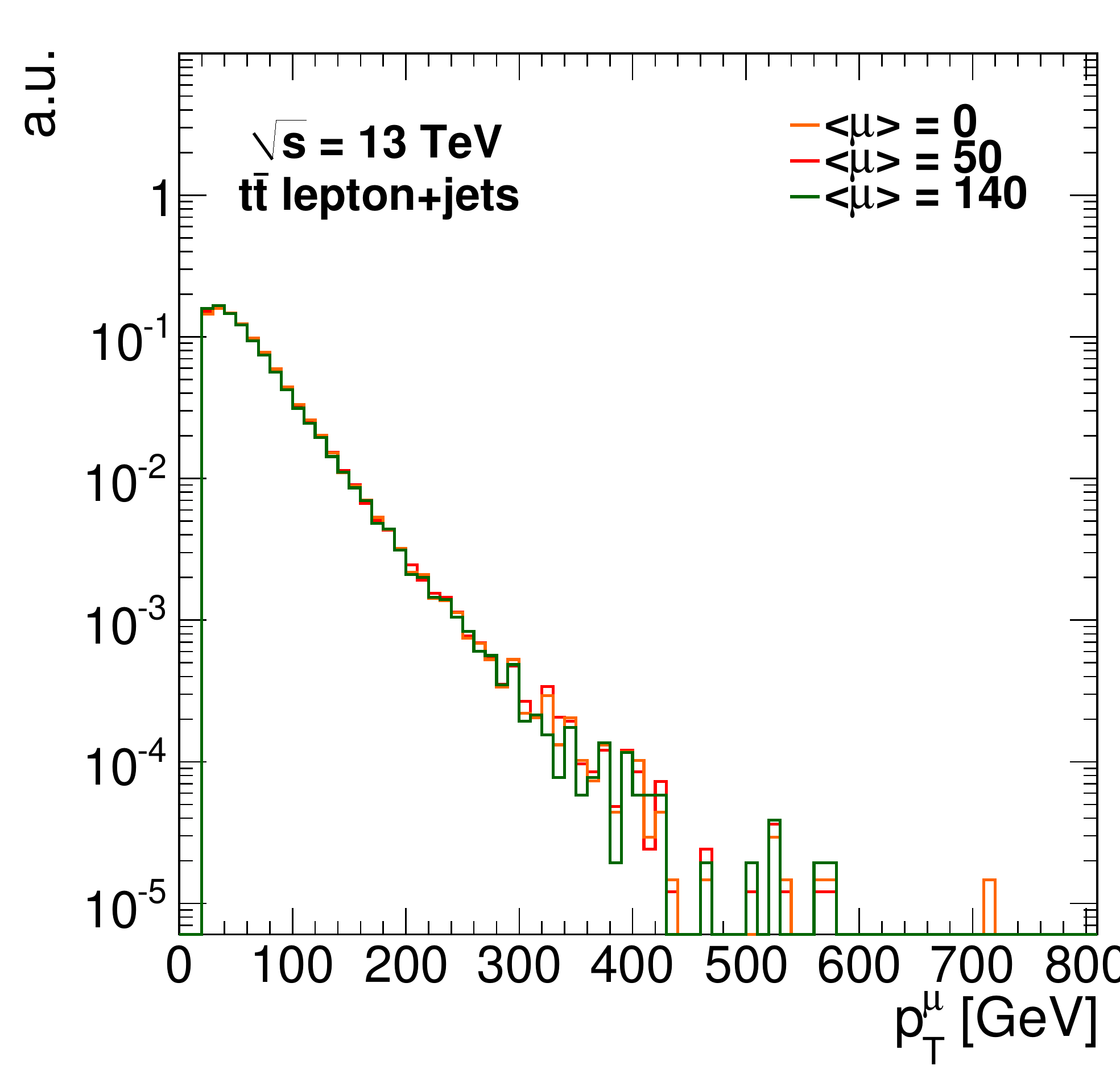}
\includegraphics[width=0.48\hsize]{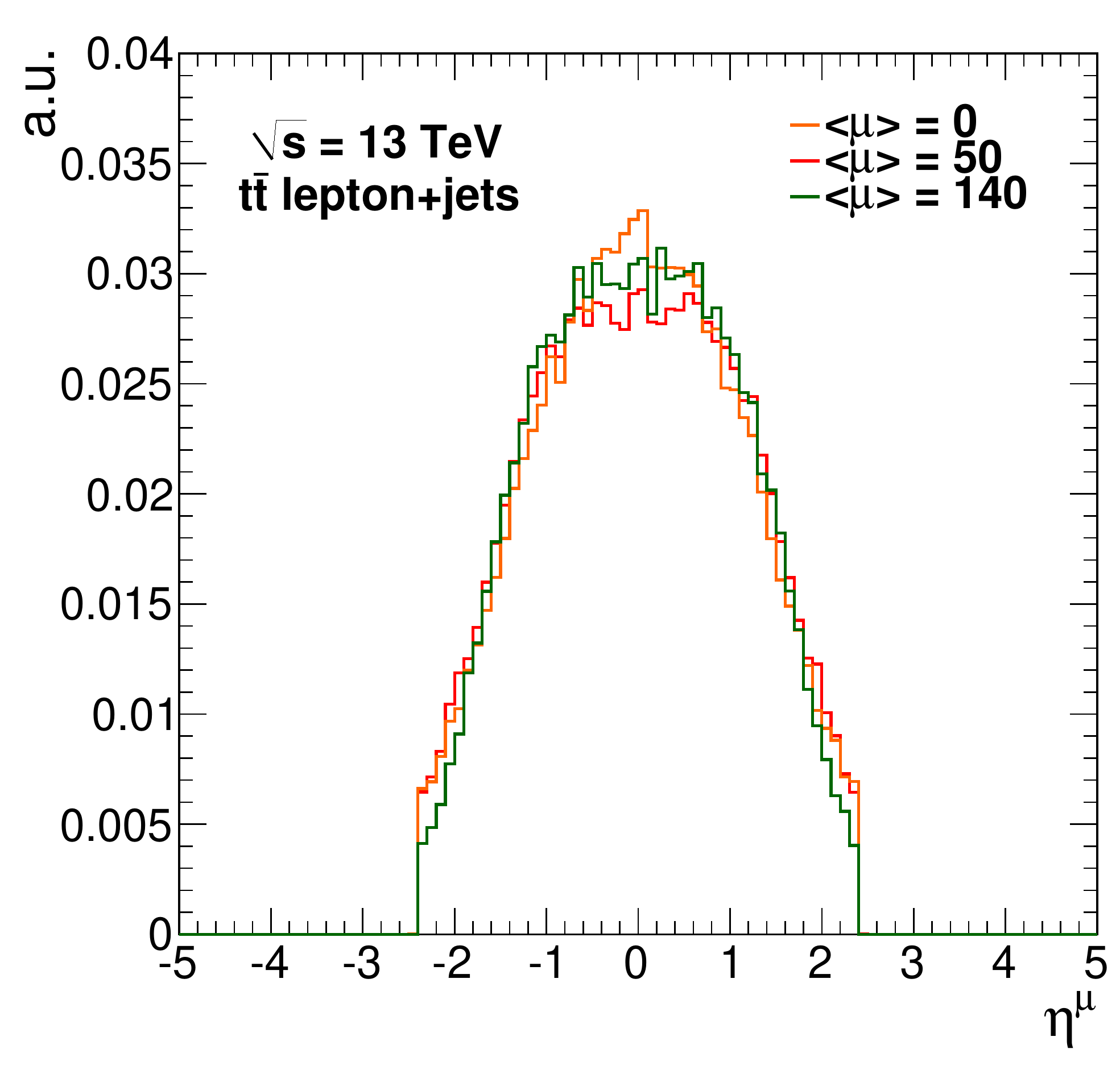}
\caption{Distributions for \pt\ (left) and $\eta$ (right) for electrons (top) and muons (bottom).}
\label{fig:leppteta}
\end{center}
\end{figure}

\begin{figure}[hbtp]
\begin{center}
\includegraphics[width=0.35\hsize]{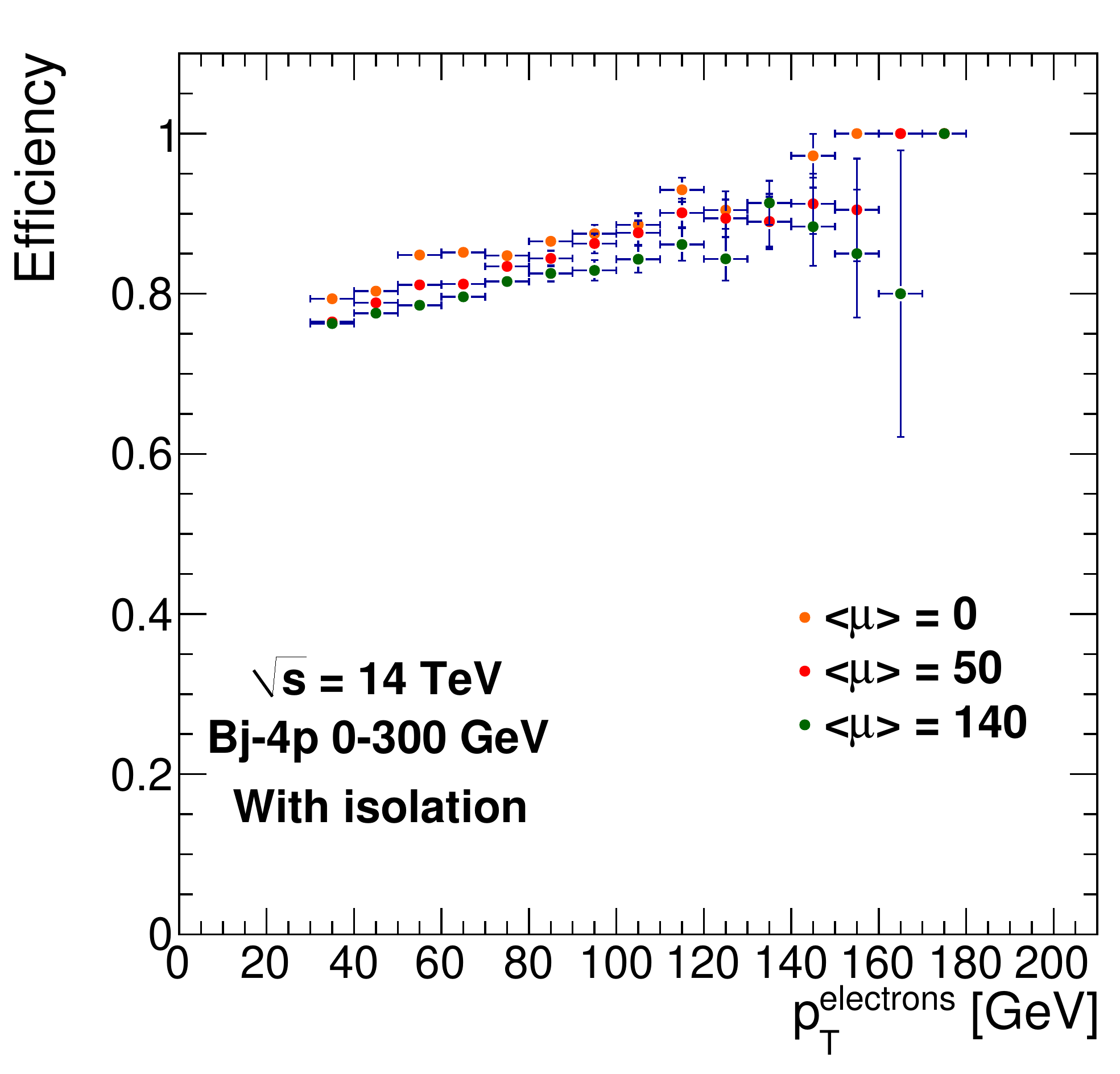}
\includegraphics[width=0.35\hsize]{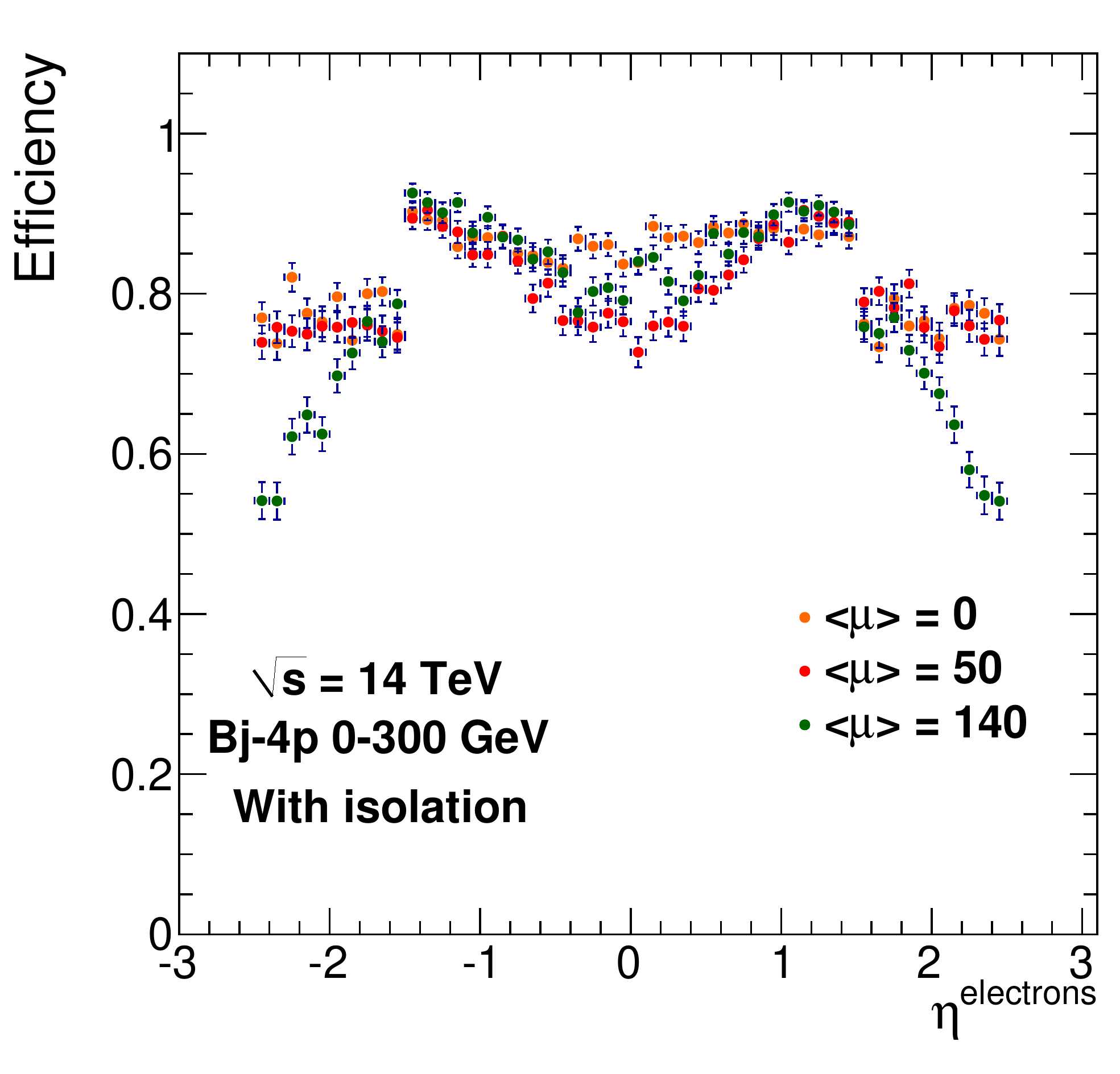}
\includegraphics[width=0.35\hsize]{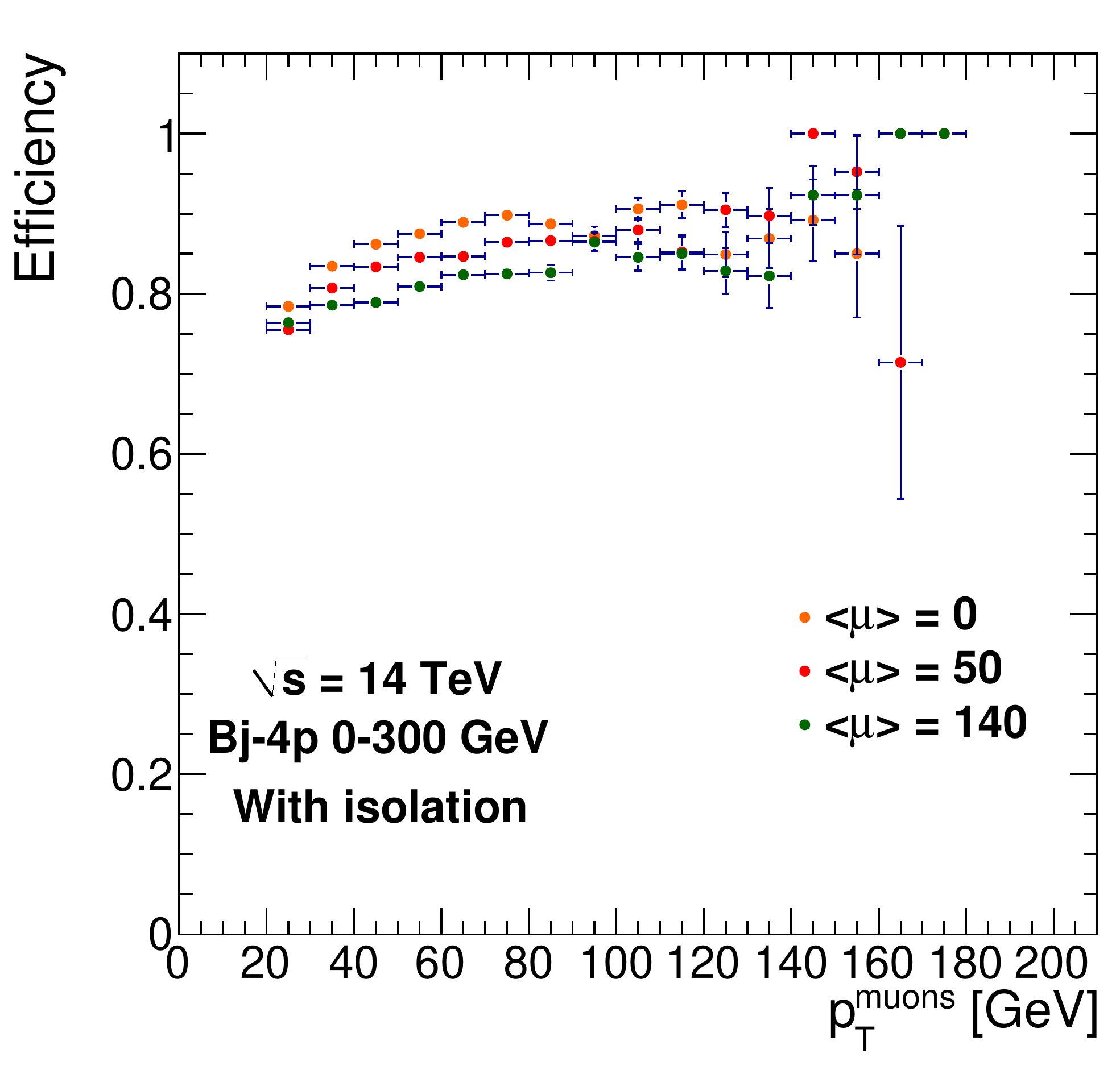}
\includegraphics[width=0.35\hsize]{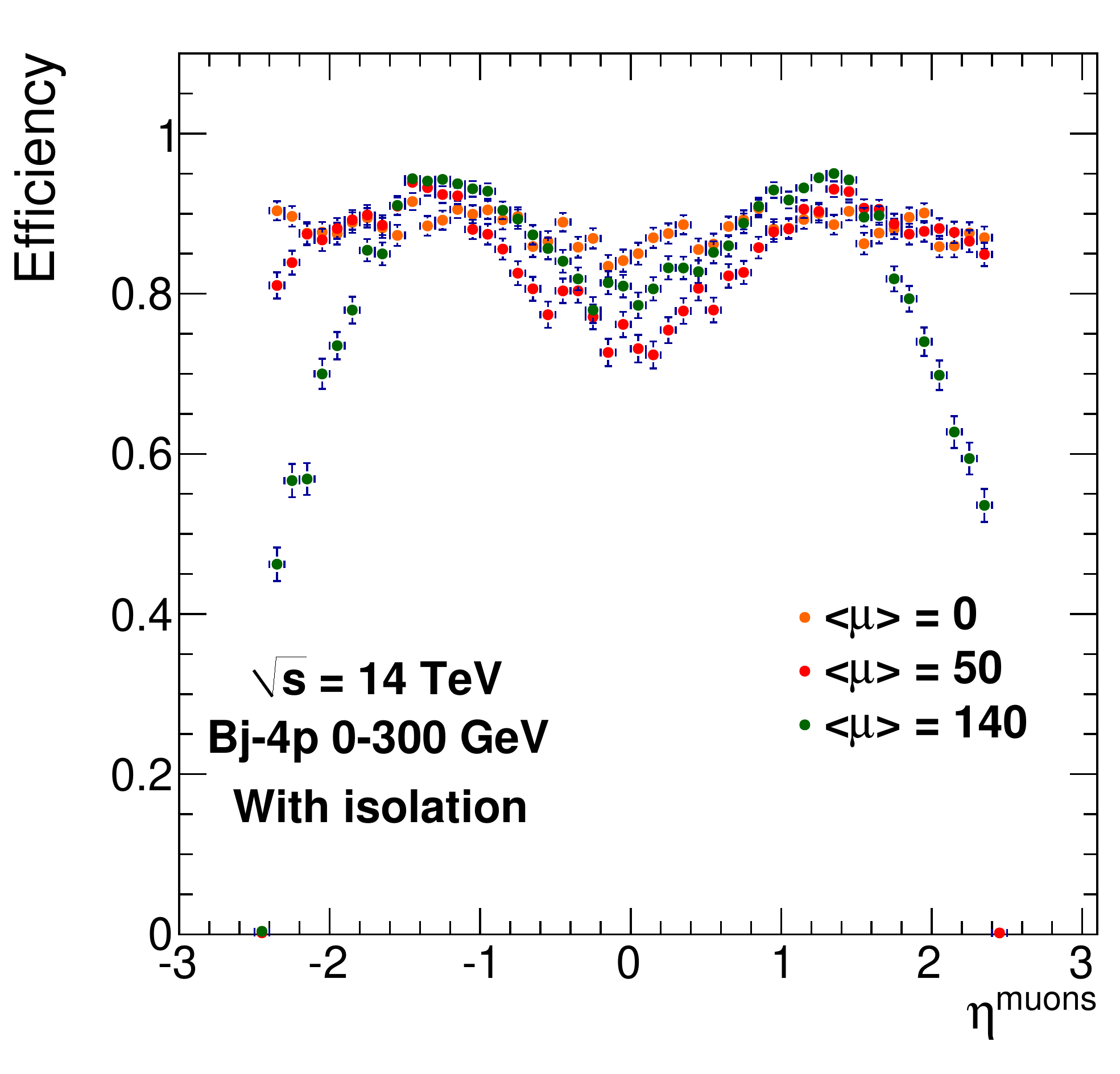}
\includegraphics[width=0.35\hsize]{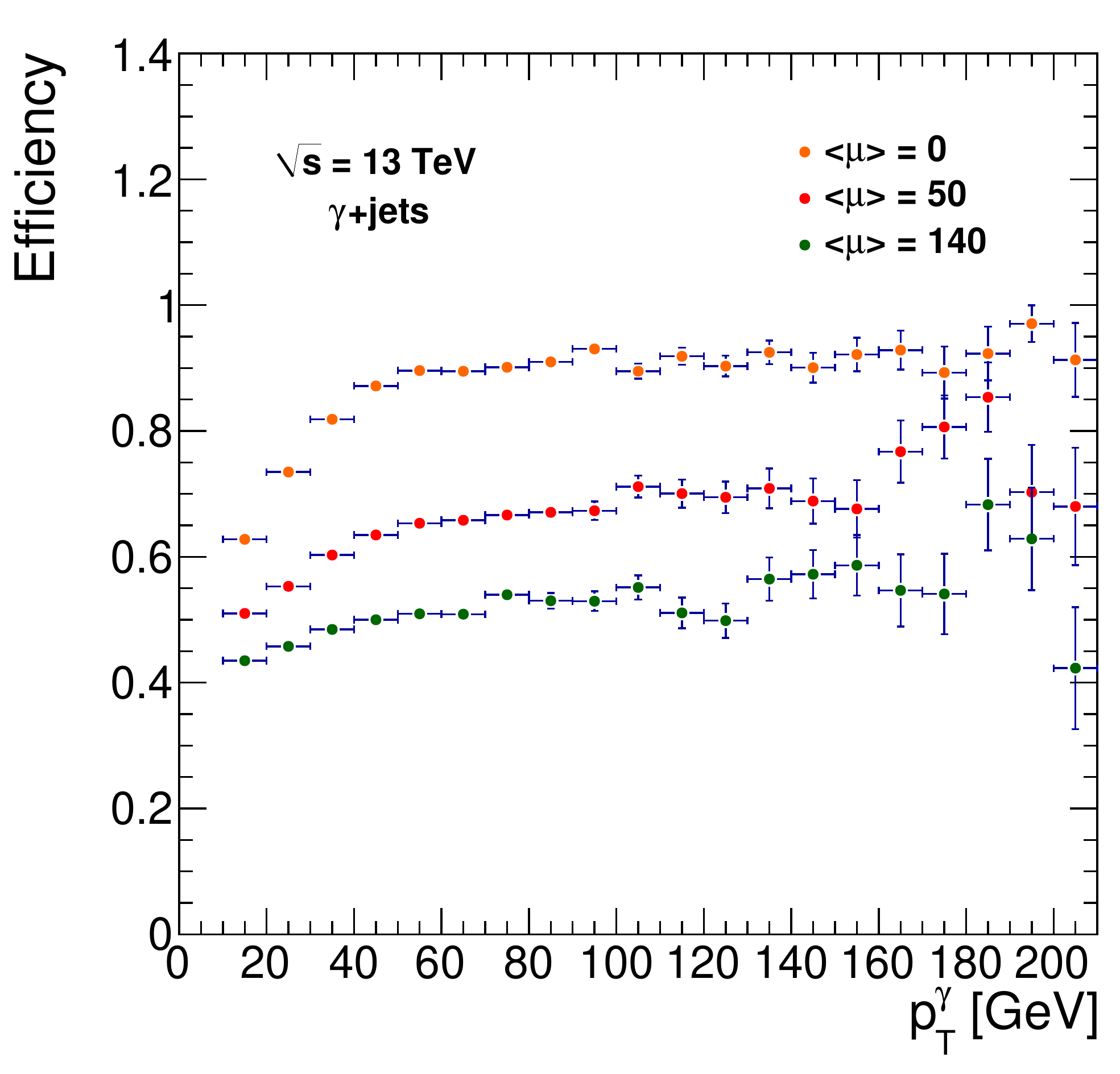}
\includegraphics[width=0.35\hsize]{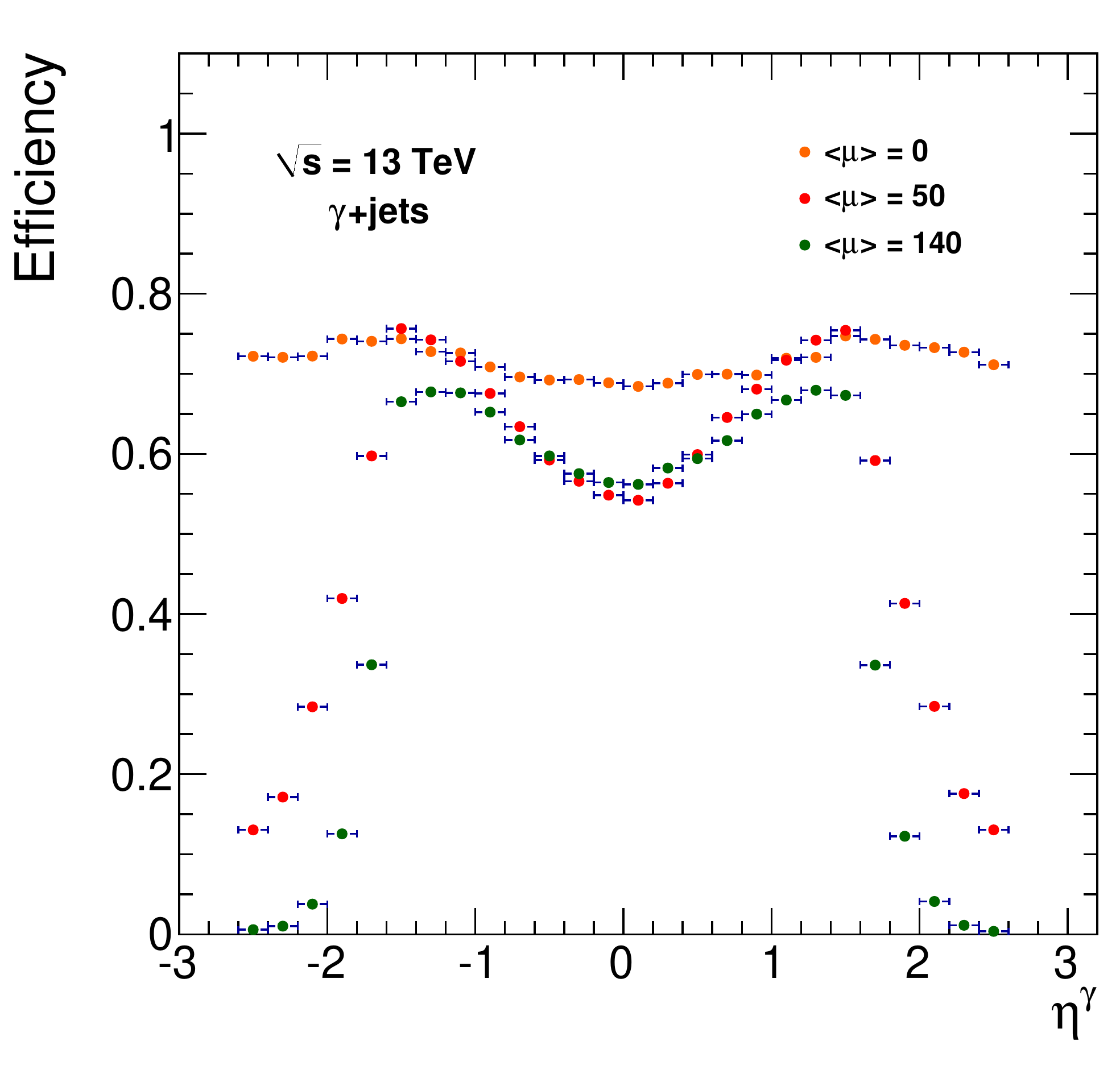}
\caption{Efficiency for reconstructing isolated electrons (top), muons (middle), and photons (bottom) as functions of \pt\ (left) and $\eta$ (right).}
\label{fig:lepeff}
\end{center}
\end{figure}

\begin{figure}[hbtp]
\begin{center}
\includegraphics[width=0.48\hsize]{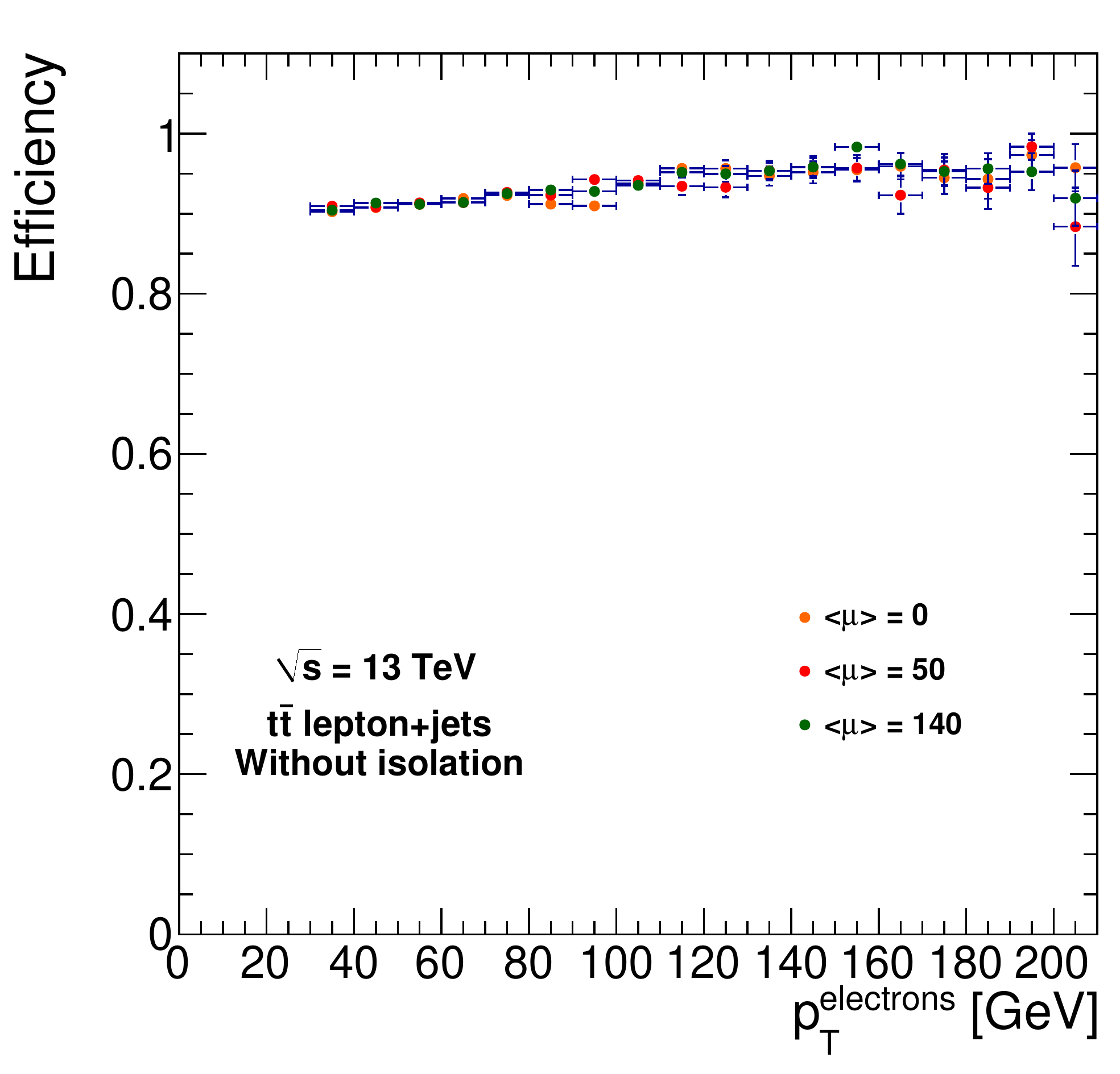}
\includegraphics[width=0.48\hsize]{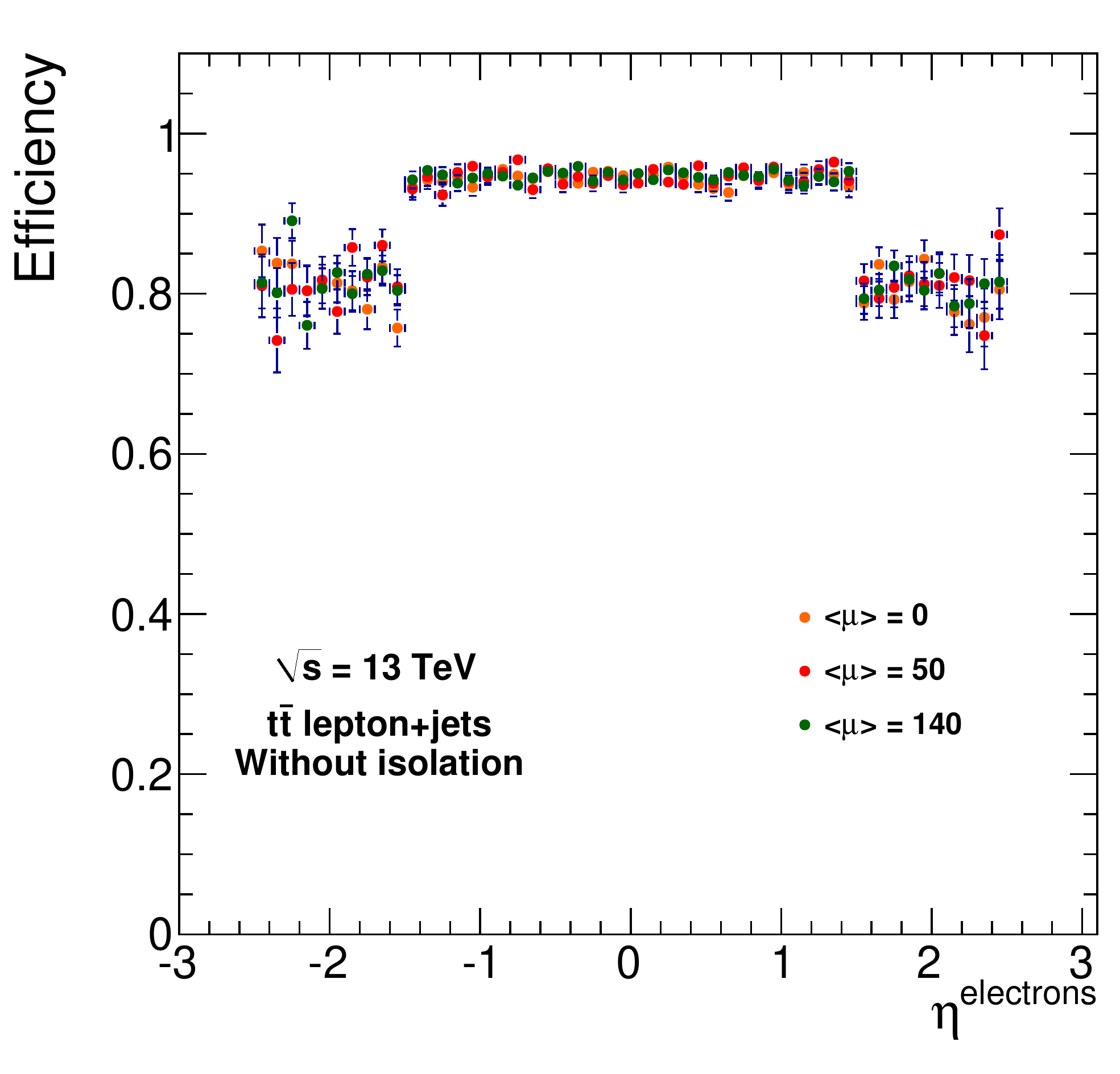}
\includegraphics[width=0.48\hsize]{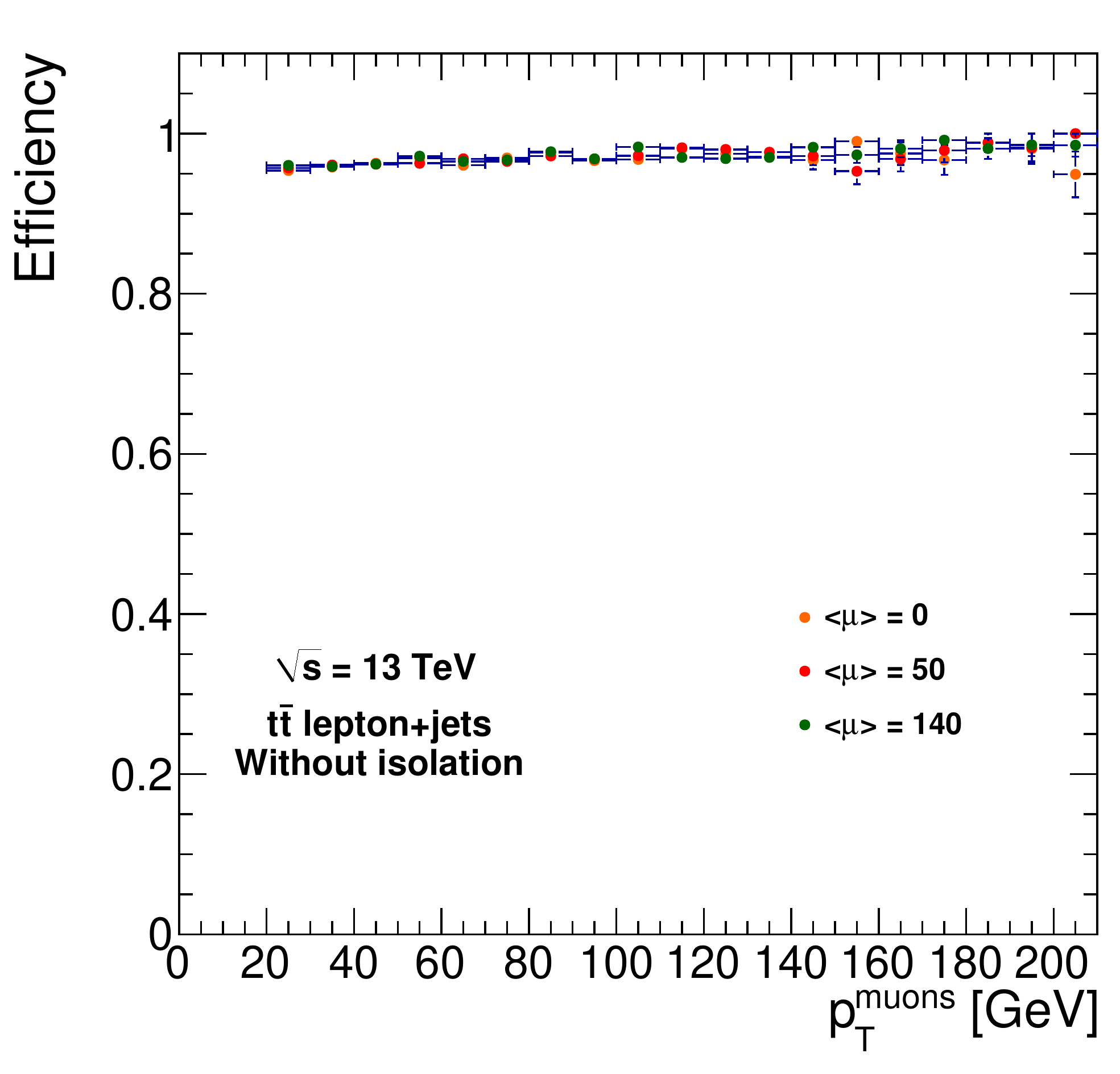}
\includegraphics[width=0.48\hsize]{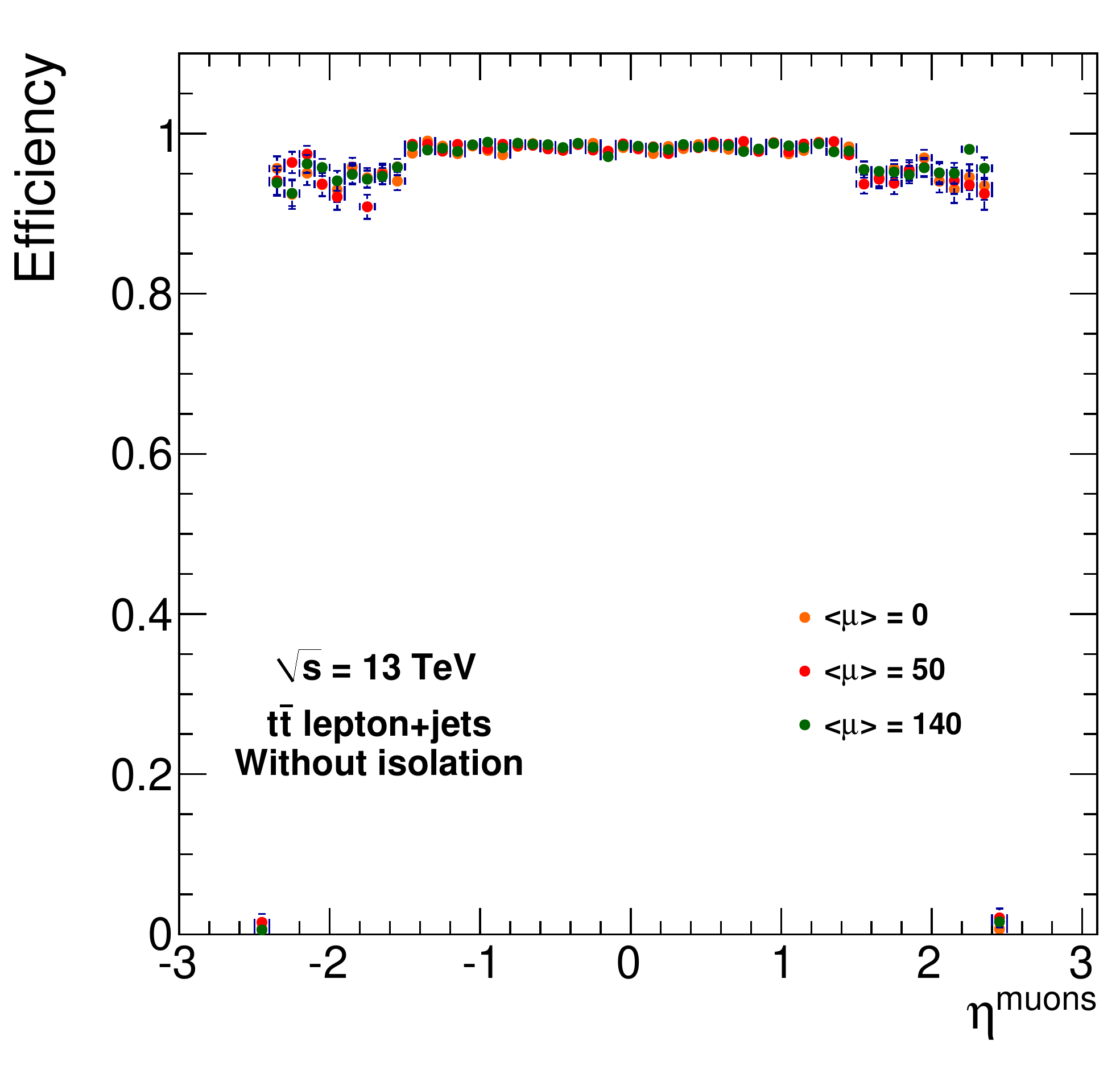}
\caption{Efficiency for reconstructing electrons (top), muons (middle) as functions of \pt\ (left) and $\eta$ (right) with no isolation requirement.}
\label{fig:lepeff_noiso}
\end{center}
\end{figure}

The lepton \pt\ and $\eta$ distributions are shown in Fig.~\ref{fig:leppteta}. They exhibit milder pile-up dependence
due to the subtraction used. Fig.~\ref{fig:lepeff} shows the 
resulting  efficiencies for leptons that pass reconstruction and isolation requirements.
%Distributions of lepton \pt\ and $\eta$ are show in Fig.~\ref{fig:leppteta},
%and the efficiencies for leptons to pass reconstruction and isolation are shown in Fig.~\ref{fig:lepeff}.  
These efficiencies result from the combined effects of tracking efficiency, reconstruction efficiency, 
resolution, and isolation.  In Fig.~\ref{fig:lepeff_noiso}, we show lepton efficiency with no isolation 
requirement. These are used as inputs with extrapolations using the LHC data [xx].

%%%%%%%%%%%%%%%%%%%%%%%%%%%%%%%%%%%%%%%%%%%
% Jet Substructure Variables
%%%%%%%%%%%%%%%%%%%%%%%%%%%%%%%%%%%%%%%%%%%

\newpage
\subsection{Jet substructure performance}

As described in Sec.~\ref{sec:jetsubstructure}, top-tagging is based on trimmed jet mass and the number of subjets, and $W$-tagging is based on trimmed jet mass and mass drop.

\begin{figure}[hbtp]
\begin{center}
\includegraphics[width=0.48\hsize]{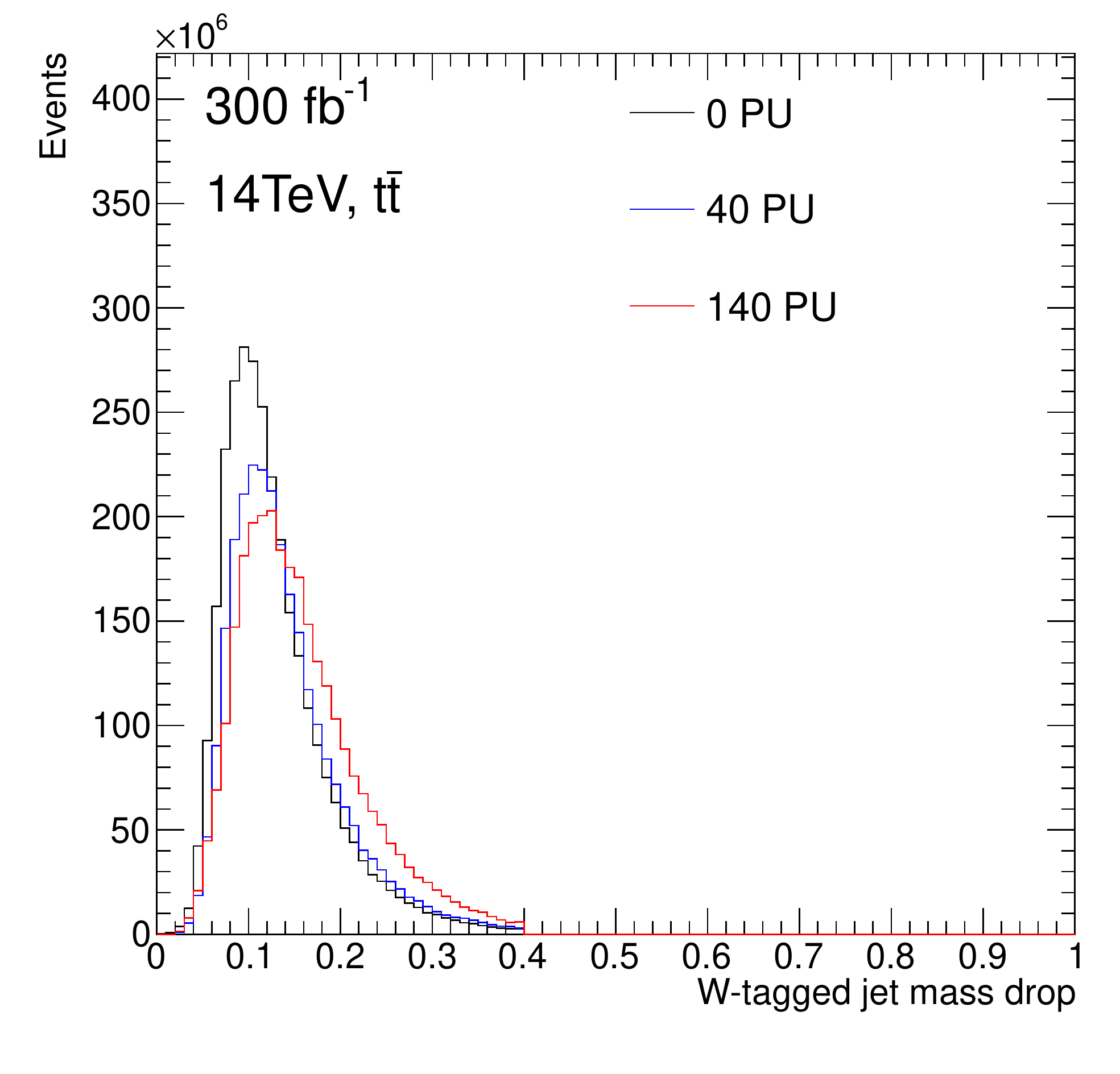}
\caption{The mass drop variable for $W-tagged $ jets; the top-tagging criteria do not include mass drop.}
\label{fig:mdrop}
\end{center}
\end{figure}

\begin{figure}[hbtp]
\begin{center}
\includegraphics[width=0.48\hsize]{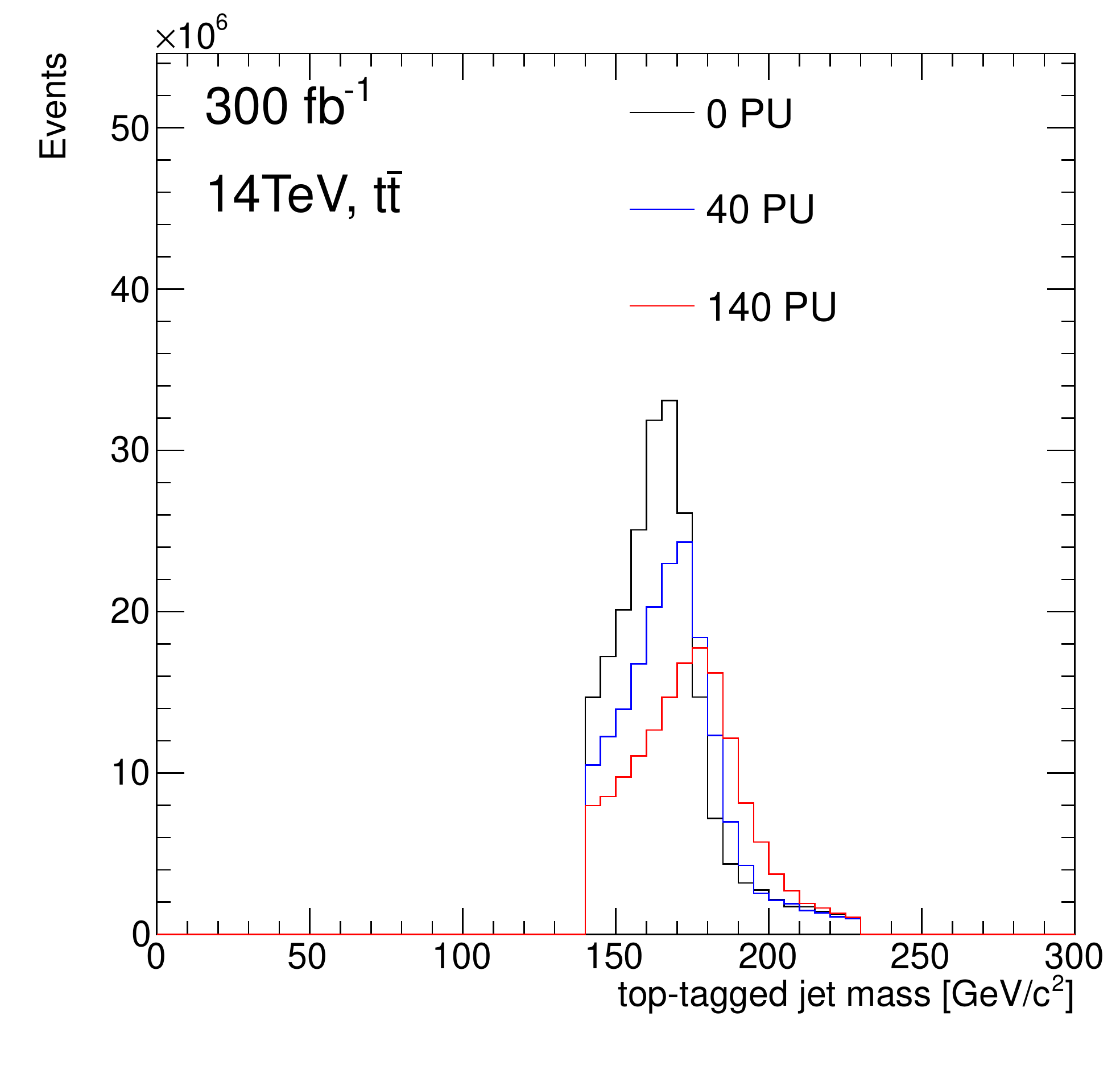}
\includegraphics[width=0.48\hsize]{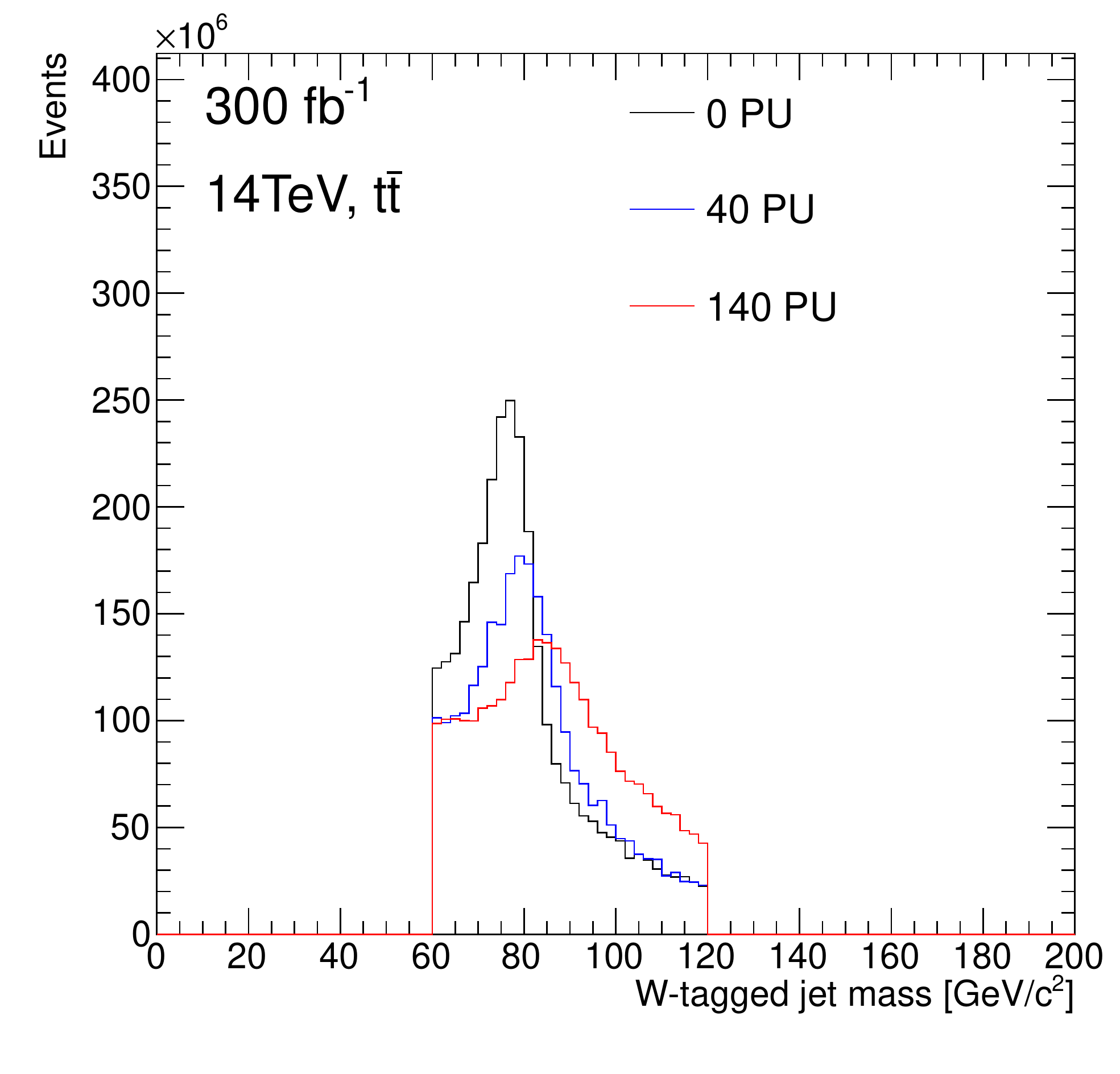}
\includegraphics[width=0.48\hsize]{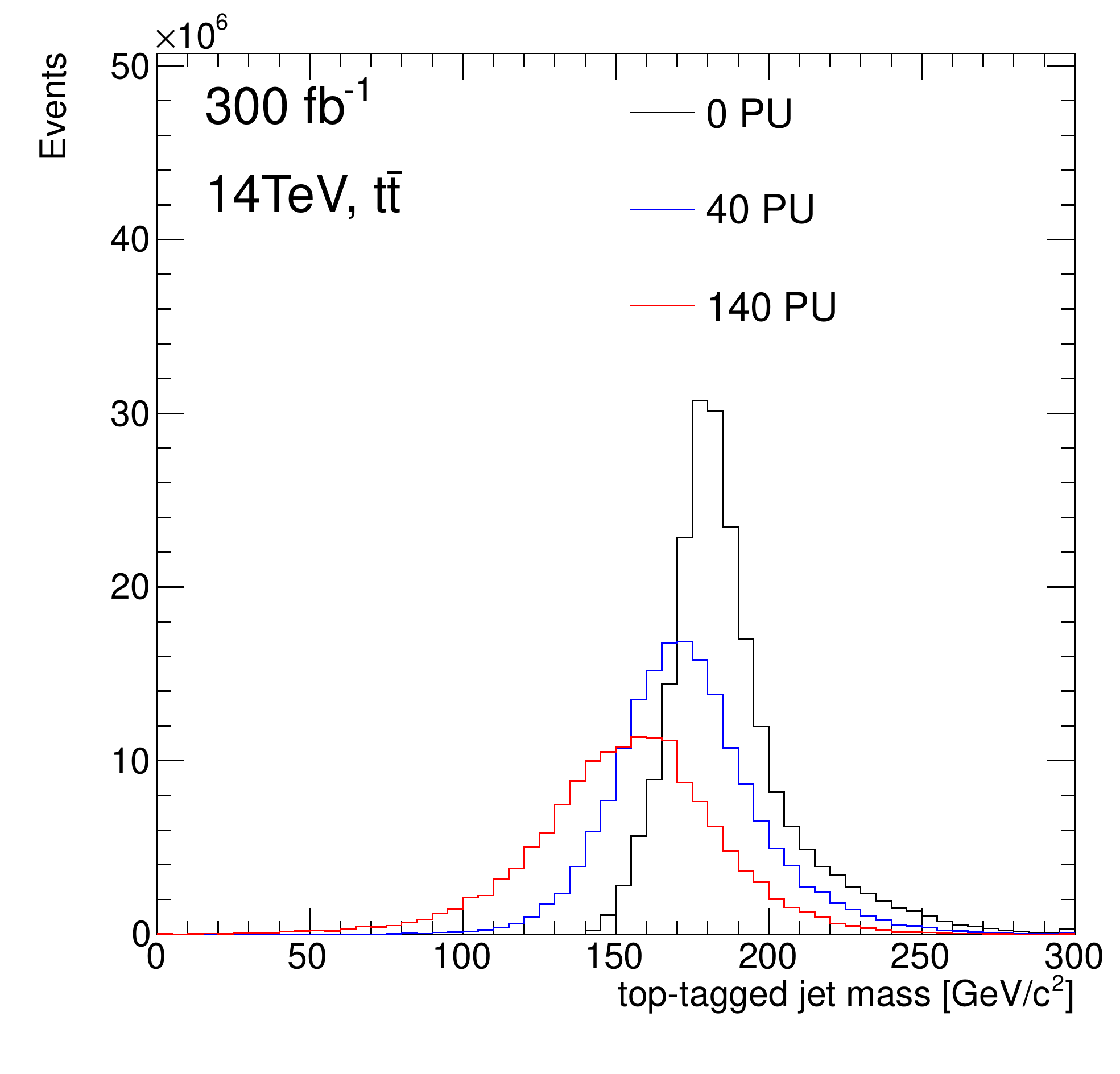}
\includegraphics[width=0.48\hsize]{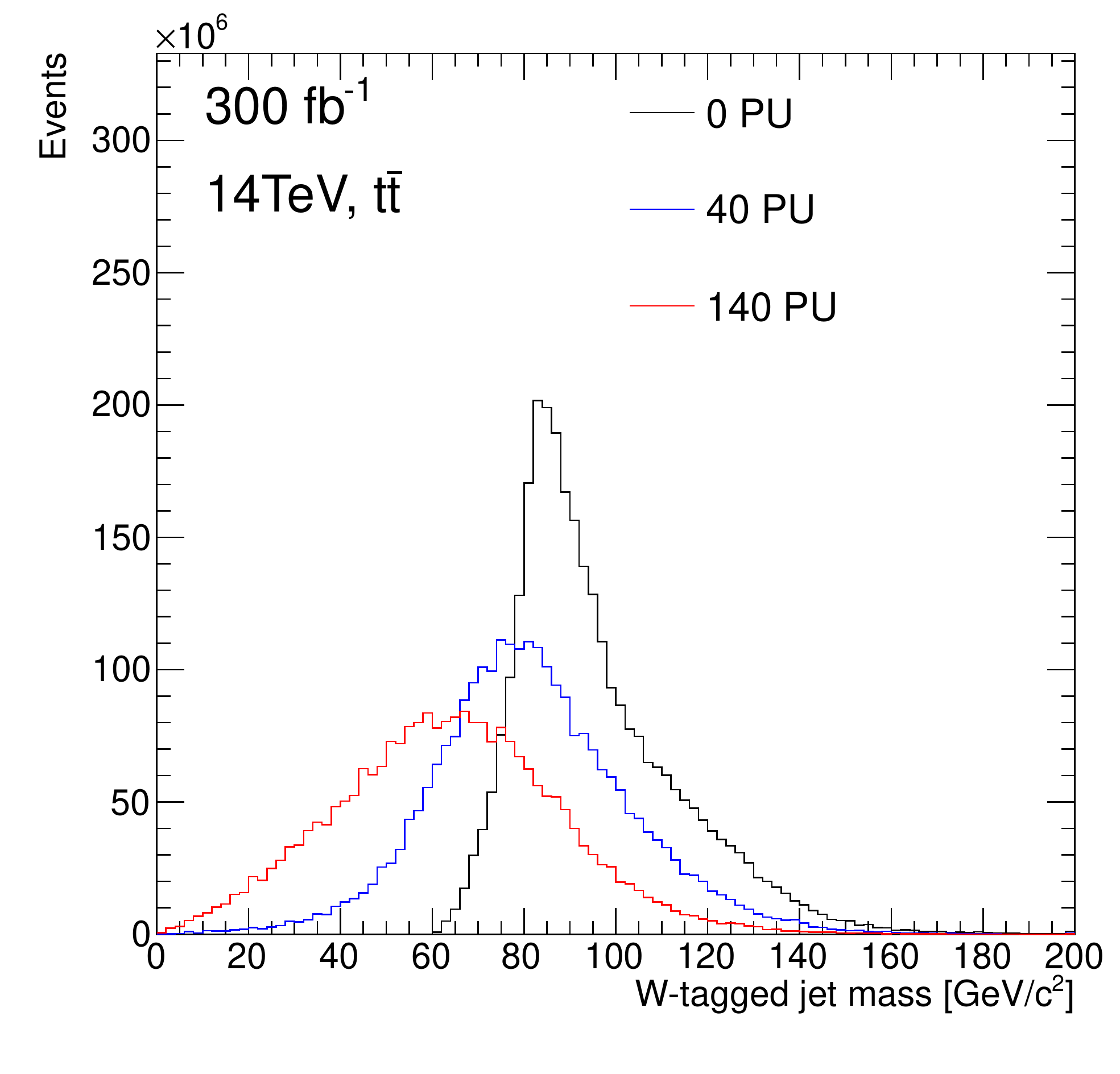}
\caption{Trimmed jet mass for top-tagged jets (top left)
and $W$-tagged jets (top right) and jet mass for top-tagged jets (bottom left) and $W$-tagged jets (bottom right).}
\label{fig:tmass}
\end{center}
\end{figure}

Distributions for the number of subjets were already shown above in 
Fig.~\ref{fig:nsubjets}.
%In Fig.~\ref{fig:nsubjet}, we show distributions for the number of %subjets for all CA8 jets with $\pt>200$ GeV and the number of subjets %for top-tagged jets taken
%from $t\bar{t}+$jets events.  
In Fig.~\ref{fig:mdrop}, we show the mass drop variable for $W$-tagged jets, and in Fig.~\ref{fig:tmass}, the trimmed jet mass for top- and $W$-tagged jets.  The effect of pile-up is apparent in the trimmed and untrimmed jet mass in Fig.~\ref{fig:tmass}: the untrimmed mass decreases with pile-up because of over-correction, while the trimmed mass increases 
slightly with pile-up.

\begin{figure}[hbtp]
\begin{center}
\includegraphics[width=0.58\hsize]{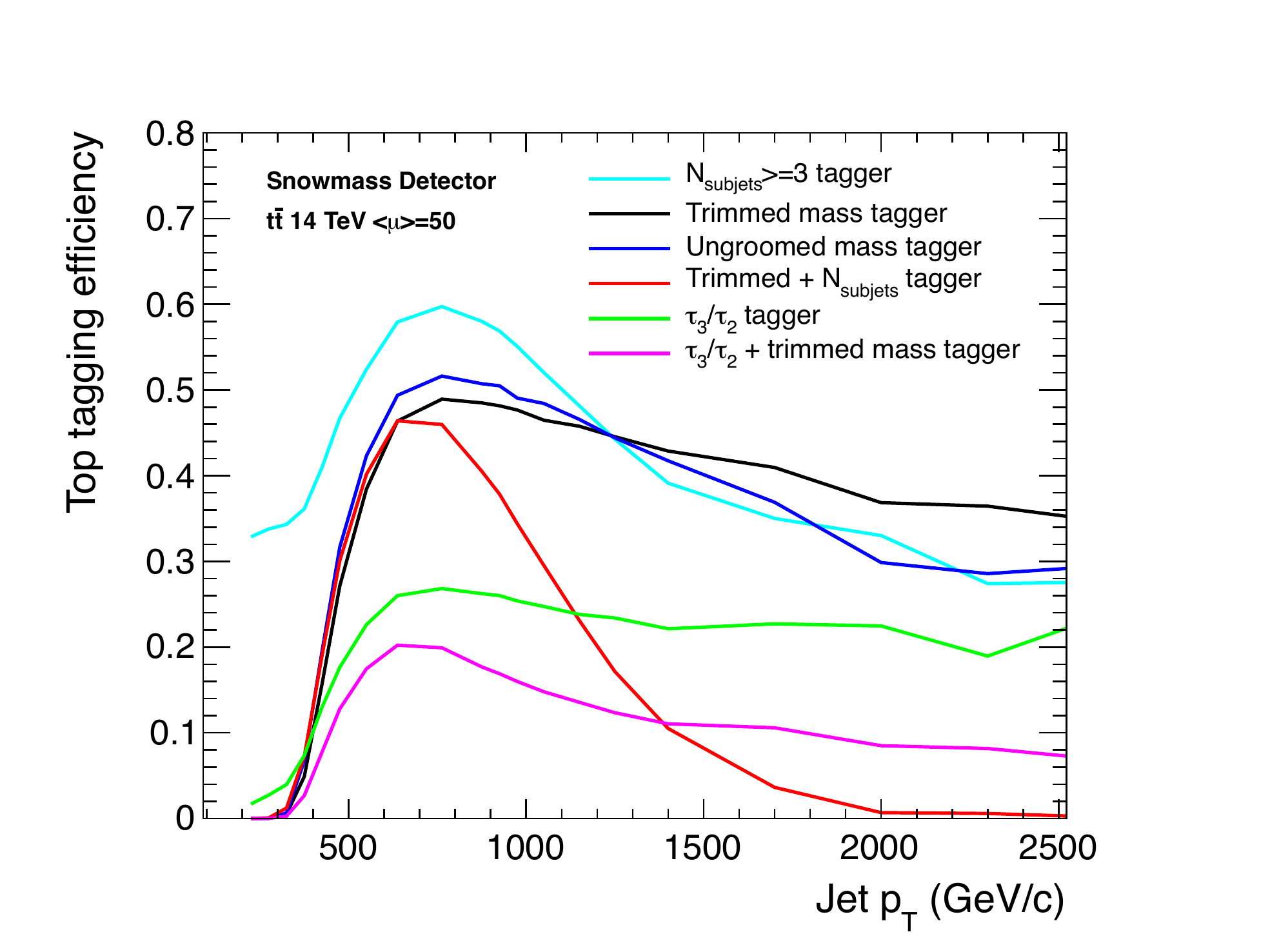}
\includegraphics[width=0.48\hsize]{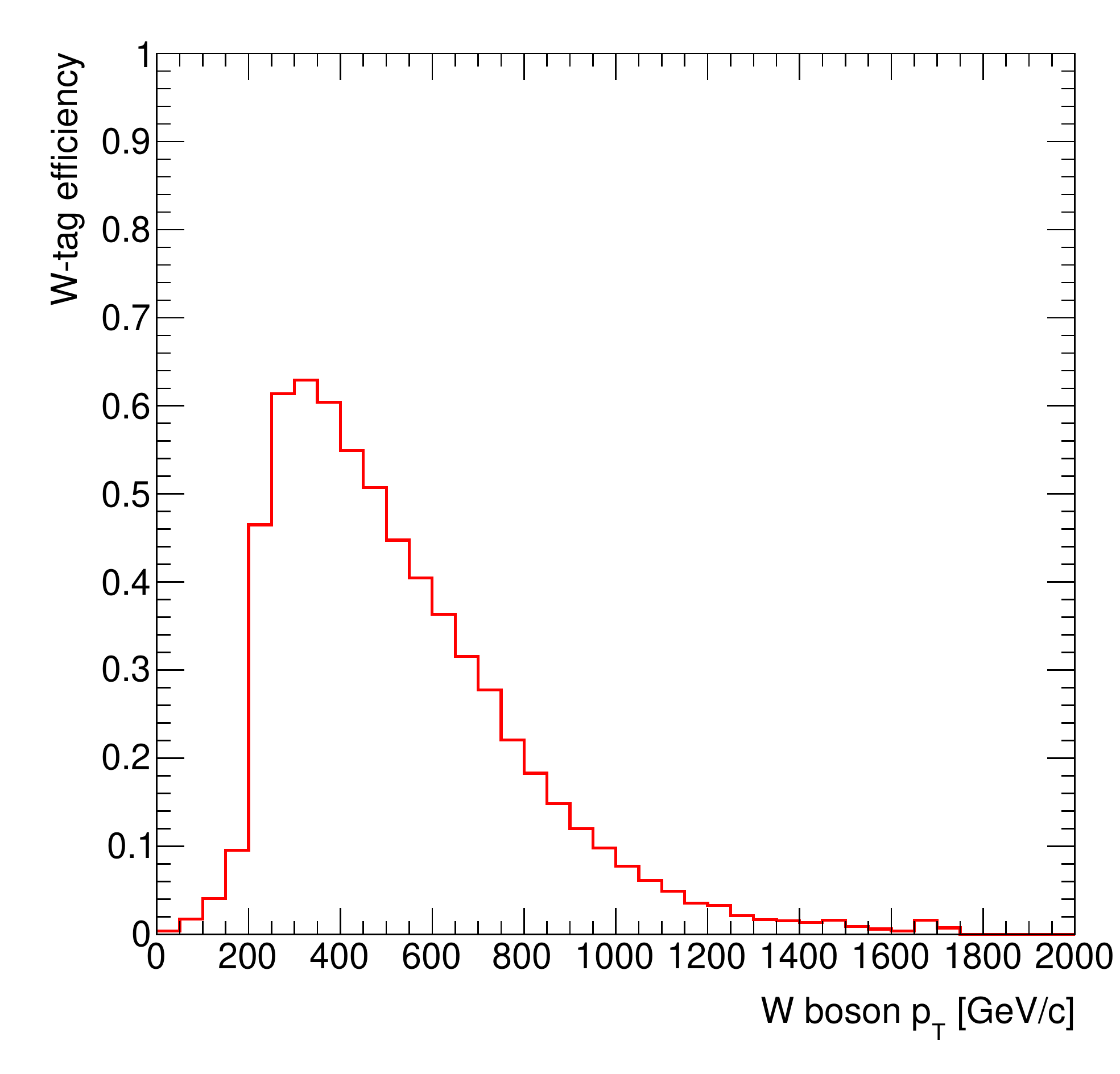}

\caption{Top-tagging efficiency versus jet \pt\ for taggers based on several combinations of substructure variables (top) and $W$-tagging efficiency for the default tagger (bottom).}
\label{fig:ttageff}
\end{center}
\end{figure}

In Fig.~\ref{fig:ttageff}, we show the top-tagging efficiency as a function of jet \pt\ for several variables.  
Though the algorithm is stable with different pile-up conditions, it is efficient up to $\pt<1000$ GeV, the top-tagging
efficiency (red curve) degrades for jet \pt\ exceeding 1 TeV.  Alternate top-tagging algorithms should be studied
in case of analyzes sensitive to very high \pt\ regions.
%\begin{figure}[hbtp]
%\begin{center}
%\includegraphics[width=0.48\hsize]{Simulations/Plots/tt_h_allCuts_CAJetN%SJ_14.pdf}
%\includegraphics[width=0.48\hsize]{Simulations/Plots/tt_h_allCuts_CATopJ%et1NSJ_14.pdf}
%\caption{Number of subjets for all s (left) and top-tagged jets (right).  %We require three or more subjets as part of the top-tagging criteria.}
%\label{fig:nsubjet}
%\end{center}
%\end{figure}

%%%%%%%%%%%%%%%%%%%%%%%%%%%%%%%%%%%%%%%%%%%
% B-tagging and tau-tagging
%%%%%%%%%%%%%%%%%%%%%%%%%%%%%%%%%%%%%%%%%%%
\clearpage\newpage
\subsection{Performance of $b$- and $\tau$-tagging}

\begin{figure}[hp]
\begin{center}
\includegraphics[width=0.33\hsize]{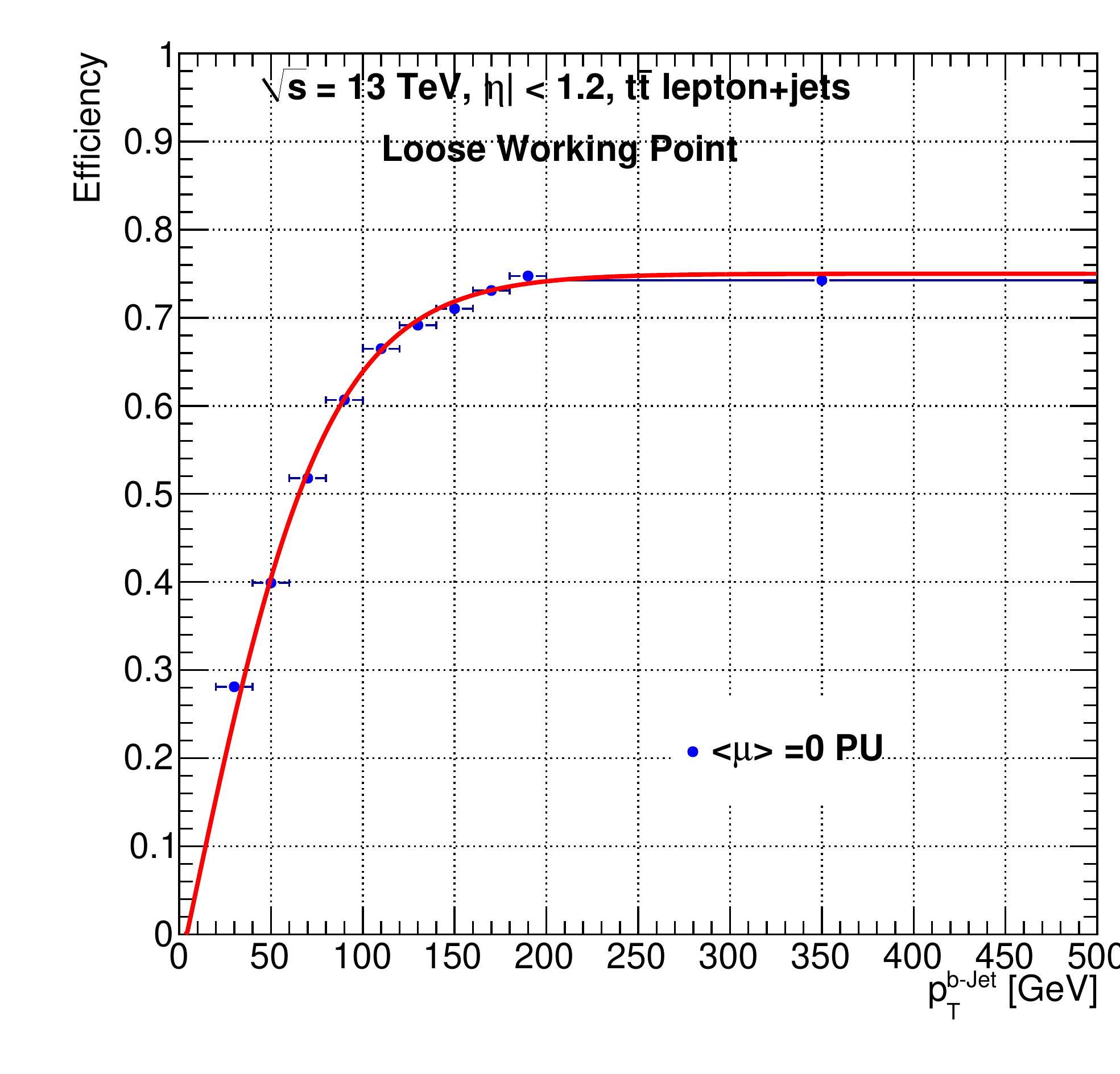}
\includegraphics[width=0.33\hsize]{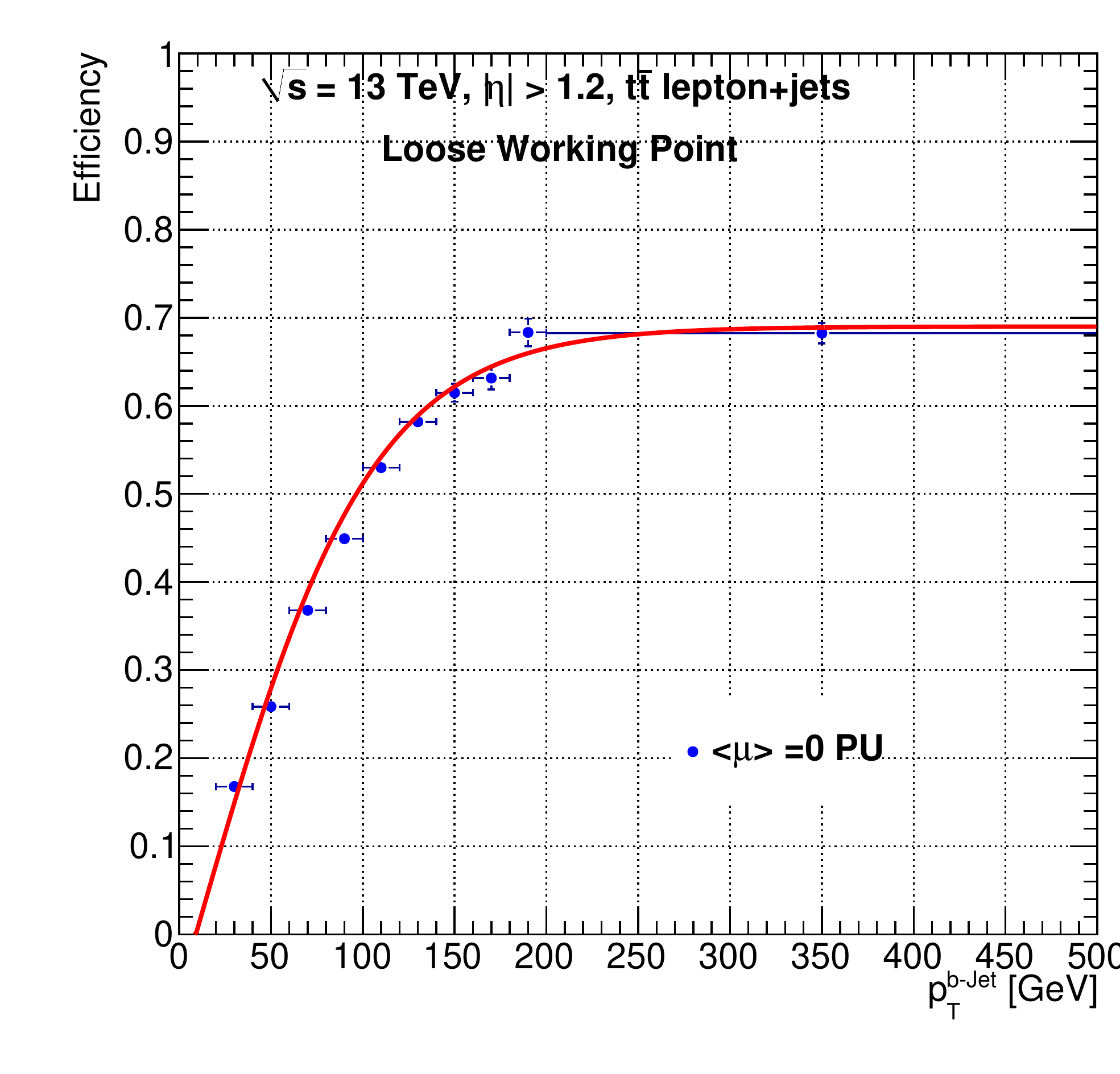}
\includegraphics[width=0.33\hsize]{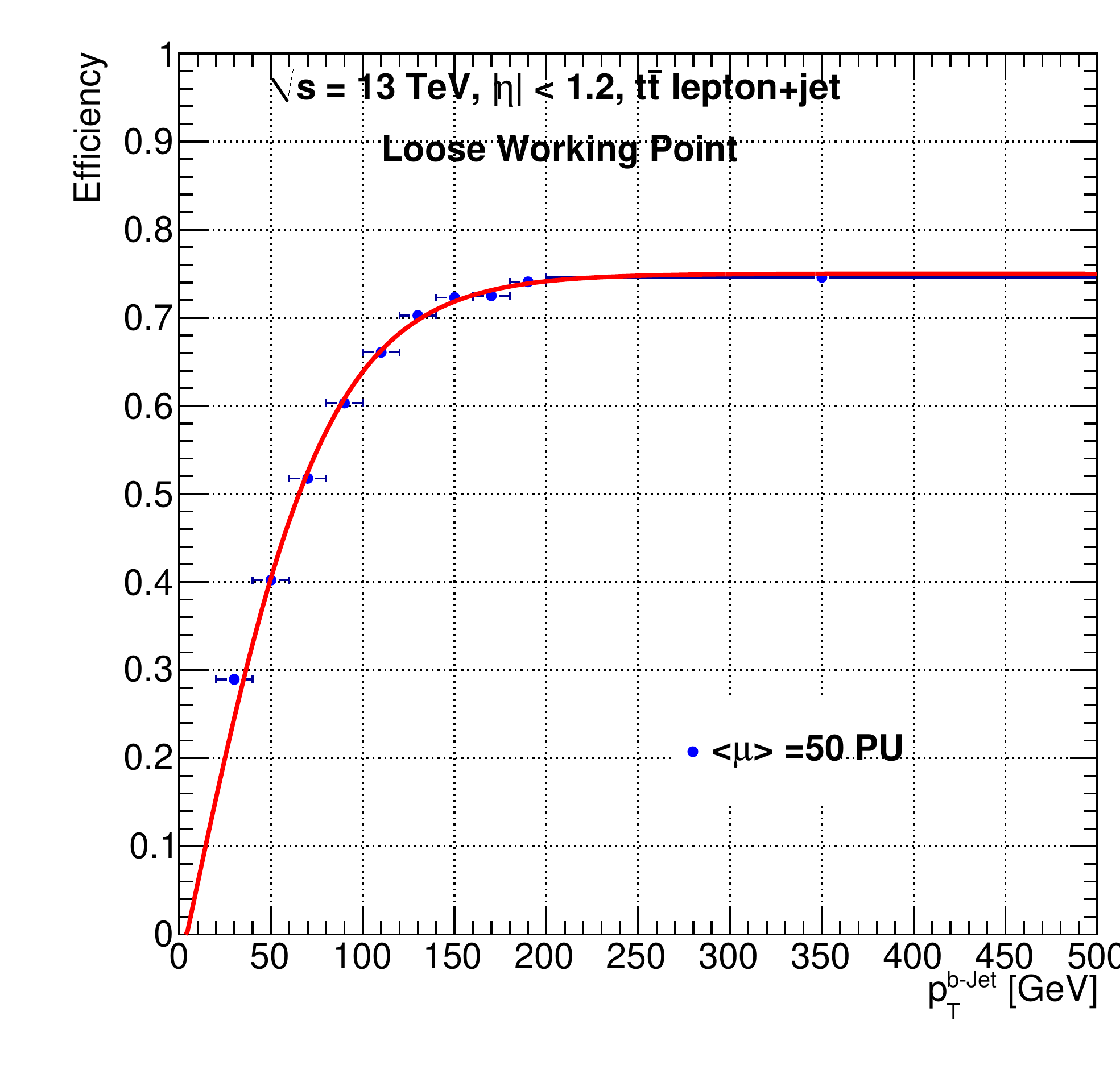}
\includegraphics[width=0.33\hsize]{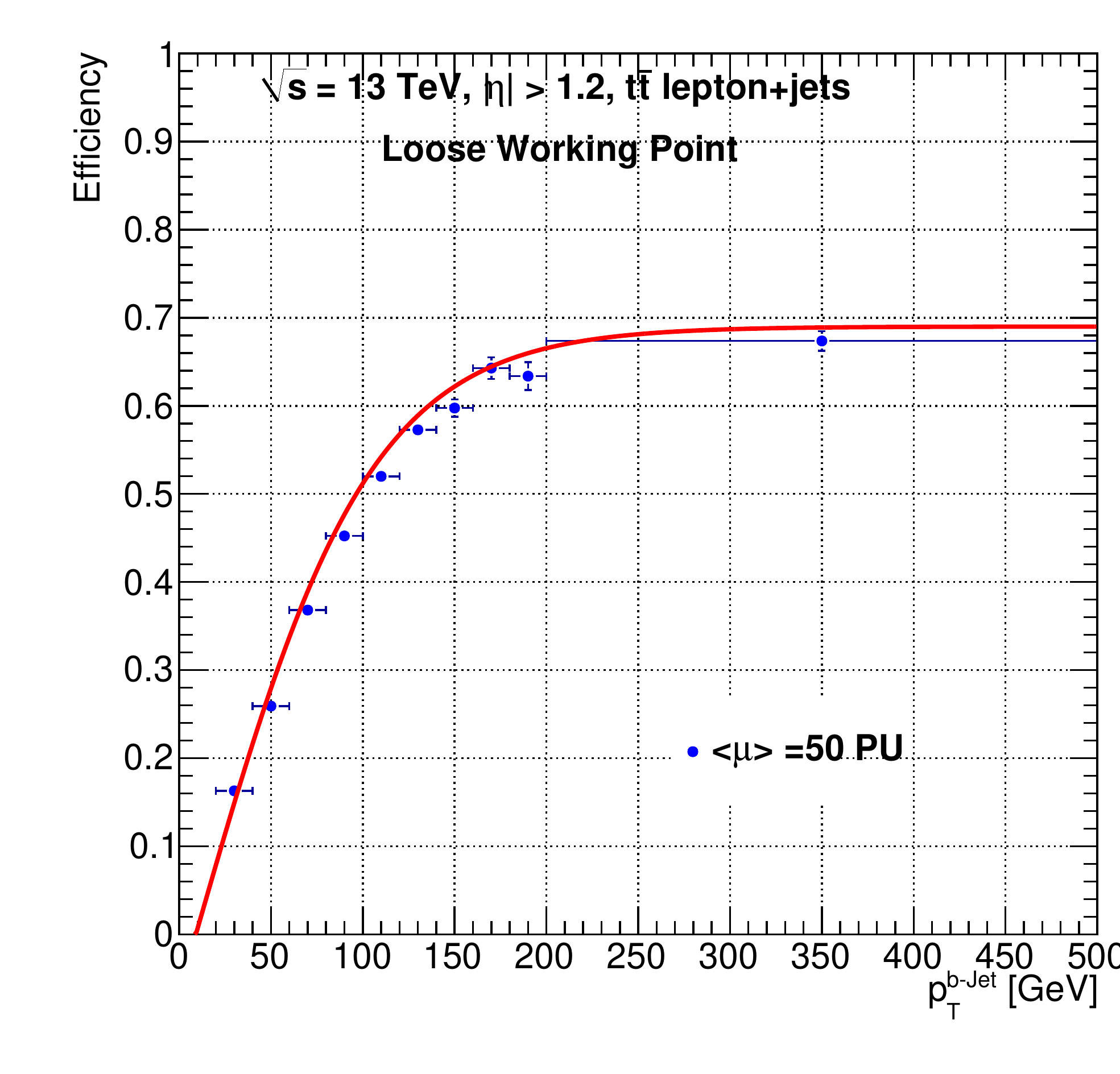}
\includegraphics[width=0.33\hsize]{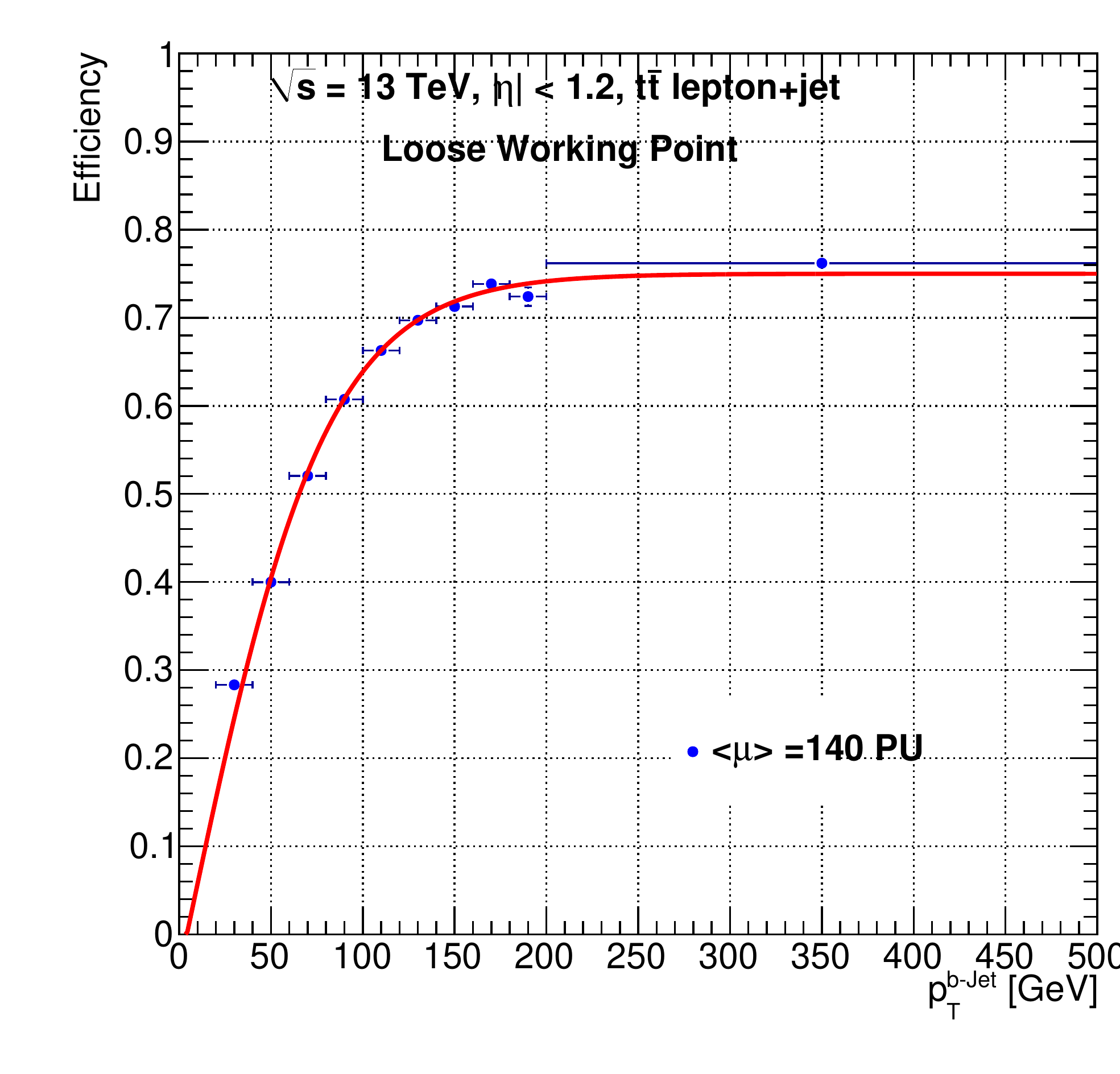} 
\includegraphics[width=0.33\hsize]{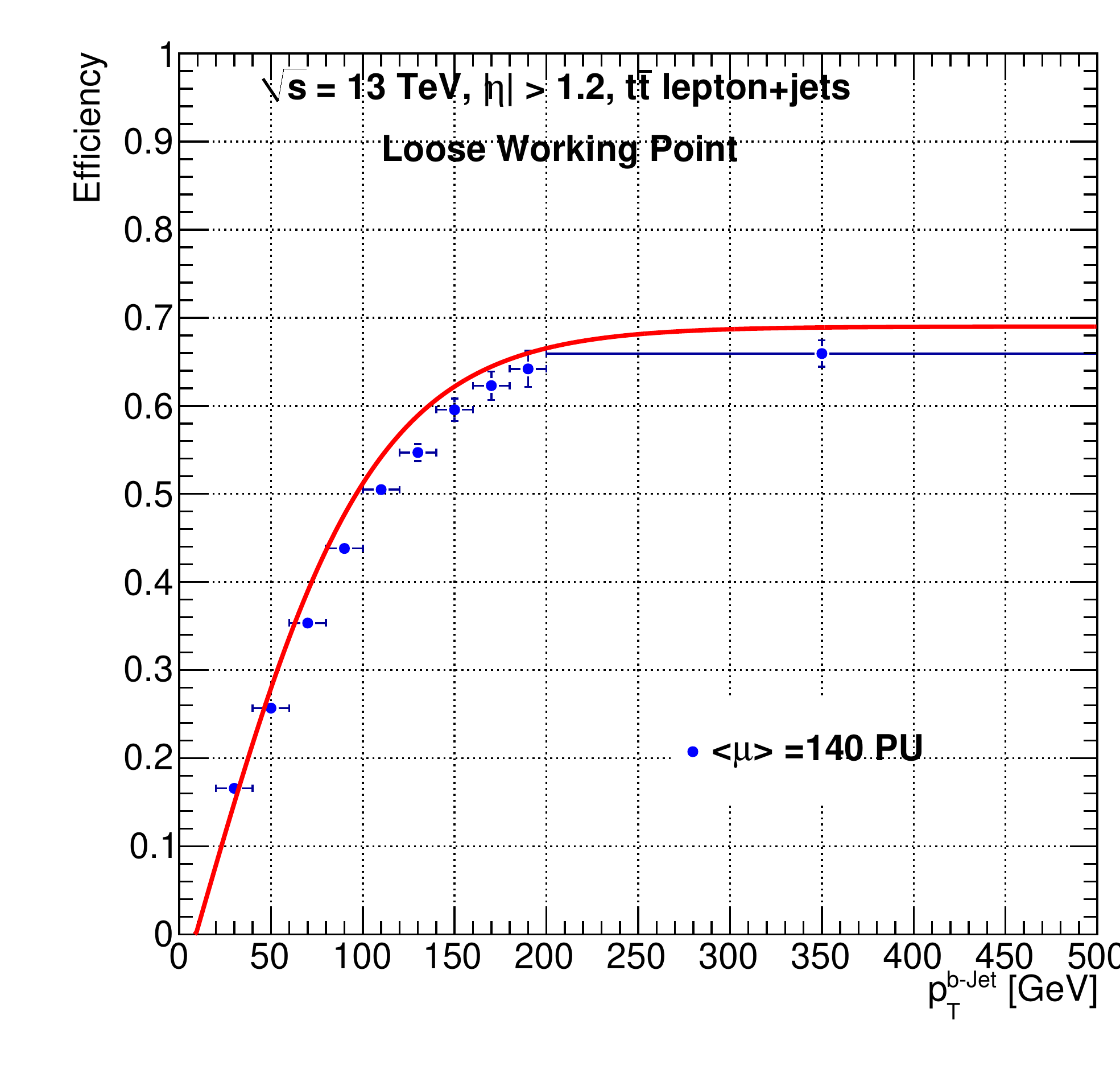} 
\caption{{\bf Loose working point}: Efficiency for correctly tagging true $b$ quark jets as a function of \pt\ for $\vert\eta\vert<1.2$ (left) and $\vert\eta\vert>1.2$ (right) and for 0 pile-up (top), 50 pile-up (middle), and 140 pile-up (bottom).  The curves are the input efficiency parametrization; the points are the measured efficiencies.}
\label{fig:btag_loose}
\end{center}
\end{figure}

\begin{figure}[hbtp]
\begin{center}
\includegraphics[width=0.35\hsize]{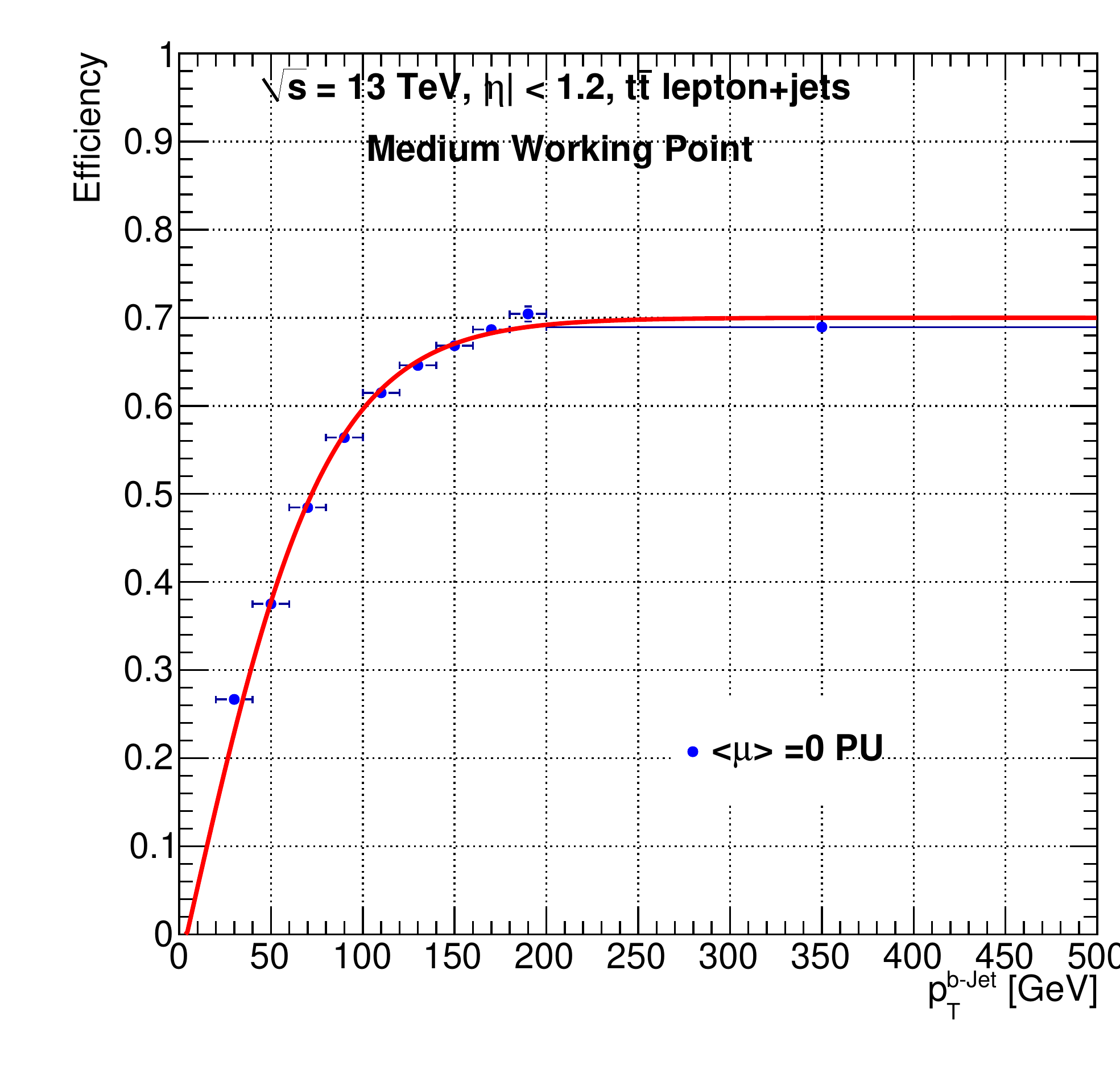}
\includegraphics[width=0.35\hsize]{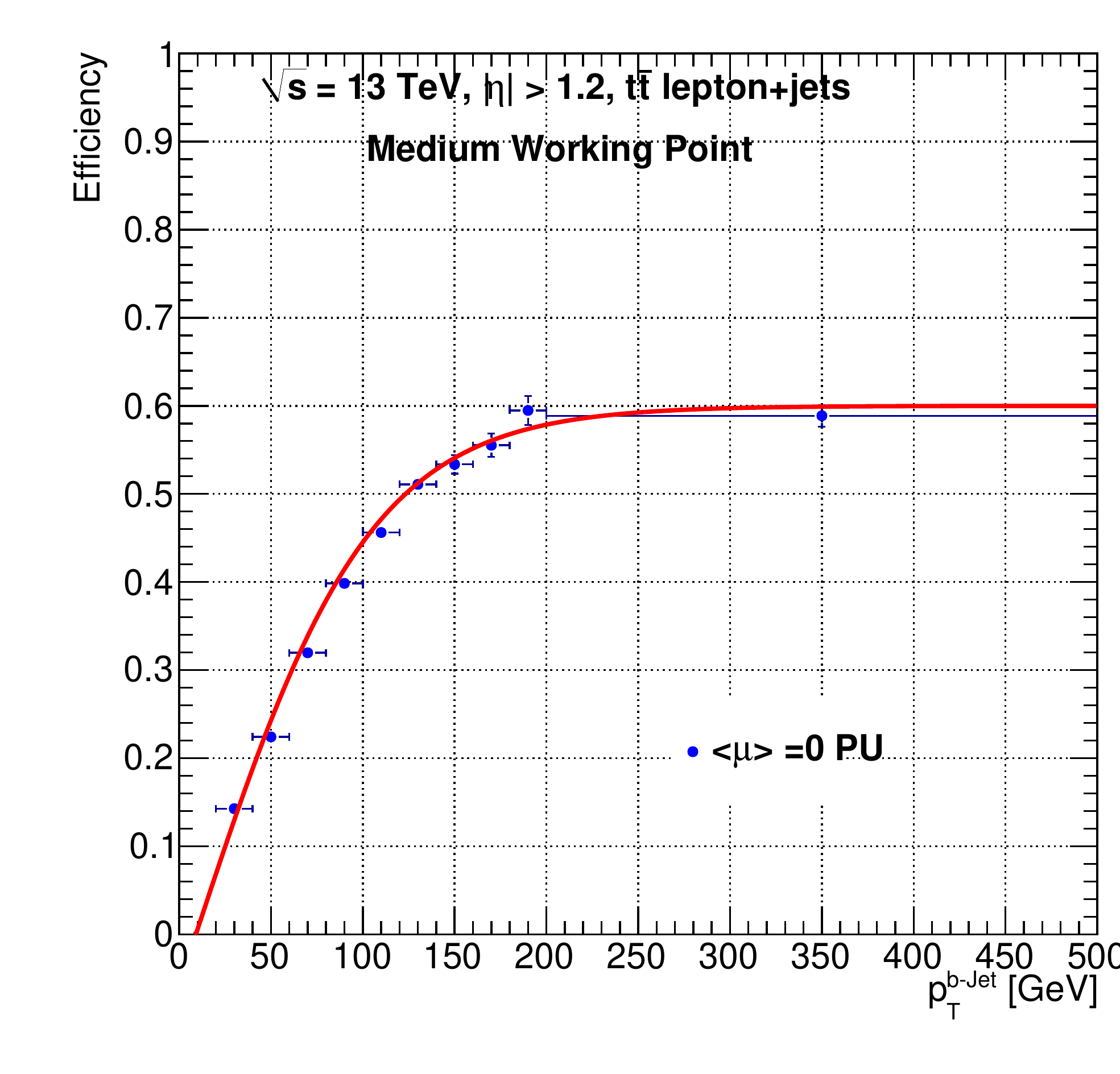}
\includegraphics[width=0.35\hsize]{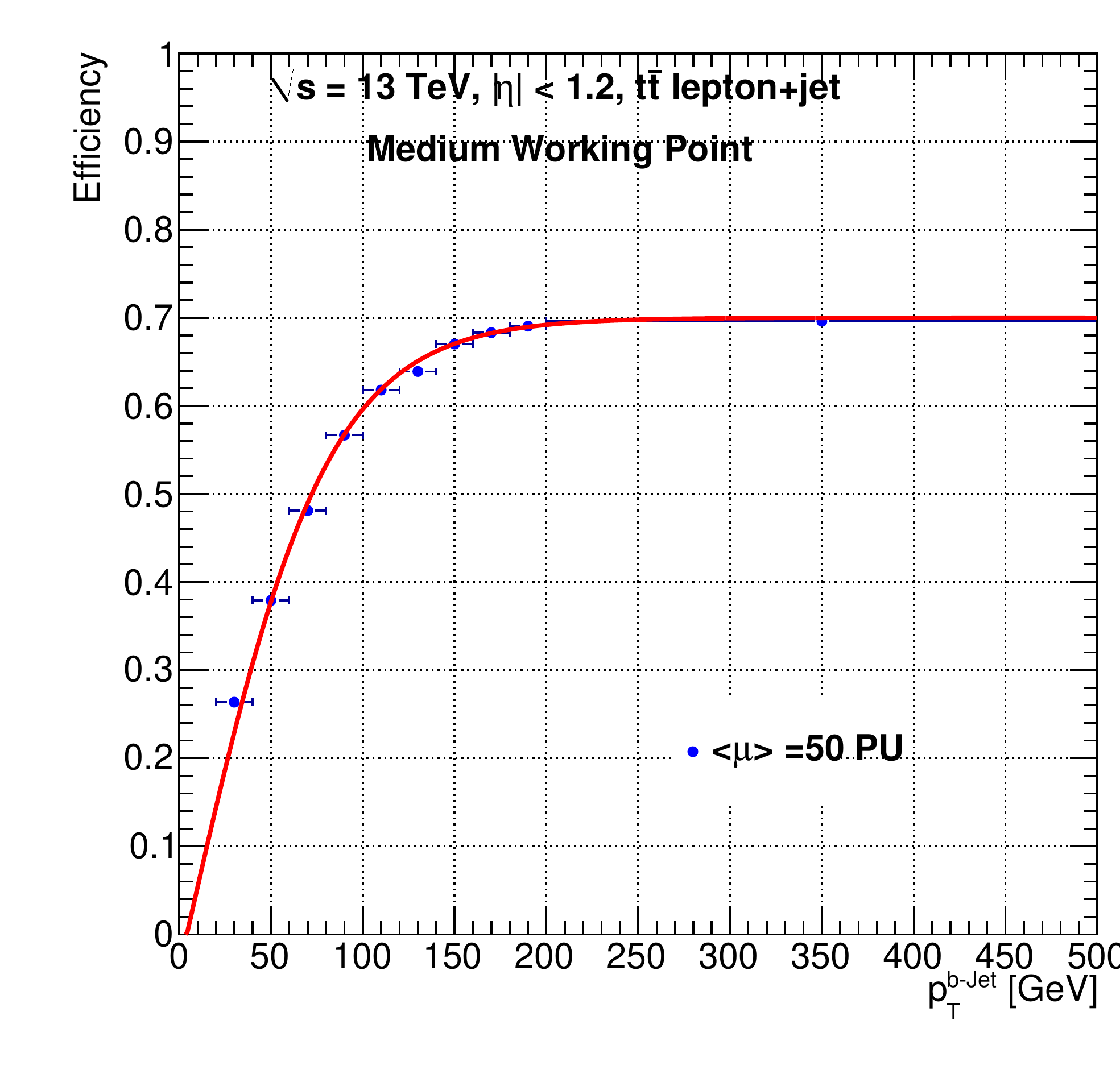}
\includegraphics[width=0.35\hsize]{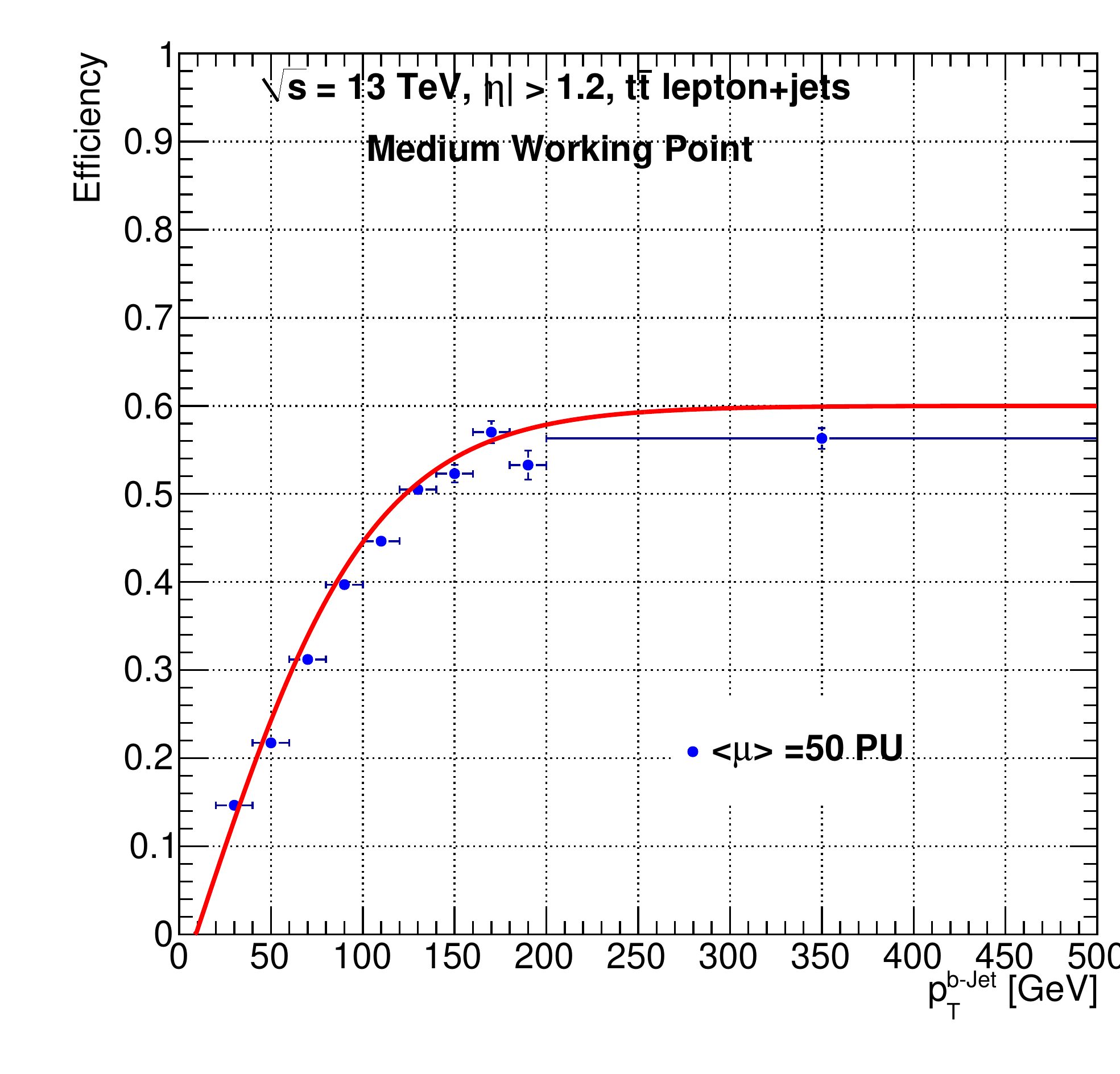}
\includegraphics[width=0.35\hsize]{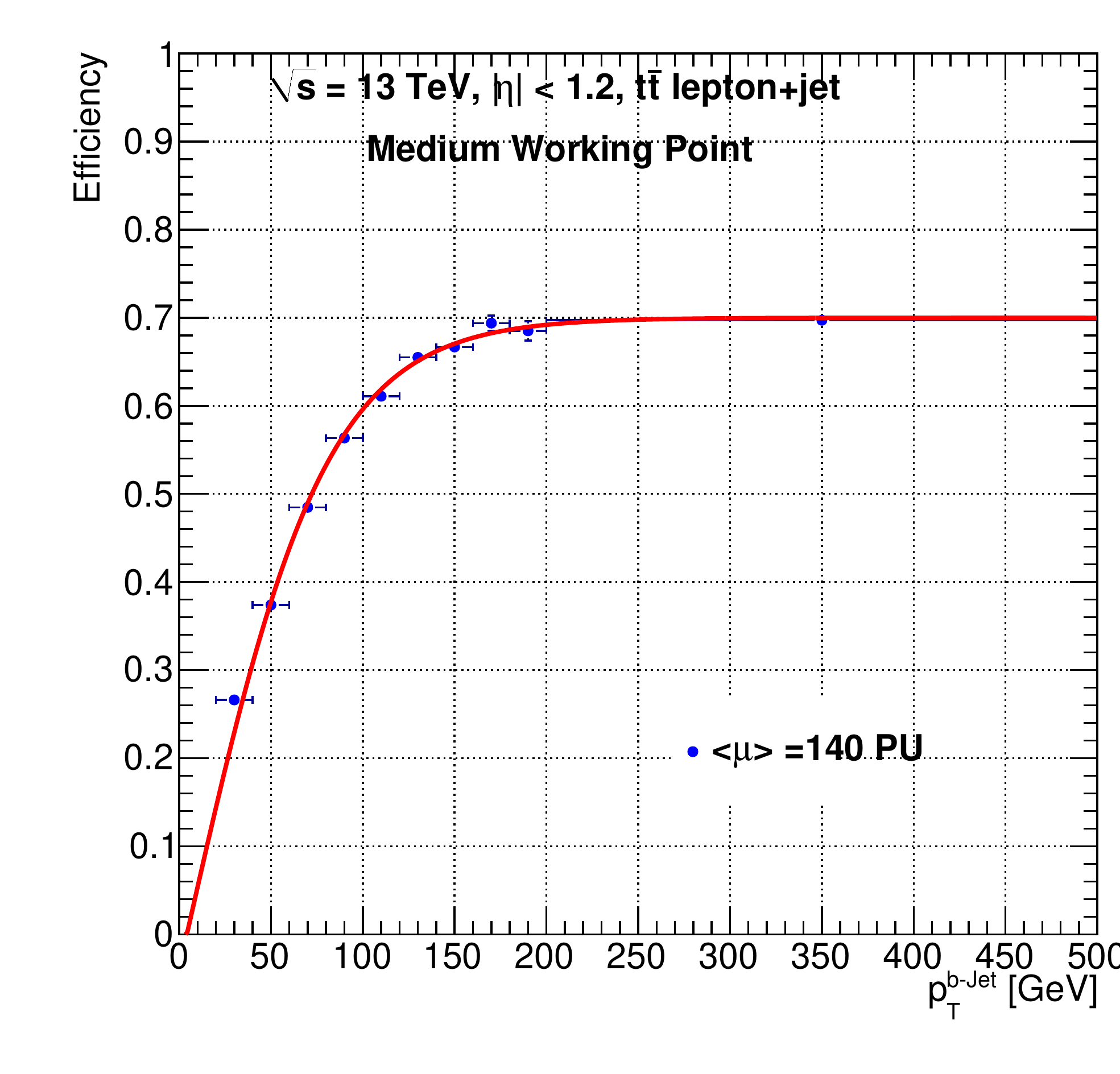}
\includegraphics[width=0.35\hsize]{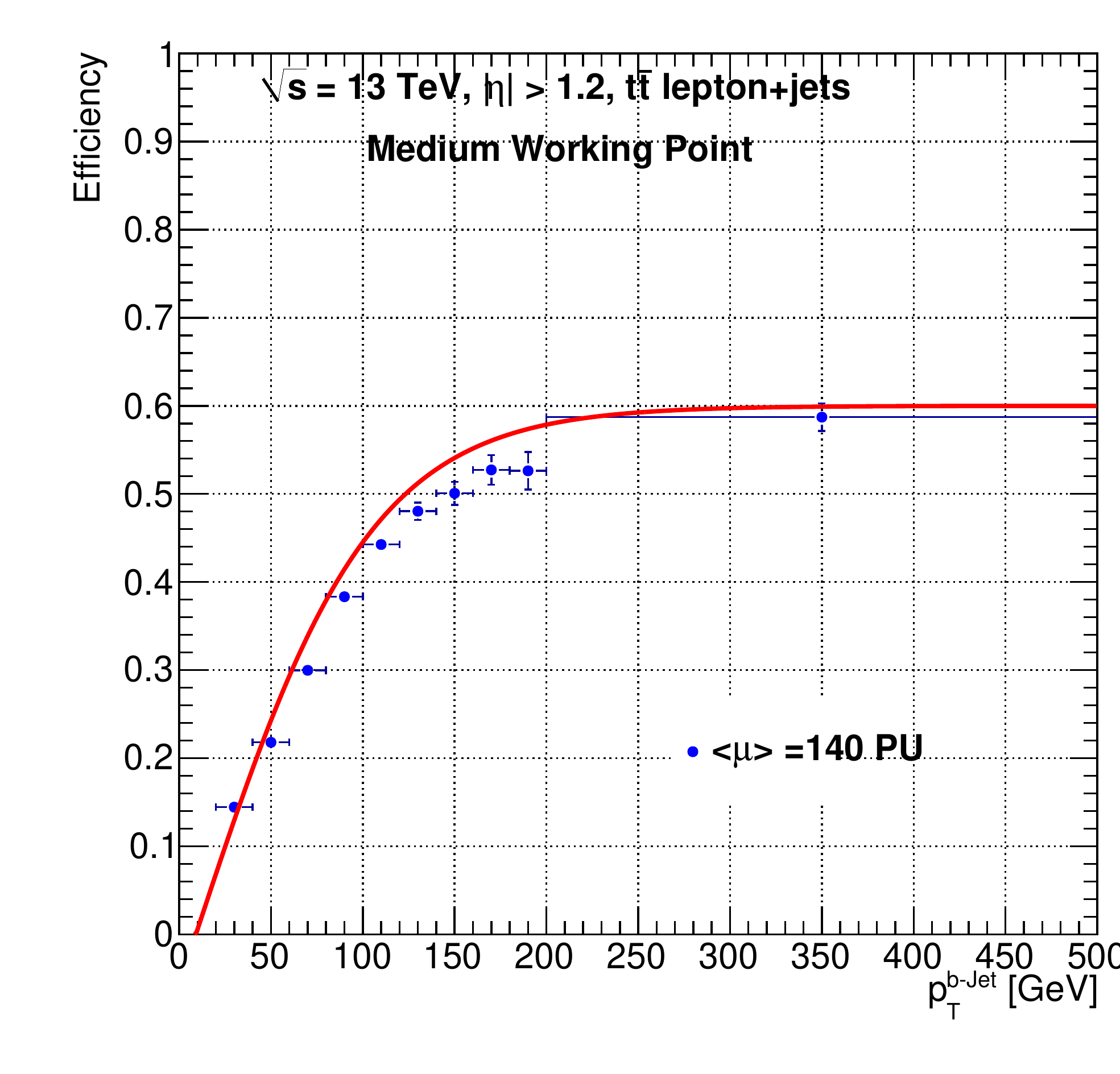}
\caption{{\bf Medium working point}: Efficiency for correctly tagging true $b$ quark jets as a function of \pt\ 
for $\vert\eta\vert<1.2$ (left) and $\vert\eta\vert>1.2$ (right) and for 0 pile-up (top), 50 pile-up (middle), and 
140 pile-up (bottom).  The curves are the input efficiency parametrization; the points are the measured efficiencies.}
\label{fig:btag_med}
\end{center}
\end{figure}

\begin{figure}[hbtp]
\begin{center}
\includegraphics[width=0.35\hsize]{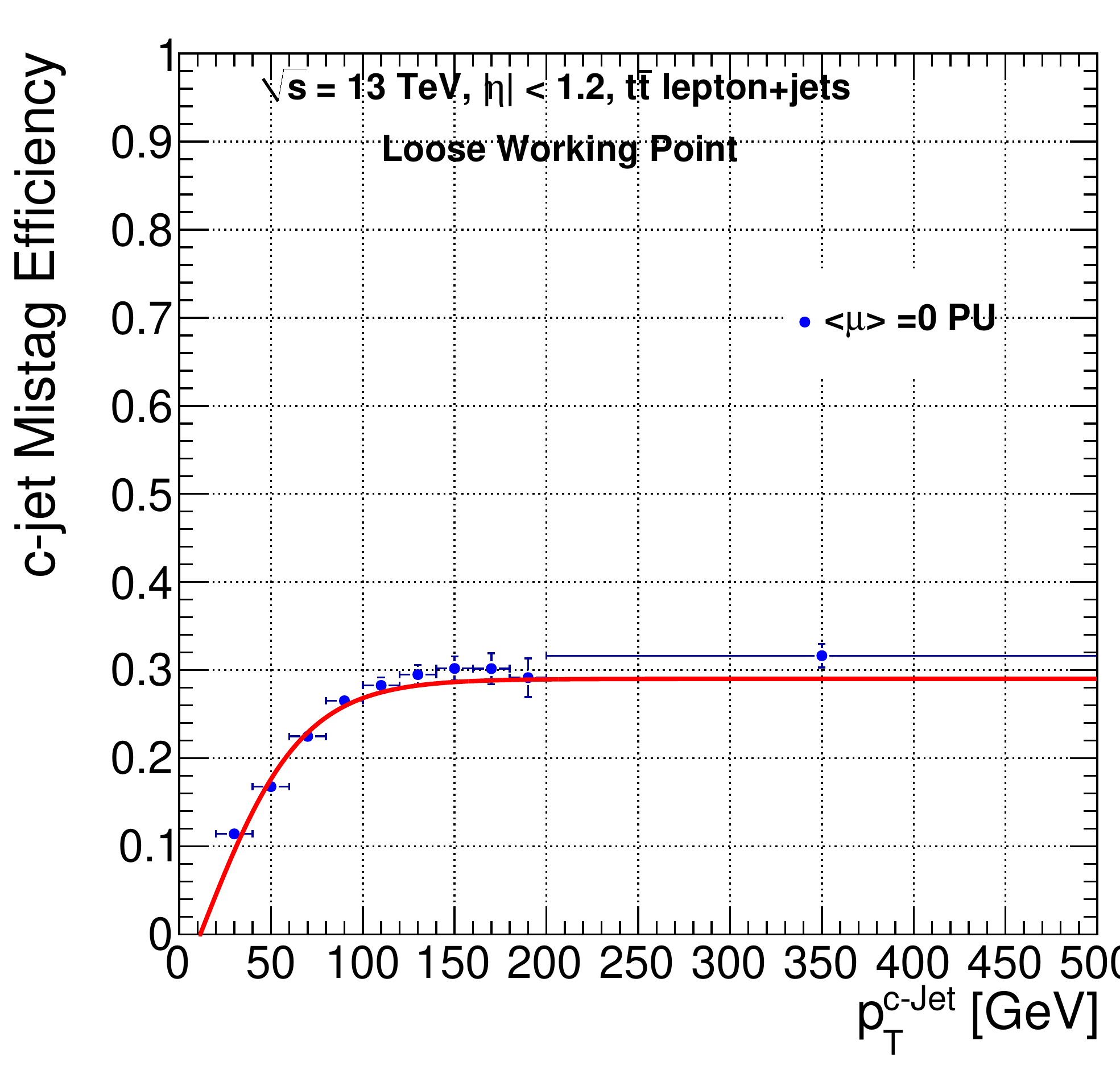}
\includegraphics[width=0.35\hsize]{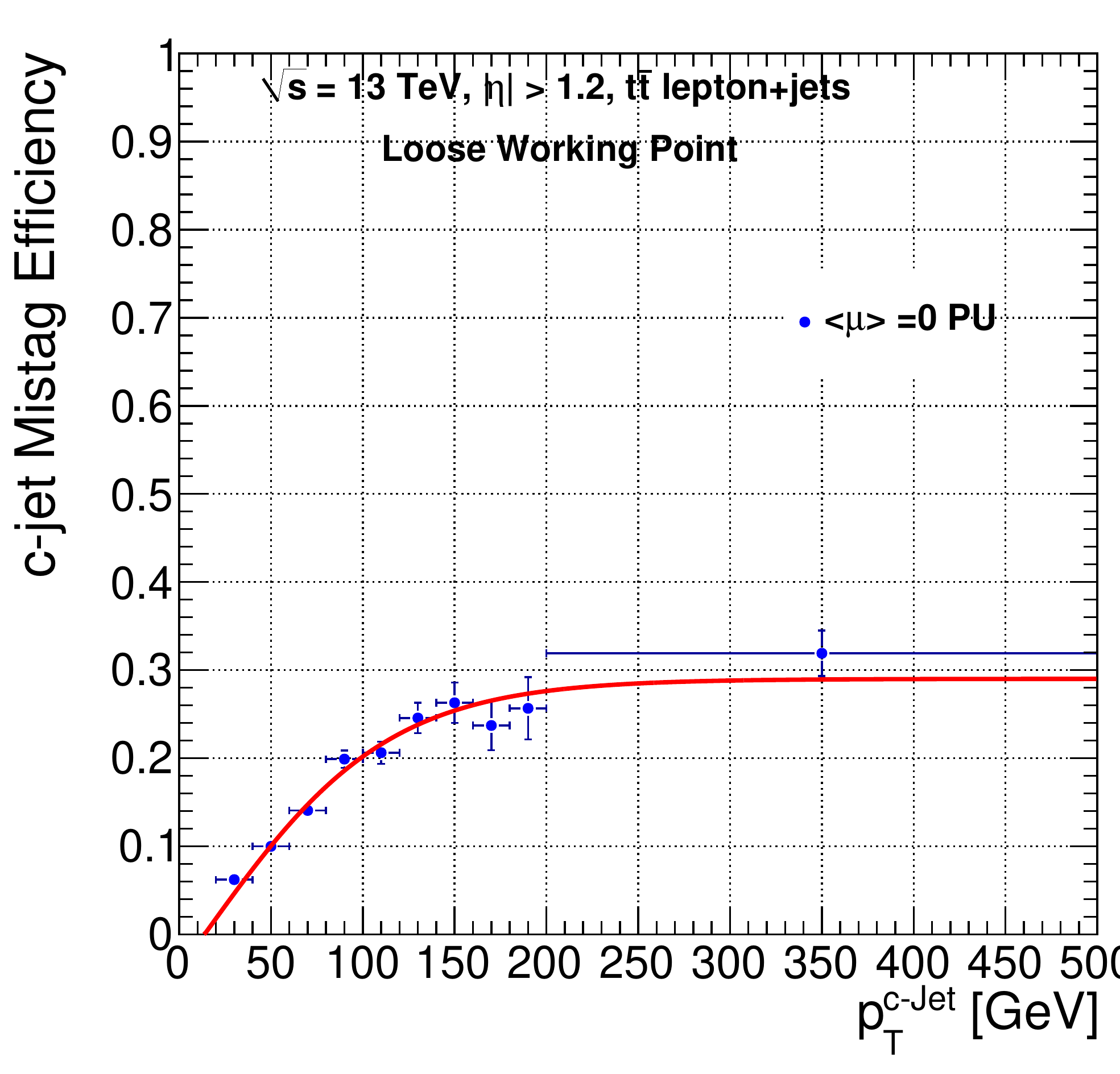}
\includegraphics[width=0.35\hsize]{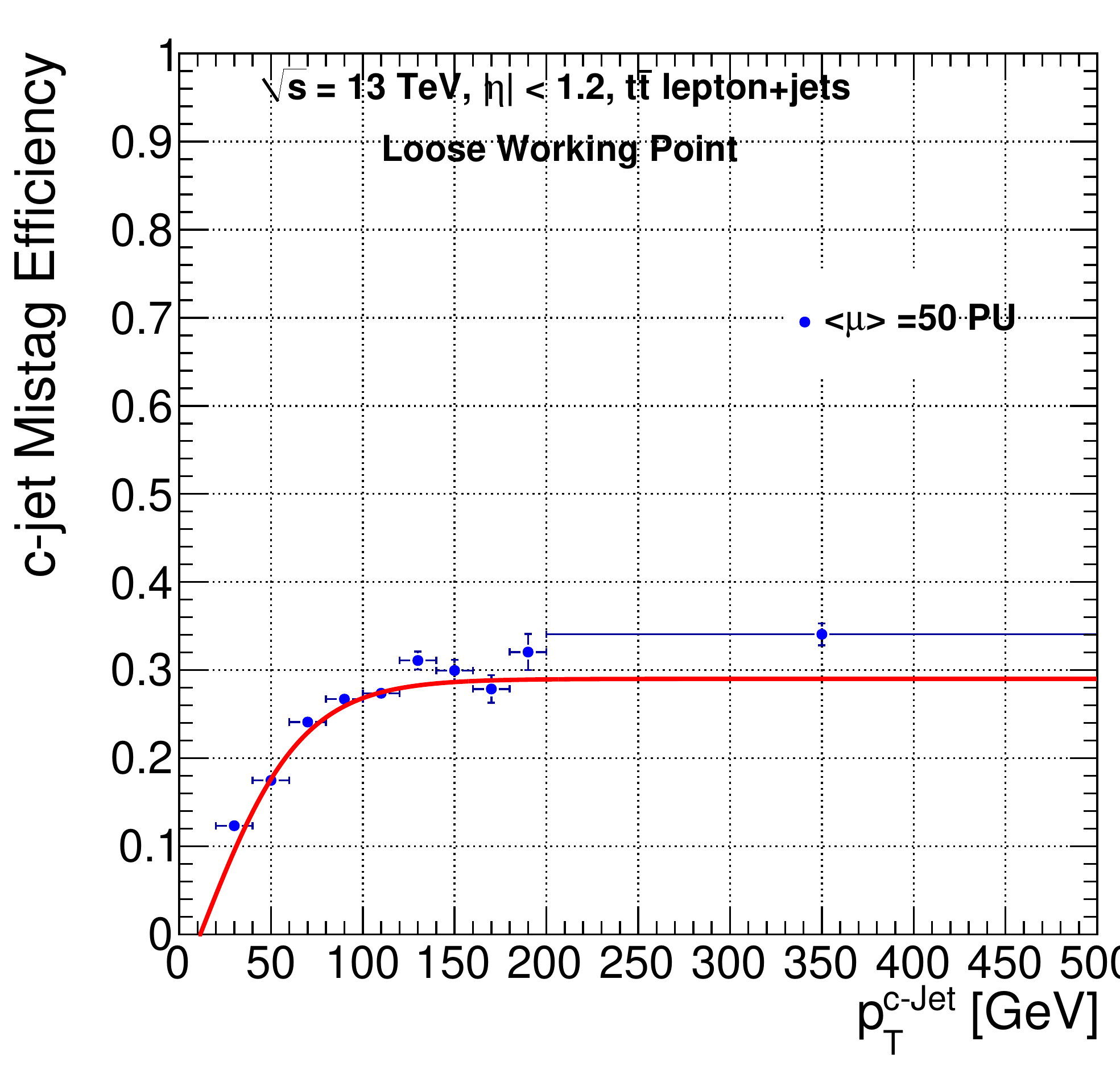}
\includegraphics[width=0.35\hsize]{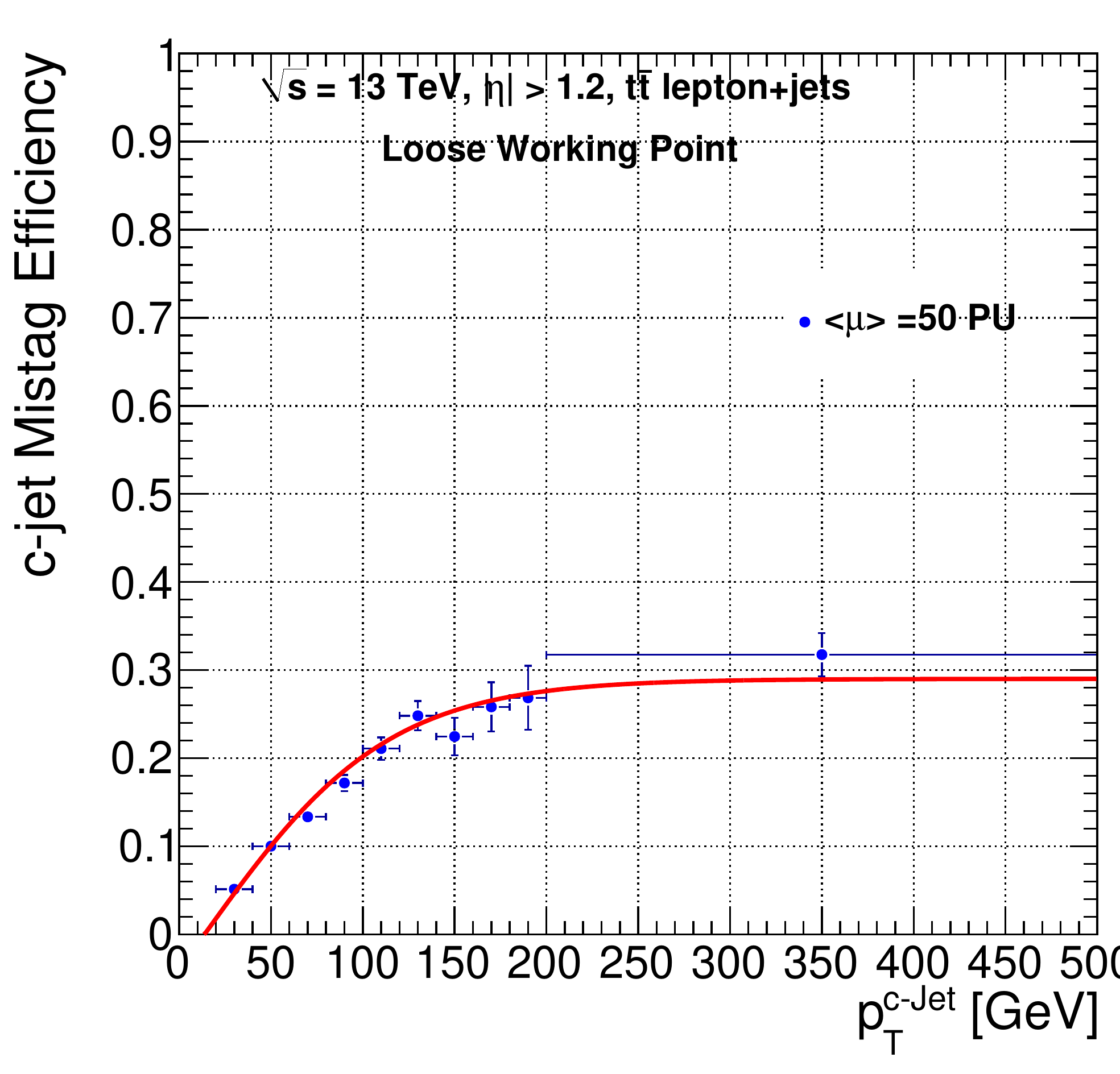}
\includegraphics[width=0.35\hsize]{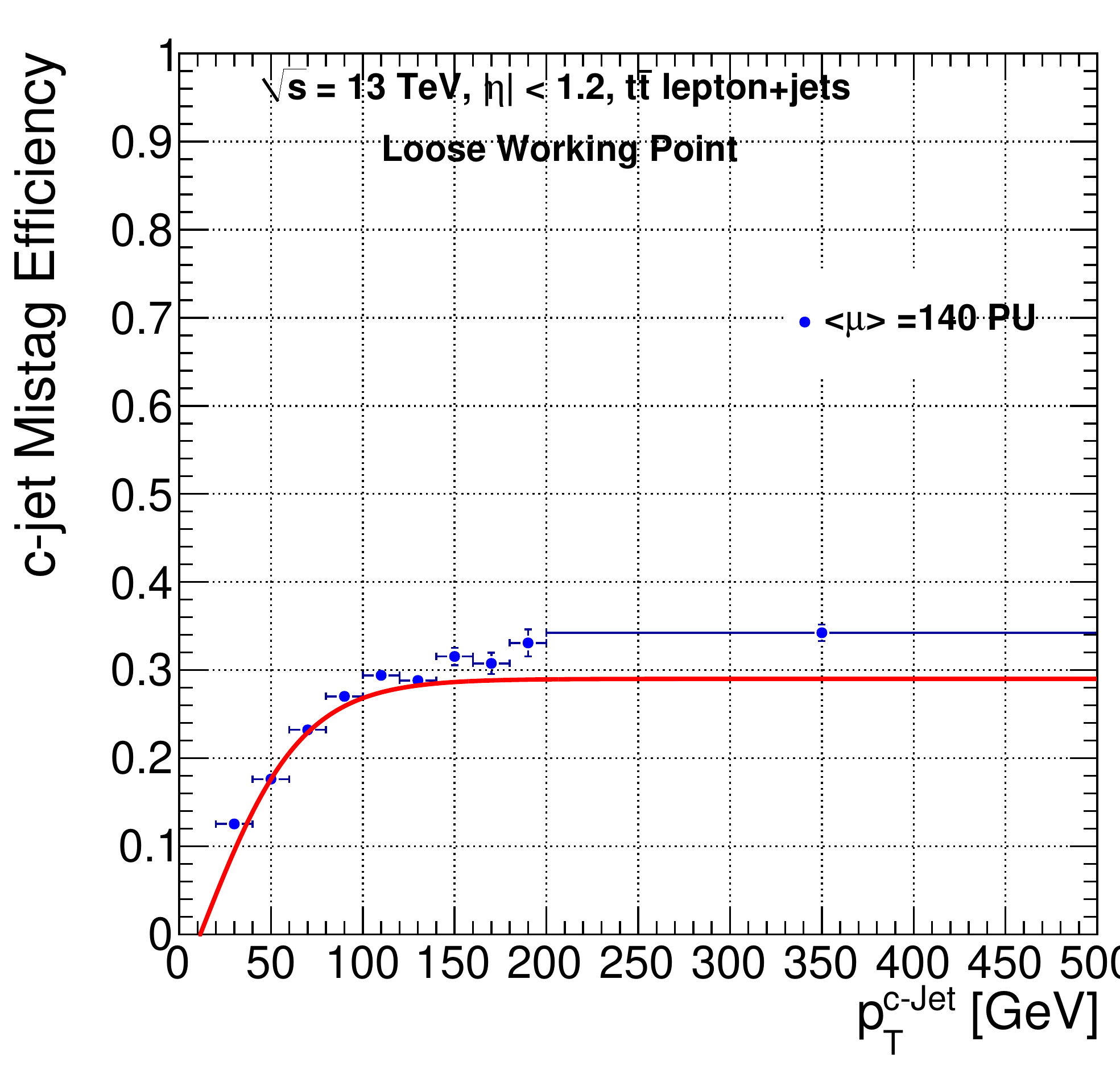}
\includegraphics[width=0.35\hsize]{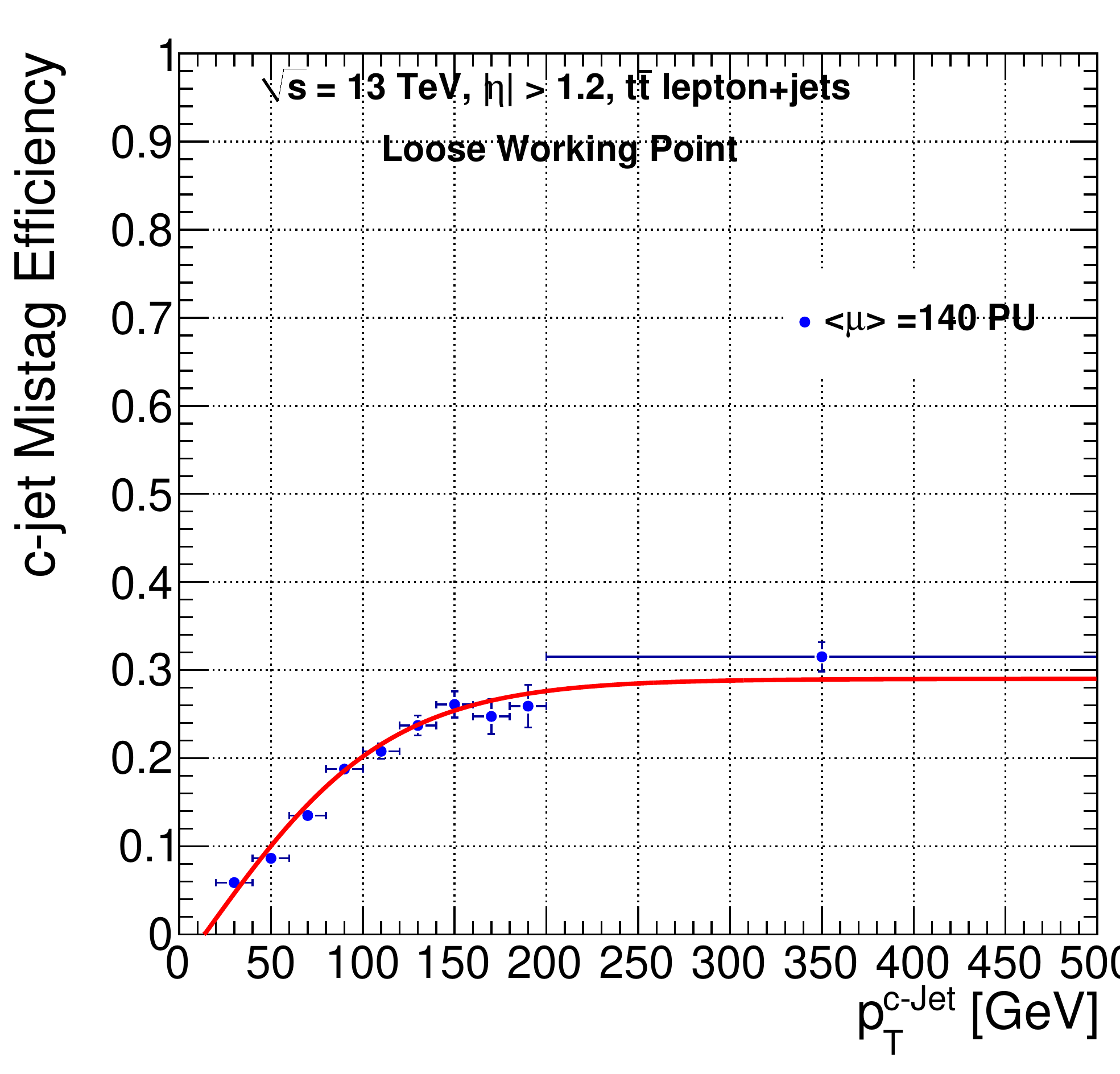}
\caption{{\bf Loose working point}: Rate to misidentify a true $c$ quark jet as a $b$ quark jet as a function of 
\pt\ for $\vert\eta\vert<1.2$ (left) and $\vert\eta\vert>1.2$ (right) and for 0 pile-up (top), 50 pile-up (middle), and 140 pile-up (bottom).  The curves are the input efficiency parametrization; the points are the measured efficiencies.}
\label{fig:bfake_loose}
\end{center}
\end{figure}

\begin{figure}[hbtp]
\begin{center}
\includegraphics[width=0.35\hsize]{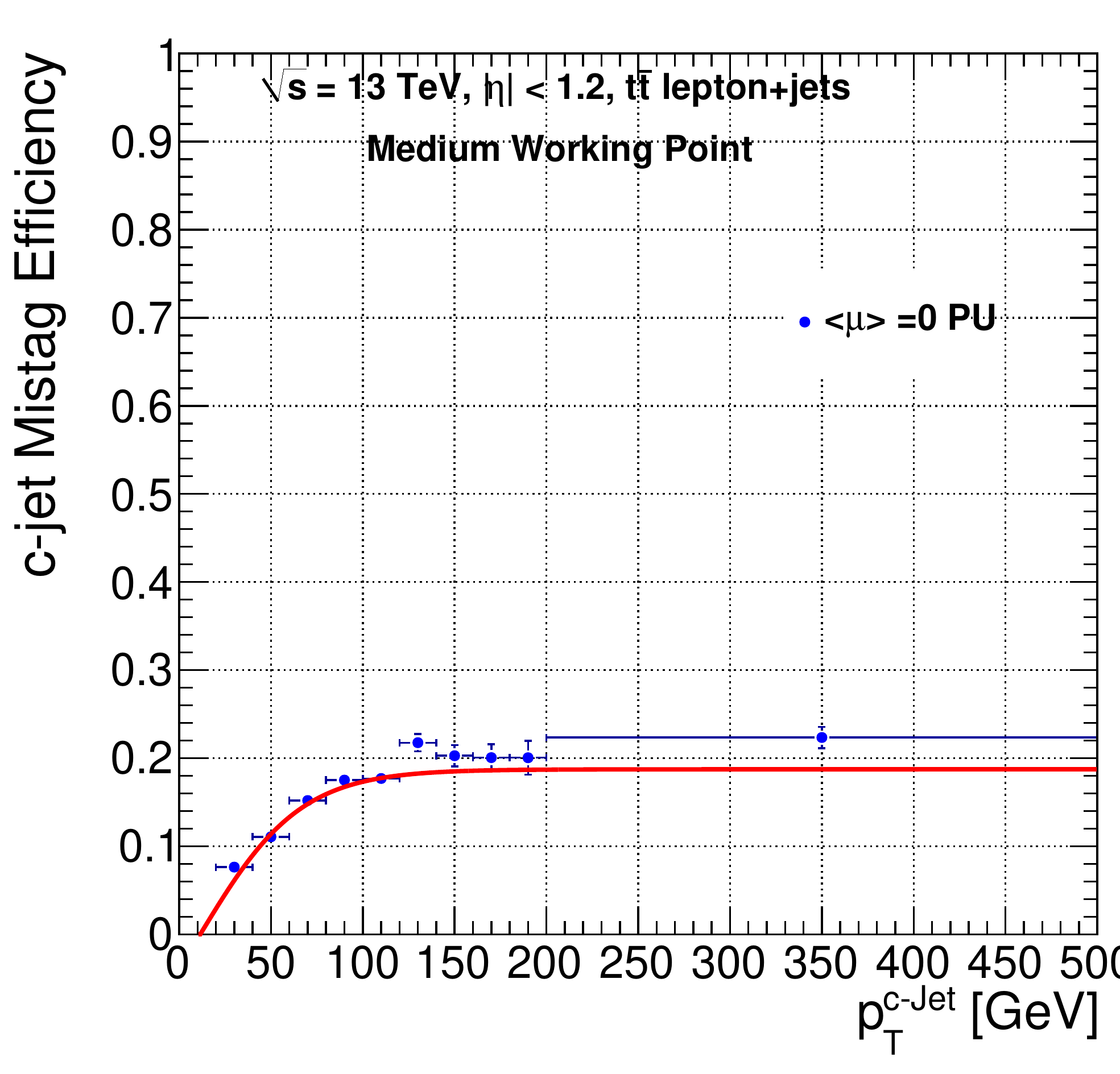}
\includegraphics[width=0.35\hsize]{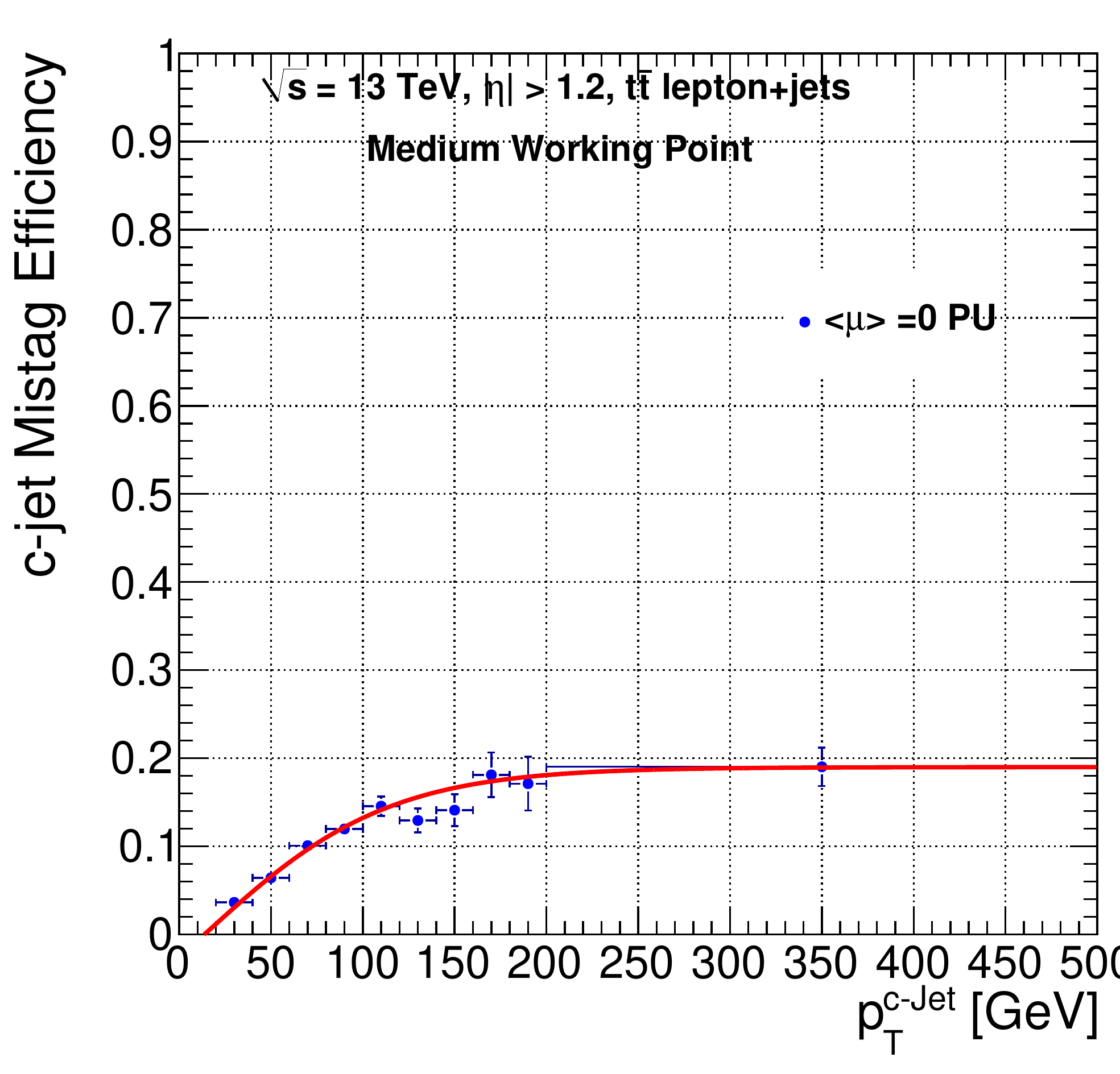}
\includegraphics[width=0.35\hsize]{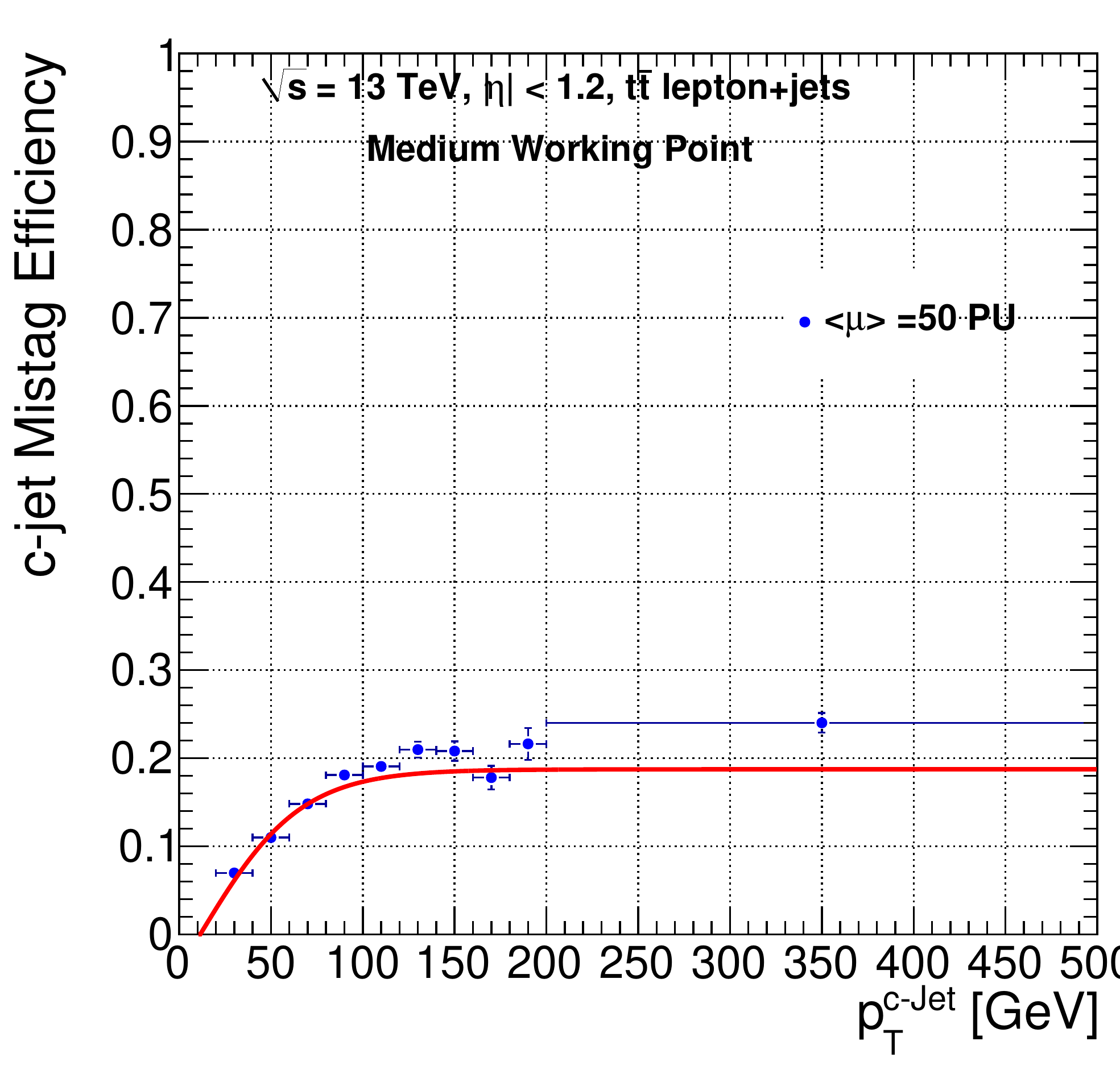}
\includegraphics[width=0.35\hsize]{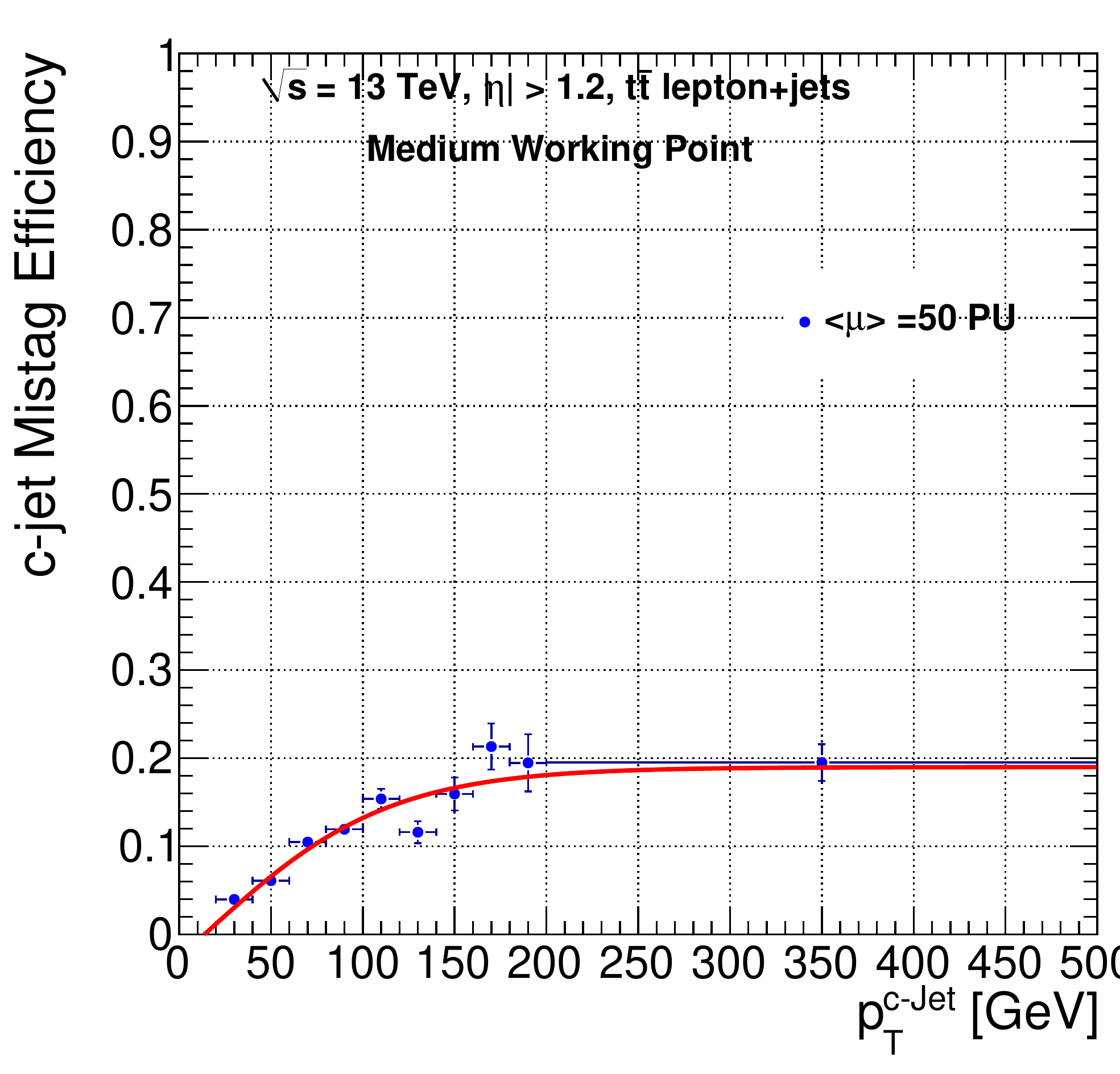}
\includegraphics[width=0.35\hsize]{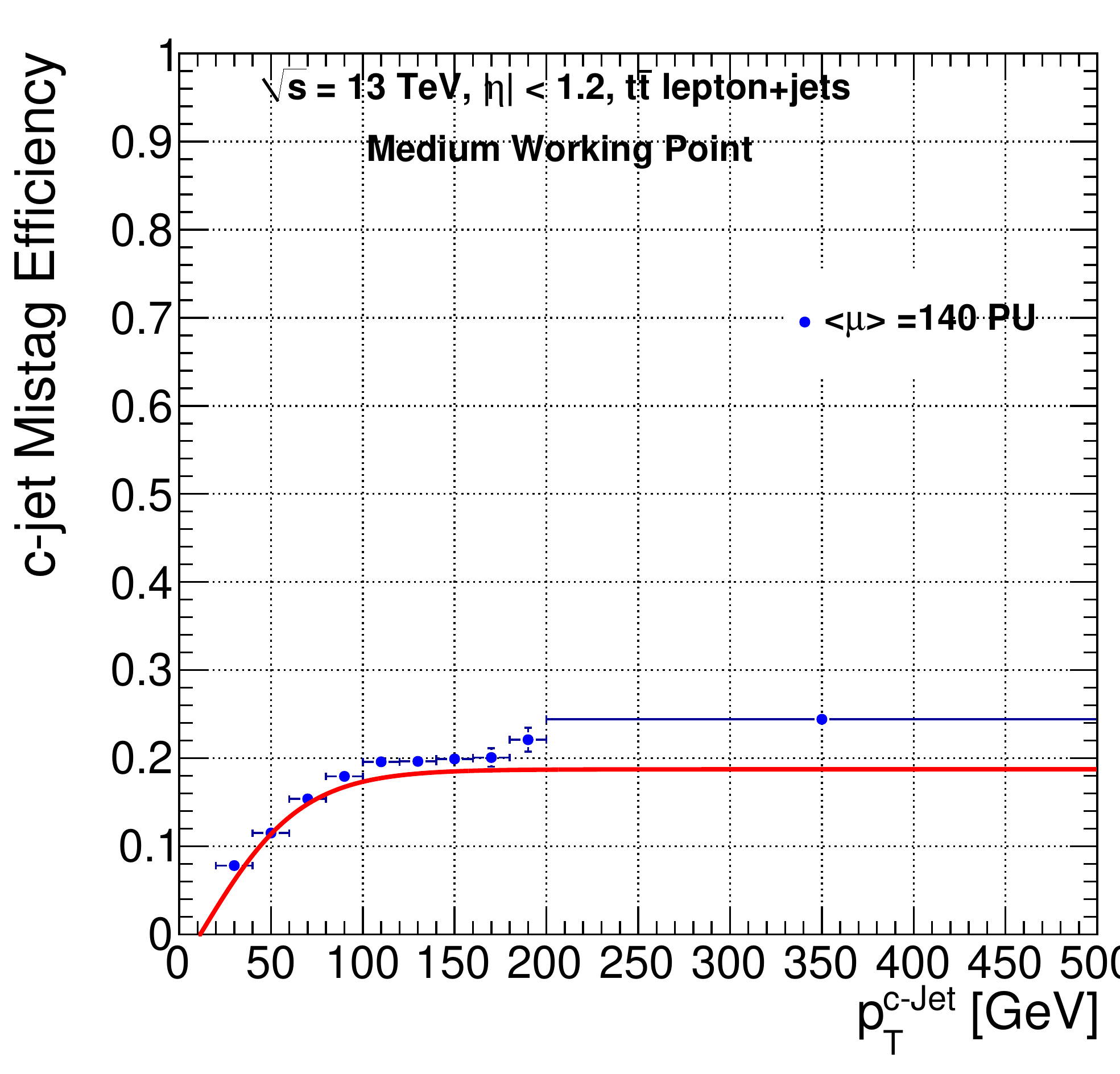}
\includegraphics[width=0.35\hsize]{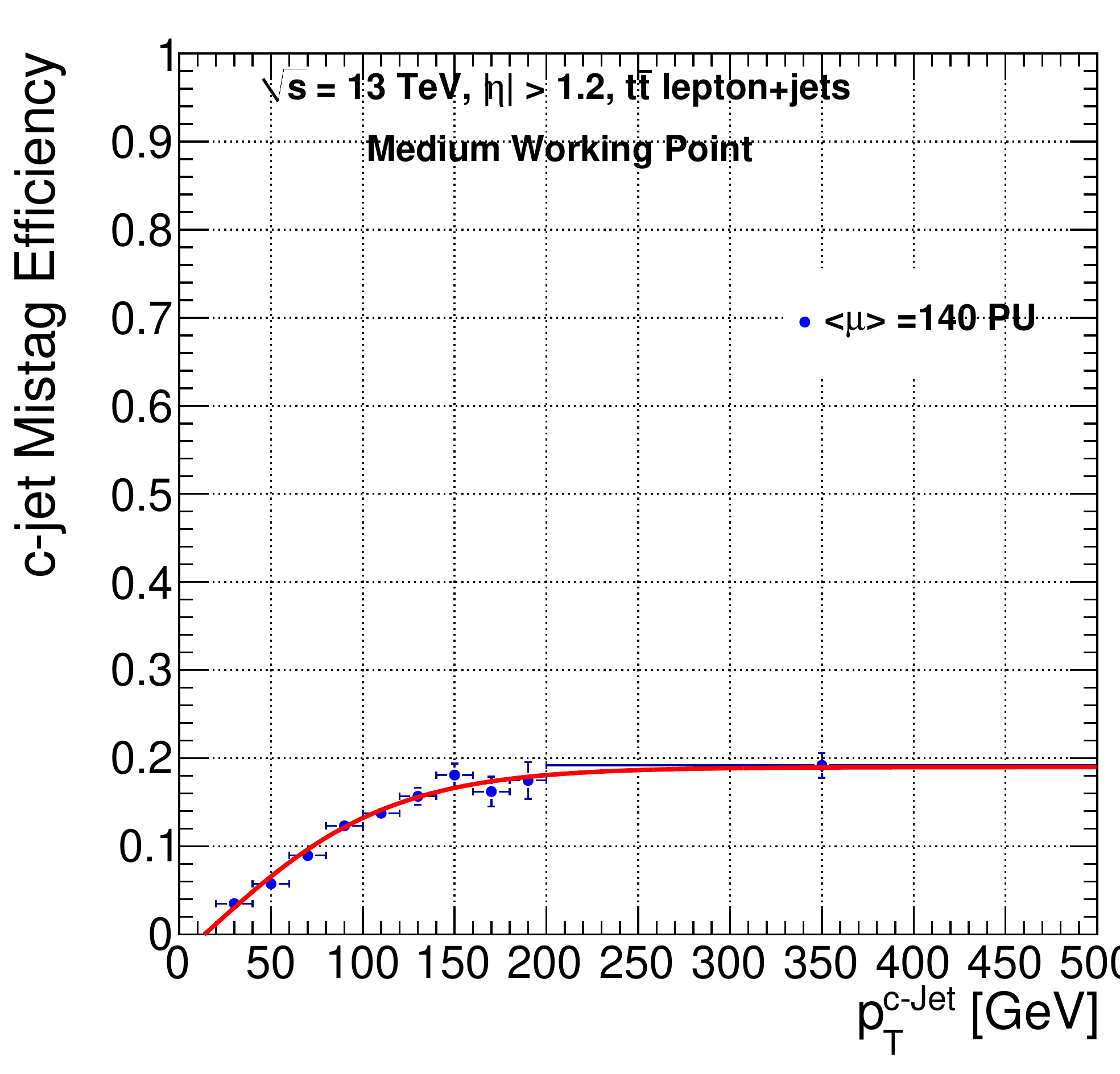}
\caption{{\bf Medium working point}: Rate to misidentify 
a true $c$ quark jet as a $b$ quark jet as a function of \pt\ for $\vert\eta\vert<1.2$ (left) and $\vert\eta\vert>1.2$ (right) and for 0 pile-up (top), 50 pile-up (middle), and 140 pile-up (bottom).  The curves are the input efficiency parametrization; the points are the measured efficiencies.}\label{fig:bfake_med}
\end{center}
\end{figure}

\begin{figure}[hbtp]
\begin{center}
\includegraphics[width=0.48\hsize]{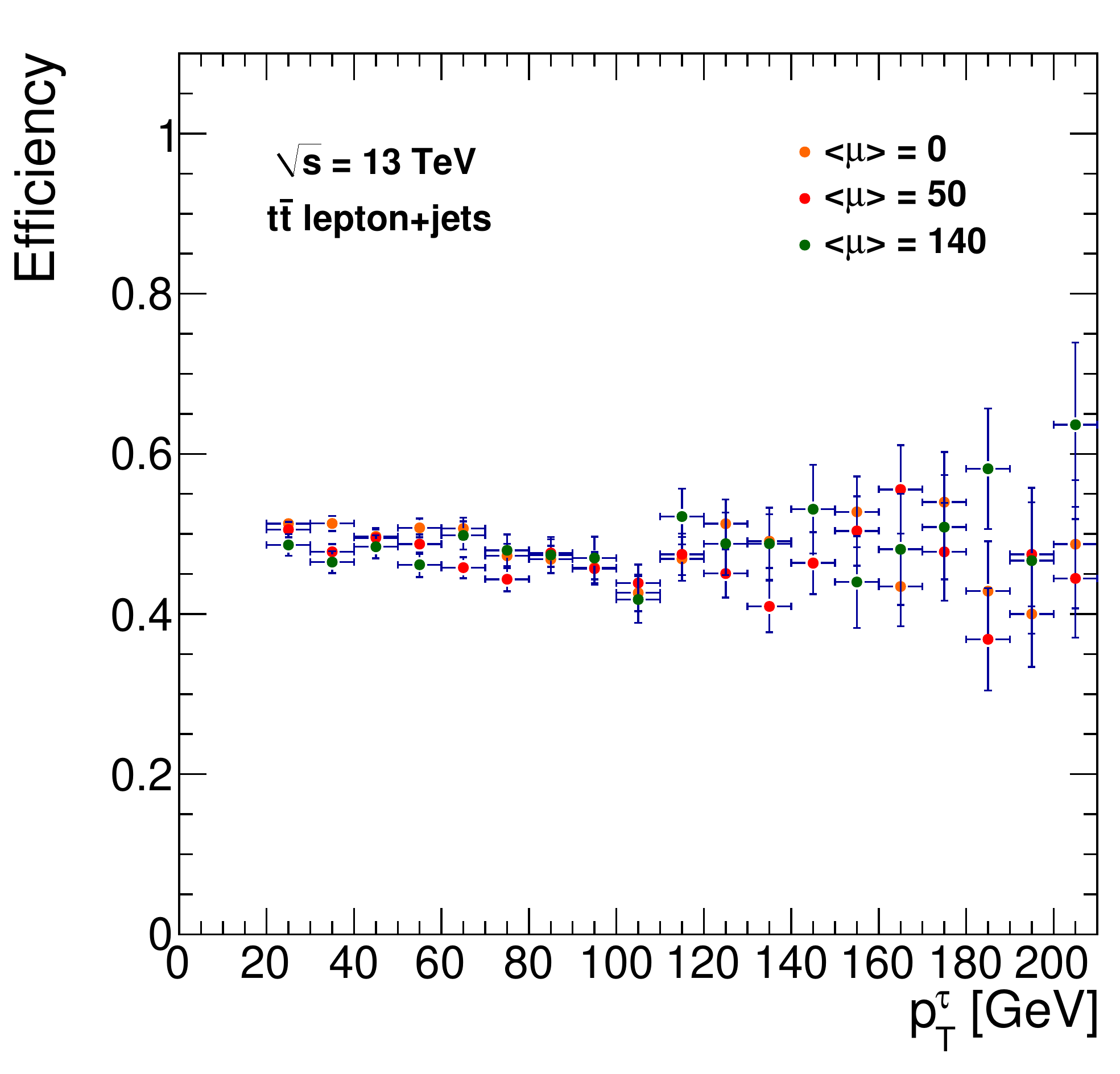}
\includegraphics[width=0.48\hsize]{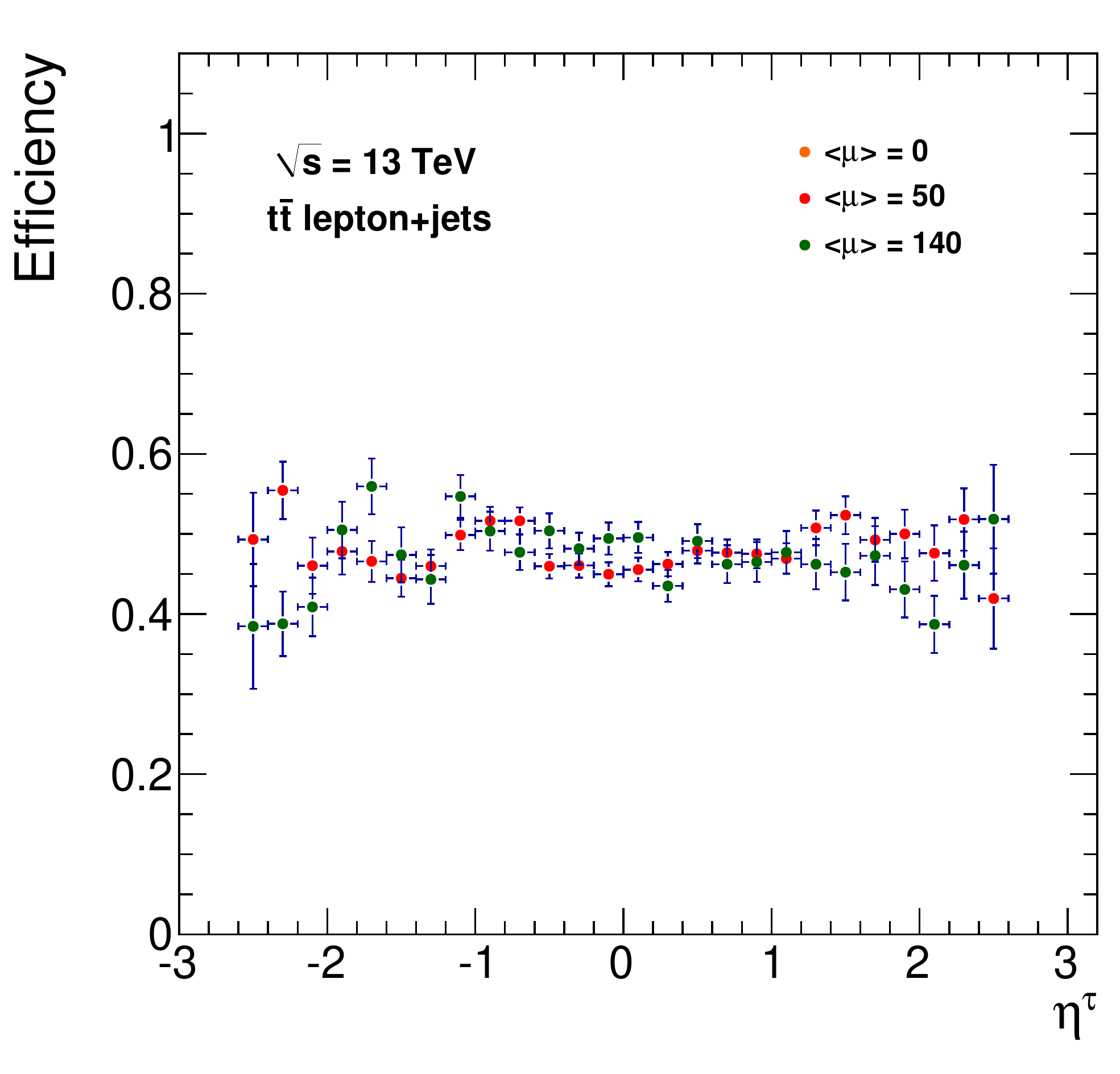}
\caption{Efficiency for correctly tagging true $\tau$ jets as a function \pt\ (left) and $\eta$ (right).}
\label{fig:tautag}
\end{center}
\end{figure}

The efficiencies to tag a true b-jet as originating from a $b$ quark or a true $\tau$-jet as originating 
from a $\tau$ lepton, along with the related rates to misidentify a light flavor jet as coming from a $b$ 
quark or $\tau$ lepton, are shown in Figs. \ref{fig:btag_loose}, \ref{fig:btag_med}, 
\ref{fig:bfake_loose}, \ref{fig:bfake_med}, and \ref{fig:tautag}. The measured efficiencies agree well with
the inputs used during parametrization.

\newpage
\section{Summary and status}
\label{sec:sim-stat}

This document describes the simulation framework and reconstructed object performance used in the Snowmass Energy 
Frontier studies for future Hadron Colliders. For the first time events with large pile-up associated with 
$pp$ interactions at center-of-mass energies $\sqrt{s}=$ 14, 33, and 100 TeV are studied. Input parametrization 
for tracks, clusters, as well as reconstructed objects such as leptons, photons, jets etc. are used based 
on publicly available detector and performance parameters. The study shows a significant  
impact on lepton, jets and \met~reconstruction due to higher pile-ups expected from the HL-LHC. 
Pile-up subtraction methods using particle flow for charged hadrons and jet 
area techniques will be vital for optimal object reconstructions needed for precision physics, as well
as searches for new physics at the LHC. Novel techniques are used to simulate large Standard Model 
backgrounds of 3 ab$^{-1}$ expected  due to luminosity evolution of the LHC.  
\Acknowledgements
We thank Beate Heinemann, Joseph Incandela and Michelangelo Manageo for comments and discussions, as well as the 
{\sc Delphes} team for support with the software and bug fixes. The studies are done using resources provided 
by the Open Science 
Grid, which is supported by the National Science 
Foundation and the U.S. Department of Energy's Office of Science.

%\input ChargedLeptons/wgreport.tex
%\input HeavyPhotons/wgreport.tex

%%%%%%%%%%%%%%%%%%%%%%%%%%%%%%%%%%%%%%%%%%%%%%%%%%
%%%%%%%%%%%%%%%%%%%%%%%%%%%%%%%%%%%%%%%%%%%%%%%%%%
%%%   Your subdirectory (here Magnetism) should include
%%%    the files:
%%%           wgreport.tex
%%%           authorlist.tex
%%%         and all needed figures in pdf format
%%%%%%%%%%%%%%%%%%%%%%%%%%%%%%%%%%%%%%%%%%%%%%%%%%%%
%%%%%%%%%%%%%%%%%%%%%%%%%%%%%%%%%%%%%%%%%%%%%%%%%%%%

\end{document}